\documentclass{aa520}

\usepackage{natbib}
\usepackage{graphicx}
\usepackage{txfonts}

\bibpunct[, ]{(}{)}{;}{a}{}{,}

\begin{document}

\bibliographystyle{aa}

\title{Non-spherical Core Collapse Supernovae}
\subtitle{I.~Neutrino-Driven Convection,
          Rayleigh-Taylor Instabilities,
          and the Formation and Propagation of Metal Clumps}

\author{
        K. Kifonidis\inst{1}  \and
        T. Plewa\inst{1,2,3}  \and
        H.-Th. Janka\inst{1}  \and
        E. M\"uller\inst{1}
       }
        
\offprints{K. Kifonidis} 
\mail{kok@mpa-garching.mpg.de}

\date{received; accepted}      
       
\institute{Max-Planck-Institut f\"ur Astrophysik,
           Karl-Schwarzschild-Stra{\ss}e 1, 
           D-85741 Garching, Germany \and
           Nicolaus Copernicus Astronomical Center,
           Bartycka 18, 00716 Warsaw, Poland \and
           Center for Astrophysical Thermonuclear Flashes,
           University of Chicago, 5640 S. Ellis Avenue, 
           Chicago, IL 60637, USA}

\abstract{We have performed two-dimensional simulations of core
          collapse supernovae that encompass shock revival by neutrino
          heating, neutrino-driven convection, explosive
          nucleosynthesis, the growth of Rayleigh-Taylor
          instabilities, and the propagation of newly formed metal
          clumps through the exploding star. A simulation of a Type~II
          explosion in a $15\,{\rm M_{\odot}}$ blue supergiant
          progenitor is presented, that confirms our earlier Type~II
          models and extends their validity to times as late as 5.5
          hours after core bounce. We also study a Type~Ib-like
          explosion, by simply removing the hydrogen envelope of the
          progenitor model. This allows for a first comparison of
          Type~II and Type~Ib evolution. We present evidence that the
          hydrodynamics of core collapse supernovae beyond shock
          revival differs markedly from the results of simulations
          that have followed the Rayleigh-Taylor mixing starting from
          ad hoc energy deposition schemes to initiate the
          explosion. We find iron group elements to be synthesized in
          an anisotropic, dense, low-entropy shell that expands with
          velocities of $\sim 17\,000$\,km/s shortly after shock
          revival.  The growth of Rayleigh-Taylor instabilities at the
          Si/O and (C+O)/He composition interfaces of the progenitor,
          seeded by the flow-structures resulting from neutrino-driven
          convection, leads to a fragmentation of this shell into
          metal-rich ``clumps''. This fragmentation starts already
          $\sim 20$\,s after core bounce and is complete within the
          first few minutes of the explosion. During this time the
          clumps are slowed down by drag, and by the positive pressure
          gradient in the unstable layers. However, at $t \approx
          300$\,s they decouple from the flow and start to propagate
          ballistically and subsonically through the He core, with the
          maximum velocities of metals remaining constant at $\sim
          3500 - 5500$\,km/s. This early ``clump decoupling'' leads to
          significantly higher $\rm{^{56}Ni}$ velocities at $t =
          300$\,s than in one-dimensional models of the explosion,
          demonstrating that multi-dimensional effects which are at
          work within the first minutes, and which have been neglected
          in previous studies (especially in those which dealt with
          the mixing in Type~II supernovae), are crucial.  Despite
          comparably high initial maximum nickel velocities in both
          our Type~II and our Type~Ib-like model, we find that there
          are large differences in the final maximum nickel velocities
          between both cases.  In the ``Type~Ib'' model the maximum
          velocities of metals remain frozen in at $\sim 3500 -
          5500$\,km/s for $t \geq 300$\,s, while in the Type~II model
          they drop significantly for $t > 1500$\,s. In the latter
          case, the massive hydrogen envelope of the progenitor forces
          the supernova shock to slow down strongly, leaving behind a
          reverse shock and a dense helium shell (or ``wall'') below
          the He/H interface. After penetrating into this dense
          material the metal-rich clumps possess supersonic speeds,
          before they are slowed down by drag forces to $\sim
          1200$\,km/s at a time of 20\,000\,s post-bounce. While, due
          to this deceleration, the maximum velocities of iron-group
          elements in SN 1987\,A cannot be reproduced in case of the
          considered $15\,{\rm M_{\odot}}$ progenitor, the ``Type~Ib''
          model is in fairly good agreement with observed clump
          velocities and the amount of mixing inferred for Type~Ib
          supernovae. Thus it appears promising for calculations of
          synthetic spectra and light curves. Furthermore, our
          simulations indicate that for Type~Ib explosions the pattern
          of clump formation in the ejecta is correlated with the
          structure of the convective pattern prevailing during the
          shock-revival phase. This might be used to deduce
          observational constraints for the dynamics during this early
          phase of the evolution, and the role of neutrino heating in
          initiating the explosion.}

\maketitle  

\keywords{hydrodynamics -- instabilities -- nucleosynthesis --
         shock waves -- stars: supernovae}

\section{Introduction}

The strong shock wave that has torn apart Sk\,$-69^{\circ}\,202$ and
in consequence has given rise to Supernova SN 1987\,A in the Large
Magellanic Cloud on February 23rd, 1987, has also shattered the at
that time widely shared belief that supernova explosions are
essentially spherically symmetric events. The avalanche of
observational data obtained since then from SN 1987\,A has
unambiguously demonstrated that the envelope of the progenitor star
had substantially fragmented during the explosion. Fast clumps of
$\rm{^{56}Ni}$ and its decay products had formed in the innermost
regions of the ejecta and propagated out to the outer layers of the
hydrogen envelope (i.e. close to the supernova's photosphere) within
only \emph{days} after core collapse \cite[compare e.g.][]{MBB01}. For
a bibliography of the observational evidence see the reviews of
\cite{Hillebrandt_Hoeflich89}, \cite{ABKW89}, \cite{MCCray93},
\cite{Nomoto+94}, \cite{Wooden97}, \cite{Mueller98}, and the
references therein.

Spherical symmetry is not only broken in case of SN 1987\,A. This is
apparently a generic feature, as is indicated by the spectra of a
number of core collapse SNe that have been observed for more than one
decade. Significant mixing and clumpiness of the ejecta were found for
the Type II explosions SN 1995\,V \citep{F+98}, SN 1988\,A
\citep{Spyromilio91} and SN 1993\,J \citep{Spyromilio94,WH94,WEWP94}.
In case of Type~Ib supernovae the indications for mixing are even
older.  For instance, \cite{Filippenko_Sargent89} using [O~I]
observations have found that the ejecta of SN 1987\,F were
clumped. Even earlier, \cite{Harkness+87} in constructing synthetic
spectra encountered the first evidence. They had to overpopulate
excited states of He~I by factors of $\sim 10^{4}$ relative to local
thermodynamic equilibrium (LTE) conditions in order to reproduce the
characteristic strong He lines of this supernova type.  It was shown
later by \cite{Lucy91} and \cite{Swartz+93} that the presence of the
He~I lines and the implied departures from LTE can be understood in
terms of impact excitations and ionizations by nonthermal
electrons. The latter are presumed to originate from
Compton-scattering of the $\gamma$-rays from $\rm{^{56}Co}$ decay and
hence this hints toward significant outward mixing of $\rm{^{56}Co}$.
Artificially mixed one-dimensional explosion models of Type~Ib
progenitors indeed yield good fits to observed Type~Ib spectra and
light curves \citep{WE97,Shigeyama+90}.

Such observations have stimulated theoretical and numerical work on
hydrodynamic instabilities, which up to now has been dichotomized into
two distinct classes.  On the one hand multi-dimensional modelling of
the explosion mechanism was attempted
\citep{HBC92,Burrows_Fryxell93,Miller+93,HBFC94,BHF95,JM96,Mezzacappa+98},
mainly with the aim to answer the question as to which extent
convective instabilities are helpful for generating neutrino-driven
explosions. Therefore, these simulations have been constrained to the
early shock propagation phase up to about one second after core
bounce.

In contrast, hydrodynamic models of the late-time shock propagation
and the associated formation of Rayleigh-Taylor instabilities in the
expanding mantle and envelope of the exploding star
\citep{AFM89,FMA91,MFA91,MFA2Ds_91,MFA3D_91,HMNS90,HMNS91,HMNS92,HMNS94,
YS90,YS91,HB91,HB92,HW94,SIHNS96,Iwamoto+97,NSS98,Kane+00} have
ignored the complications of the explosion mechanism.  Due to
considerable computational difficulties in resolving the relevant
spatial and temporal scales \emph{all} of these latter investigations
relied on ad hoc procedures to initiate the explosion. This usually
meant that some simple form of energy deposition (by a piston or
``thermal bomb'') was used to generate a shock wave in a progenitor
model. The subsequent propagation of the blast wave was followed in a
one-dimensional simulation until the shock had reached the (C+O)/He or
the He/H interface. Then, the 1D model was mapped to a
multi-dimensional grid, and an assumed spectrum of seed perturbations
was added to the radial velocity field in order to break the spherical
symmetry of the problem and to trigger the growth of the instability.
The studies of \cite{YS90,YS91}, \cite{NHSY97,NSS98}, and the recent
3D calculations of \cite{Hungerford+03}, who investigated the effects
of metal mixing on X-ray and $\gamma$-ray spectra formation, differ
from this universally adopted approach by making use of parameterized,
aspherical shock waves.

All Rayleigh-Taylor calculations, that have been carried out to date
for reproducing the mixing in SN 1987\,A, share the same problem: With
at most 2000\,km/s their maximum $\rm{^{56}Ni}$ velocities are
significantly smaller than the observed values of
$3500-4000$\,km/s. \cite{HB92} dubbed this problem the ``nickel
discrepancy''. They also speculated that it should disappear when the
``premixing'' of the ejecta during the phase of neutrino-driven
convection is taken into account. On the other hand \cite{NSS98}
do not agree with \citeauthor{HB92}. They claim that the nickel
discrepancy is resolved if the supernova shock is (mildly) aspherical.
Similar conclusions have been reached by \cite{Hungerford+03}. While
\citeauthor{NSS98} as well as \citeauthor{Hungerford+03} do not rule
out neutrino emission from the neutron star as an explanation for the
assumed asphericity of their shocks, \cite{Khokhlov+99},
\cite{Wheeler+02} and \cite{Wang+02} questioned the neutrino-driven
mechanism. Instead, they speculated on ``jet-driven'' explosions,
which might originate from magneto-hydrodynamic effects in connection
with a rapidly rotating neutron star.

The above controversy demonstrates the need for models which link
observable features at very large radii to the actual energy source
and the mechanism of the explosion. Without such models,
interpretation of observational data requires caution.  With the
present work, which is the first in a series of papers on
non-spherical core collapse supernova evolution, we attempt to provide
this link for the standard paradigm of neutrino-driven supernovae.  We
have performed the first one and two-dimensional hydrodynamic
calculations of Type~II and Type~Ib-like explosions of blue supergiant
stars that reach from 20\,ms up to 20\,000\,s (i.e. 5.5 hrs) after
core bounce. Thus, we follow the evolution well beyond the time of
shock eruption from the stellar photosphere. Our models include a
detailed treatment of shock revival by neutrinos, the accompanying
convection and nucleosynthesis and the growth of Rayleigh-Taylor
instabilities at the composition interfaces of the progenitor star
after shock passage. We employ high spatial resolution, which is
mandatory to follow the episodes of clump formation, mixing, and clump
propagation, by making use of adaptive mesh refinement techniques.
Preliminary results of this work, which covered the first five minutes
in the evolution of a Type~II supernova, were reported by
\cite{Kifonidis+00}.

This paper is organized as follows: We start with an account of the
physical assumptions and numerical methods implemented in our computer
codes in Sect.~\ref{sect:codes}.  We then describe our stellar model
in Sect.~\ref{sect:star}. Our boundary conditions along with some
technical details of our computational strategy are discussed in
Sect.~\ref{sect:setup}. This is then followed in
Sect.~\ref{sect:first_sec} with our description of the early evolution
up to 0.82 seconds after core bounce. To focus the discussion, we
restrict ourselves to one basic shock-revival model, from which we
start all subsequent calculations. To illustrate its physics as
clearly as possible, we first present the one-dimensional case in
Sect.~\ref{sect:first_sec_1D} and then the two-dimensional results in
Sect.~\ref{sect:first_sec_2D}.  In the same manner we discuss the
subsequent evolution of a Type~II supernova model in
Sect.~\ref{sect:AMR_models}.  In Sect.~\ref{sect:SN_Ib_model}, we
consider the problem of a Type~Ib explosion by simply removing the
hydrogen envelope of our progenitor.  Section~\ref{sect:conclusions}
finally contains our conclusions.

\section{The codes}
\label{sect:codes}

There is as yet no numerical code that could satisfactorily handle all
the computational difficulties encountered in the modelling of core
collapse supernovae over the time scales that we are interested in in
this paper. Fortunately, however, such a code is not mandatory for
studying the effects that are responsible for the formation and
propagation of the nickel clumps. A crucial simplification arises from
the fact that the physical character of the explosion changes
fundamentally from a neutrino-hydrodynamic to a purely hydrodynamic
problem once the shock has been revived and launched successfully by
neutrino heating. A simulation of the long-time evolution of the
ejecta can therefore be split into two parts. A neutrino-hydrodynamic
calculation that encompasses the challenges of modelling
neutrino-driven explosions but does not require an extremely high
dynamical range of the spatial resolution, since only the innermost
stellar core needs to be resolved adequately. In contrast, in the
second, hydrodynamic part, the physics of the problem simplifies to
the solution of the hydrodynamic equations because (to a good
approximation) the central neutron star influences the dynamics of the
ejecta only via its gravitational attraction and can otherwise be
disregarded. It is this part of the calculation, however, where a high
spatial resolution is essential in order to resolve the growth and
expansion of Rayleigh-Taylor instabilities.

The separation into two largely independent problems might not be
appropriate, if one is interested in an accurate determination of
quantities that are influenced by the long-time hydrodynamic evolution
of the layers near the neutron star. Among these quantities are the
yields of r-process elements and the amount of fallback, which depend
on the properties of the neutrino-driven wind as well as on the
propagation of reverse shocks through the inner ejecta.  We will not
address these problems in the present work. Instead, we have developed
two different codes to solve each of the sub-tasks described above: A
modified version of the hydrodynamics code of \cite{JM96} (henceforth
JM96) that includes neutrino effects, and the adaptive mesh refinement
(AMR) code \textsc{AMRA}, a FORTRAN implementation of the AMR
algorithm of \cite{BC89}. Here we discuss only the most important
features of these codes and give some details of our hydrodynamic
advection scheme. \textsc{AMRA} is described in length in \cite{PM01}.

Our goal is to study the main observational consequences of
hydrodynamic instabilities in different layers of a supernova.  We do
not attempt to compete with the contemporary, highly sophisticated
modelling of neutrino-driven explosions (for which an accurate
treatment of neutrino transport is indispensable). For our purpose, a
simple neutrino light-bulb algorithm, as the one of JM96, has the
advantage to save enormous amounts of computer time as compared to
detailled transport codes \citep[see][and the references
therein]{Rampp_Janka02}. Omitting the core of the neutron star and
replacing it by a boundary condition, that parameterizes its
contraction and the radiated neutrino luminosities and spectra, we
make use of the freedom to vary the neutrino properties. Thus we are
able to control the explosion time scale and the final explosion
energy in our simulations. By computing models with different values
of the parameters, we can explore the physical possibilities as a
consequence of neutrino-driven explosions. The local effects of
neutrino heating and cooling by all relevant processes are treated
reasonably well with our light-bulb scheme. However, a light-bulb
description, of course, neglects all forms of back-reactions to the
neutrino fluxes and spectra that result from neutrino absorption in
the heating layer behind the supernova shock, and from neutrino
emission associated with the accretion of matter onto the neutron
star. We have recently developed a gray, characteristics-based scheme
for the neutrino transport that accounts for spectral and luminosity
modifications due to such effects. This new treatment is currently
applied in new two and three-dimensional simulations. The calculations
in this paper, however, were still performed with the light-bulb code.

\begin{figure*}
\centering
\begin{tabular}{cc}
\includegraphics[width=6.5cm]{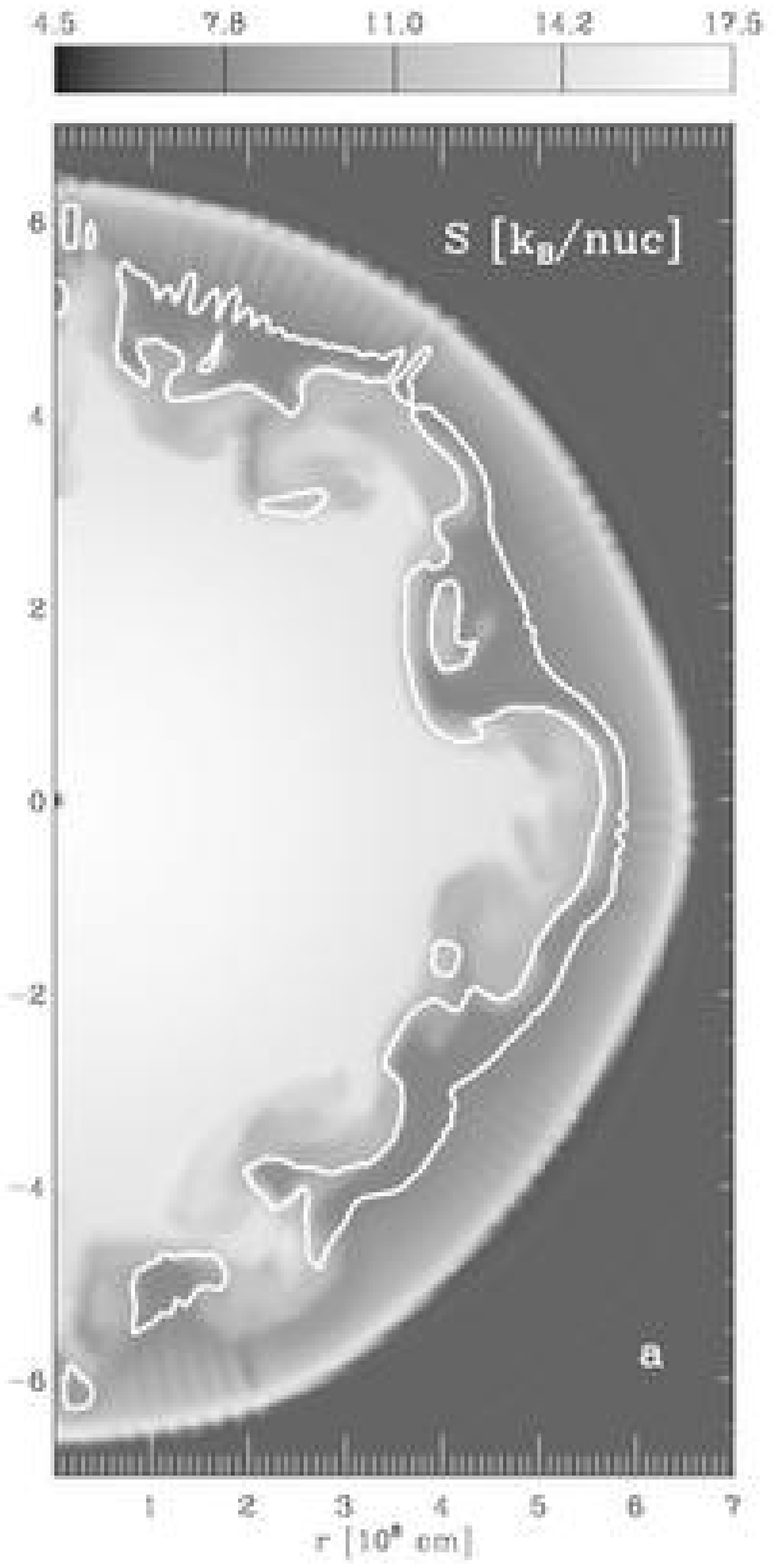} &
\includegraphics[width=6.5cm]{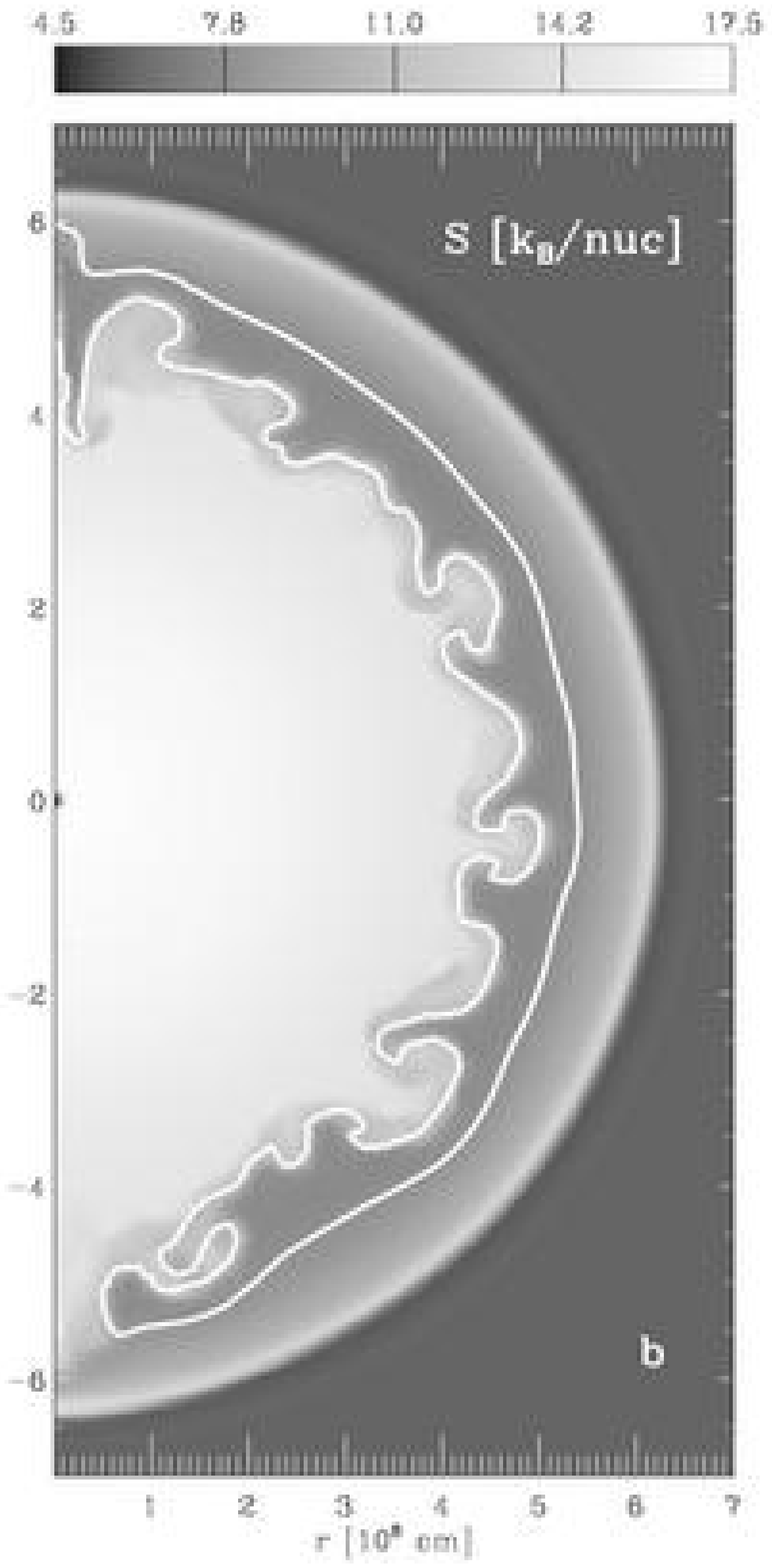}
\end{tabular}
\caption{Entropy (in units of $k_{\rm B}$ per nucleon) in Model T310
         at a time of 420\,ms after core bounce. The white contour
         line encloses the region where the $\rm{^{56}Ni}$ mass
         fraction is $\geq 20\%$.  a) Calculation performed with the
         original \textsc{PROMETHEUS} code showing odd-even
         decoupling.  Note the deformation of the shock. b)
         Calculation performed with \textsc{HERAKLES}.}
\label{fig:odd-even}
\end{figure*}

The neutrino physics is combined with an enhanced version of the PPM
hydrodynamics scheme (see below) and the equation of state (EOS) of
JM96.  The baryonic component of this EOS consists of 4 nuclei
(neutrons, protons, $\alpha$-particles and a representative nucleus of
the iron group) in nuclear statistical equilibrium (NSE). These four
species are also used to compute the energy source terms resulting
from nuclear transmutations.  In addition to this small NSE
``network'', but without feedback to the EOS and the hydrodynamics, we
also evolve a 14 species nuclear reaction network to approximately
calculate the products of explosive nucleosynthesis, whose spatial
distribution we wish to follow.  The latter network consists of the 13
$\alpha$-nuclei from $\rm{^{4}He}$ to $\rm{^{56}Ni}$ and an additional
tracer nucleus to which we channel the flow resulting from the
reaction $\rm ^{52}Fe(\alpha,\gamma)^{56}Ni$ in case the electron
fraction $Y_{\rm e}$ drops below 0.490 and $\rm{^{56}Ni}$ ceases to be
the dominant nucleus synthesized in the iron group
\cite[e.g.][]{TNH96}. In this way we can ``mark'' material that
freezes out from NSE at conditions of neutron excess and distinguish
it from $\rm{^{56}Ni}$ whose yield would otherwise be
overestimated.  The 14 species network is solved for temperatures
between $10^8$\,K and $8\times10^9$\,K.  Above $8\times10^9$\,K, we
assume that nuclei have been disintegrated to $\alpha$-particles. Of
course, the 14 species network is a simplification of the
nucleosynthesis processes in a supernova, since it neglects important
isotopes and production channels of the considered nuclei. Moreover,
in our current code version the heating and composition changes due to
explosive silicon, oxygen, neon, and carbon burning as described by
the 14 species network are not taken into account in the solution of
the hydrodynamics, which implies an approximation of the
thermodynamical history of the regions where nuclear burning
occurs. However, the hydrodynamic evolution should hardly be affected
in the simulations presented in this paper. This can be seen by
estimating the energy release due to nuclear transmutations. In
core-collapse supernovae three thermodynamic regimes can be
distinguished.  Below $2.1\times10^{9}$~K (the minimum temperature
required for explosive Ne/C burning; \citealt{TNH96}) nuclear burning
time scales are large compared to the hydrodynamic expansion time
scale of the layers that contain the nuclear fuel, and no appreciable
nucleosynthesis occurs.  For temperatures $2.1\times10^{9}~{\rm K}
\leq T \leq 5\times10^{9}~{\rm K}$ the burning time scales are
comparable to the expansion time scale, and the abundance and energy
changes need to be followed by solving a nuclear reaction
network. Above $5\times10^{9}$~K NSE holds, i.e. the composition
adjusts immediately to temperature and density changes caused by the
hydrodynamics.  In our code, energy source terms due to explosive
burning are not accounted for only in the second regime (in the NSE
regime the energy source term is obtained from the 4 species network).
For the $15~{\rm M_{\odot}}$ \cite{WPE88} progenitor that we employed
(see Sect.~\ref{sect:star}), we estimated that thermonuclear burning
in layers with temperatures $2.1\times10^{9}~{\rm K} \leq T \leq
5\times10^{9}~{\rm K}$ releases $\sim 5\times10^{49}$~ergs. This
energy is small compared to the explosion energies of our models
(about $1.8\times10^{51}$~ergs), and is therefore negligible with
respect to the dynamics. This may not be true, however, in case of
other (e.g. more massive) progenitors.

Gravity is taken into account in our neutrino-hydrodynamics code by
solving Poisson's equation in two spatial dimensions with the
algorithm of \cite{MS95}. To avoid the formation of transient shock
oscillations in case general relativistic post collapse models are
used as initial data (compare JM96), we add 1D (i.e. spherical)
general relativistic corrections to the 2D Newtonian gravitational
potential as in \cite{Keil+96} (see \citealt{Rampp_Janka02} for
details of the implementation and compare also to
\citealt{VanRiper79}). These corrections are especially important in
the early phase of shock revival, since the deeper relativistic
potential well keeps the shock at significantly smaller radii than in
the Newtonian case. However, Newtonian gravity becomes an excellent
approximation at large distances from the neutron star. Therefore it
can be used once the shock has emerged from the iron core and
propagates through the oxygen shell. This is also the time when we map
our explosion models to follow the late evolution with
\textsc{AMRA}. In the AMR calculations Poisson's equation is solved in
one spatial dimension with an angular average of the density. The
equation of state that we employed for these simulations includes
contributions from non-relativistic, non-degenerate electrons,
photons, electron-positron pairs \citep[the formulae for pressure and
energy density due to pairs can be derived from information given
by][]{Witti+94} and the 14 nuclei (treated as Boltzmann gases) that
are included in the reaction network of our neutrino code. Separate
continuity equations are solved for each of these nuclei in order to
determine the mixing of the elements.

The neutrino-hydrodynamics code as well as \textsc{AMRA}, make use of
the \textsc{HERAKLES} hydrodynamics solver, an implementation of the
direct Eulerian version of the Piecewise Parabolic Method (PPM) of
\cite{CW84}.  \textsc{HERAKLES} originated from the
\textsc{PROMETHEUS} code of \cite{FMA91}. It includes the entire
functionality of \textsc{PROMETHEUS}, but differs from its predecessor
in the following aspects. The handling of multifluid flows of
\cite{FMA91} is replaced by the Consistent Multifluid Advection scheme
(CMA) of \cite{PM99}. This allows for a significant reduction of
numerical diffusion of nuclear species (to the level that is
characteristic for the PPM advection scheme) without suffering from
errors in local mass conservation \cite[see][]{PM99}.  Further
improvements in reducing the numerical diffusivity of the code are
achieved by adopting the elaborate procedure for the flattening of
interpolated profiles for the (primitive) state variables that is
suggested in the Appendix of \cite{CW84}. Additionally, the algorithm
for the computation of the left and right input states for the Riemann
problem is revised.  The version of \textsc{HERAKLES} that is used as
hydrodynamic solver in our AMR calculations has been further extended
to achieve good computational performance even on small grid
patches. For this purpose a new memory interface was written, that
allows for optimal pipelining of the one-dimensional hydrodynamic
sweeps that result from dimensional splitting in multi-dimensional
calculations, while due attention is paid to optimal cache reuse. With
this memory interface the code can make efficient use of both
superscalar as well as vector computer architectures.

The most important modifications as compared to \textsc{PROMETHEUS}
(or other standard implementations of the PPM scheme) are, however,
the inclusion of new algorithms for the artificial viscosity and for
multi-dimensional shock detection, along with the addition of
\citeauthor{Einfeldt88}'s HLLE Riemann solver
\citep{Einfeldt88}. Following \cite{Quirk94,Quirk94_reprinted} we use
the \citeauthor{Einfeldt88} solver for zones inside strong
grid-aligned shocks, while we retain the (less diffusive) Riemann
solver of \cite{CG85} for all remaining grid cells. The latter changes
are necessary to eliminate the odd-even decoupling instability
\citep{Quirk94,Quirk94_reprinted,Liou2000}, from which the original
\citeauthor{CG85} solver suffers \citep{Kifonidis+00}. This numerical
failure shows up if a sufficiently strong shock is (nearly) aligned
with one of the coordinate directions of the grid, and if, in
addition, the flow is slightly perturbed. Many Riemann solvers allow
these perturbations to grow without limit along the shock surface,
thus triggering a strong rippling of the shock front and the
post-shock state. In supernova simulations, these perturbations
strongly enhance the growth of hydrodynamic instabilities, since their
amplitudes can exceed those of the intentionally introduced seed
perturbations by several orders of magnitude.  In case of
neutrino-driven convection this leads to faster growth of convective
instabilities and angular wavelengths of convective bubbles which are
significantly larger than in a ``clean'' calculation.  This is
demonstrated in Fig.~\ref{fig:odd-even} where we compare entropy plots
of two simulations of neutrino-driven convection that were conducted
using spherical coordinates and the neutrino parameters given in
Sect.~\ref{sect:first_sec_1D}. The first of these
(Fig.~\ref{fig:odd-even}a) was done with our original
\textsc{PROMETHEUS} version and a resolution of $400 \times 180$ zones
and is affected by odd-even decoupling, while the second
(Fig.~\ref{fig:odd-even}b) was obtained with the improved scheme
implemented in \textsc{HERAKLES} using $400 \times 192$ zones.

Odd-even decoupling and the associated carbuncle phenomenon
\citep[i.e. a local occurrence of odd-even decoupling for shocks that
are only partly aligned with the
grid;][]{Quirk94,Quirk94_reprinted,LeVeque} have gone largely
unnoticed in the astrophysics literature and have marred almost all
multi-dimensional supernova simulations conducted to date on
cylindrical or spherical grids with hydrodynamic codes that are based
on Riemann solver type schemes \citep[e.g.][]{Kifonidis+00,
Mezzacappa+98,Iwamoto+97,JM96,BHF95,HMNS94,Burrows_Fryxell93,
HMNS92,MFA2Ds_91}.

\section{The stellar model}
\label{sect:star}

As in JM96 and \cite{Kifonidis+00} the initial model adopted for our
study originated from the simulations of \cite{Bruenn93}, who followed
core collapse and bounce of the $15\,{\rm M_{\odot}}$ blue supergiant
progenitor of \cite{WPE88}.  We have set up our shock revival
calculations using his model WPE15~LS~(180) at a time of 20\,ms after
core bounce. However, Bruenn's data set extends only to layers within
the stellar He core and the original progenitor model of \cite{WPE88}
is no longer available.  To follow the late-time propagation of the
shock in our subsequent AMR calculations we thus had to reconstruct
the outer envelope of the \citeauthor{WPE88} star.  For this purpose
we used a new 15\,${\rm M_{\odot}}$ blue supergiant model
(S.~E.~Woosley, private communication) that was smoothly joined to
Bruenn's data by choosing a matching point at $8.4 \times 10^{9}$\,cm
($M_r = 1.94\,{\rm M_{\odot}}$), i.e. within the He
core. Table~\ref{tab:star} summarizes the location of the composition
interfaces and the initial position of the $Y_{\rm e}$ discontinuity
\citep[that defines the boundary of the iron core according
to][]{Woosley_Weaver95} of the resulting ``hybrid'' $15\,{\rm
M_{\odot}}$ progenitor. Except for the He/H interface, the listed
radii were taken from the post-collapse data of \cite{Bruenn93}.

Since the velocity of the supernova shock and the associated growth of
Rayleigh-Taylor instabilities are very sensitive to the density
profile, we have verified in a number of one-dimensional test
calculations that the hybrid stellar model had no adverse effects on
shock propagation. In these calculations we cut out the model's iron
core and induced the explosion artificially by depositing $\sim
10^{51}$\,ergs in the form of thermal energy at the inner boundary of
the silicon shell. This procedure enabled us to compare the
hydrodynamic evolution of our hybrid model with induced explosions of
newer SN 1987\,A progenitor models computed by \cite{WHLW97} (for
which no collapse calculations including neutrino transport were
available at the time we performed our simulations). No differences
pointing to numerical artifacts were found.

\begin{table}[h]
\begin{center}
\caption{Location of chemical interfaces in our progenitor model
         at 20\,ms after core bounce.
         \label{tab:star}}
\begin{tabular}{cccccc}
\hline
\hline
    & $Y_{\rm e}$ disc. & Fe/Si  & Si/O  & (C+O)/He & He/H \\
\hline
$r\,{\rm [km]}$ &  260    &  1376  &   6043  &    29\,800  &  708\,000 \\
$M_{\rm r}\,{\rm [{\rm M_{\odot}}]}$ & 1.25  &  1.32  &   1.50    &  1.68  & 4.20
\\
\hline
\end{tabular}
\end{center}
\end{table}

\section{Computational setup}
\label{sect:setup}

\subsection{Explosion models (hydrodynamic calculations with neutrinos)}

To set up a shock revival simulation we adopt spherical coordinates.
We omit the innermost 0.848 ${\rm M_{\odot}}$ of the core of the
neutron star and replace it by the gravitational potential of a point
mass and an inner boundary which acts as a neutrino source. This
boundary is placed at a radius of $3.17\times 10^{6}\,{\rm cm}$, which
is somewhat below the electron neutrinosphere in Bruenn's model. The
outer radial boundary is located at $r = 1.7\times10^9$\,{\rm cm},
inside the C+O-core of the star.  Our grid consists of 400
non-equidistant radial zones that yield a maximum resolution of
1.2\,km at the neutron star and 200\,km at the outer grid boundary. In
our two-dimensional models 192 angular zones are distributed uniformly
between $0 \leq \theta \leq \pi$ and axial symmetry is assumed around
the polar axis.  Reflecting boundaries are imposed at $\theta = 0$ and
$\theta = \pi$ and a random initial seed perturbation is added to the
velocity field on the entire grid with a modulus of $10^{-3}$ of the
(radial) velocity of the post-collapse model. In the immediate
upstream region of the accretion shock, this corresponds to velocity
perturbations with a modulus of $\sim 5\times10^{-3}$ of the local
sound speed, while further outward this value decreases with the
decreasing infall velocities to $\sim 10^{-3}$.

In order to mimic the contraction of the cooling and deleptonizing
proto-neutron star, the inner boundary is moved inward during the
computations, approximating the motion of the corresponding mass shell
in Bruenn's calculations. This is achieved by making use of the moving
grid implemented in our code (cf. JM96 for details). Free outflow is
allowed for across the outer radial boundary. With this setup, the
time-step resulting from the Courant-Friedrichs-Lewy stability
condition is typically of the order of a few $10^{-5}$\,s in
one-dimensional calculations, and several $10^{-6}$\,s in
two-dimensional simulations. Starting at 20\,ms after core bounce the
computations are continued until $820$\,ms post-bounce, when the
explosion energy has saturated and all nuclear reactions have frozen
out.

\subsection{Models for the Rayleigh-Taylor mixing (AMR calculations)}

By performing test calculations we have found that, in order to enable
a sufficiently detailed study of the growth of all relevant
Rayleigh-Taylor instabilities, a spatial resolution of $\Delta r \leq
10^{-6} R$ (with $R$ being the stellar radius) is required for at
least some fraction of the time that it takes the shock to erupt from
the surface of the star. The above estimate holds for the case of a
blue supergiant like Sk\,$-69^{\circ}\,202$. In case of a red
supergiant an even smaller ratio $\Delta r / R$ would be
necessary. Without adaptive mesh refinement and/or a moving grid such
calculations are currently unfeasible, especially in more than one
spatial dimension. While this range of spatial scales appears
formidable, for explicit codes like PPM an even more severe difficulty
is posed by the vastly differing time scales involved in the
problem. Even if the inner boundary is placed far away from the
neutron star, and the gravitational attraction of the omitted stellar
layers is taken into account by a point mass, the resulting Courant
time step is still too small to allow one to follow the evolution over
hours. The small time step is caused by fallback, which leads to high
sound speeds in the central zones because matter that falls back
toward the inner boundary is compressed and heated.  The only remedies
to this problem that we have found to be practical are either a
coarsening of the grid or a shift of the inner boundary with time
towards larger radii, where smaller temperatures are
encountered. While the former is useful in 1D calculations, where the
remaining grid resolution that we achieved was still reasonable (see
below), the latter proved to be the better choice for our
two-dimensional calculations.

\begin{table}[t]
\begin{center}
\caption{Computational setup for the 2D AMR calculation T310a.
         (see Sect.~\ref{sect:AMR_2D}). The initial and final times
         over which the inner and outer grid boundaries were held 
         fixed at $r_{\rm in}$ and $r_{\rm out}$,
         are denoted by $t_{\rm i}$ and $t_{\rm f}$, respectively. 
         $\Delta r$ is the (absolute) resolution and 
         $N^{\rm eff}_{r}$ the number of equidistant radial zones
         that would have been required to cover the entire star
         in order to achieve a resolution of $\Delta r$.
         Exponents are given in brackets.}
\label{tab:2D_refinement}
\begin{tabular}{rrrrrr}
\hline
\hline
 $t_{\rm i}$ [s] & $t_{\rm f}$ [s] & $r_{\rm in}$ [cm] & 
           $r_{\rm out}$ [cm] &  $\Delta r$ [km]  & $N^{\rm eff}_{r}$\\ 
\hline
 $0.8(+0)$  & $2.3(+0)$ & $1.0(+8)$    &
            $4.8(+9)$ &  1.5(+1)    & 2\,599\,933\\ 

 $2.3(+0)$ & $8.0(+0)$ & $1.7(+8)$    &
            $1.4(+10)$ &  4.6(+1)    & 847\,804\\ 

 $8.0(+0)$ & $3.4(+1)$ & $3.9(+8)$    &
            $4.3(+10)$ &  1.4(+2)   & 280\,568\\ 

 $3.4(+1)$  & $1.2(+2)$ & $9.9(+8)$  &
            $1.3(+11)$ & 4.2(+2)   & 93\,299\\ 

 $1.2(+2)$ & $2.8(+2)$ & $2.3(+9)$  &
            $2.6(+11)$ & 8.4(+2)  &  46\,650\\ 

 $2.8(+2)$ & $6.5(+2)$ & $3.9(+9)$  &
            $5.2(+11)$ & 1.7(+3) &  23\,282\\ 

 $6.5(+2)$  & $1.5(+3)$ & $6.8(+9)$ &
            $1.0(+12)$ & 3.4(+3) & 11\,634\\ 

 $1.5(+3)$ & $3.6(+3)$ & $1.2(+10)$ &
            $2.1(+12)$ & 6.7(+3) & 5810\\ 

 $3.6(+3)$ & $2.0(+4)$ & $2.0(+10)$ &
            $3.9(+12)$ & 1.3(+4) & 3072\\
\hline
\end{tabular}
\end{center}
\end{table}

In all of our AMR simulations we have adopted spherical coordinates
and a pre-chosen AMR grid hierarchy to which we mapped the data of our
explosion models \emph{without adding any extra seed
perturbations}. At radii where data from the explosion models are not
available, the structural information is taken from the progenitor
model. The inner regions of an explosion model are cut out by placing
the inner boundary of the AMR base level grid at a radius of
$10^{8}$\,cm, which is well inside the neutrino-driven wind of the
neutron star. This baryonic mass flow is therefore neglected in the
subsequent evolution. Instead we allowed for free outflow through the
inner (as well as the outer) boundary.  This is certainly a gross
simplification for the evolution of the innermost regions of the flow
during the first minute of the explosion (but becomes more realistic
later when the neutrino wind is expected to cease).  By ignoring the
neutrino wind one obtains early fallback of matter through the inner
boundary. In addition, the evolution of the reverse shock, which
results from the slow-down of the main shock in the oxygen core, is
affected. This reverse shock separates the wind from the outer ejecta
at times later than $\sim 500$\,ms after bounce (compare
Sect.~\ref{sect:first_sec}). Without the pressure of the wind, the
reverse shock propagates inward and is (partially) reflected at the
inner boundary because the latter is not perfectly transmitting for
numerical reasons.  Although both the inward motion and reflection of
the reverse shock are expected to happen once the wind weakens (with
the reverse shock being reflected at the steep density gradient near
the neutron star ``surface''), our approximations do not allow to
model these phenomena reliably in space and time. Except for resulting
in these limitations (which are not relevant for studying clump
formation and propagation), the mapping from one code to the other
worked very well and did not produce any noticeable artifacts.

For our 1D AMR calculation of Sect.~\ref{sect:AMR_1D}, we employed a
grid hierarchy with a base grid resolution of 512 zones and 7 levels
of refinement. We use two criteria to flag zones on a given level for
refinement. First, we estimate the local truncation error for the
(mass) density, momentum density, total energy density, and the
partial densities of nuclei, $\rho X_i$, using Richardson
extrapolation. Whenever the error for one of these quantities is found
to be $\geq 10^{-2}$ (or $\geq 10^{-3}$ in case of the $\rho X_i$), we
flag the corresponding zone for refinement.  Independent of the
truncation error estimate, we also flag zones in which density jumps
$\geq 100\%$ are encountered as compared to neighbouring zones.  To
refine between different levels, we use factors of 4 in resolution
until level 5 is reached, where we switch to refinement factors of
2. With this scheme we achieve a maximum resolution equivalent to
524\,288 equidistant zones. Since the outer boundary of the
computational domain is kept at the stellar radius of
$3.9\times10^{12}$\,cm, this translates to an absolute resolution of
$\sim 75$\,km. Using this setup, we measure speed-up factors (as
compared to using a uniform grid) that are as large as $10^3$. Still,
however, the small time step resulting from the high sound speed in
the innermost zones leads to long computing times. We therefore remove
the 7th level of the grid hierarchy after 1870\,s of evolution and
thereby reduce the maximum resolution to 149\,km. This zoning is kept
until $t = 4770$\,s, when also level 6 is discarded and the rest of
the evolution followed with a maximum resolution of 298\,km. It should
be noted that AMR allows one to introduce this coarsening in a rather
straightforward fashion by manually resetting the refinement flags
returned from the error estimation and zone flagging modules of the
code. The grid can even be coarsened only locally if required. Our 1D
simulation is stopped at $t = 8500$\,s post bounce, 1700 seconds after
the shock has emerged from the photosphere.

\begin{figure}[t]
\resizebox{\hsize}{!}{\includegraphics{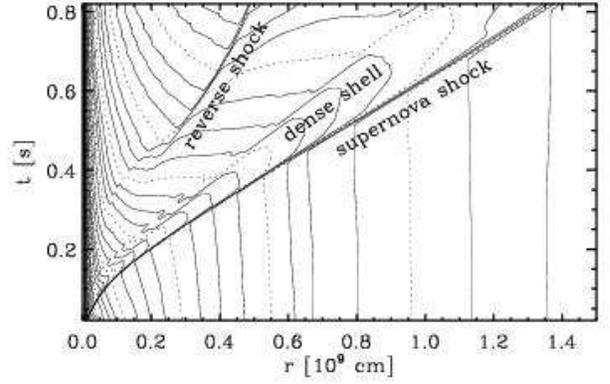}}
\caption{Spacetime plot for the evolution of the logarithm of the
         density for our one-dimensional explosion, Model O310.  The
         supernova shock is visible as the outermost discontinuity
         that extends diagonally through the plot. Around 500\,ms a
         reverse shock forms in the inner ejecta that originates from
         the (hardly visible) slight deceleration of the main shock in
         the oxygen core. The reverse shock separates the ejecta from
         the neutrino-driven wind.}
\label{fig:dens_cont}
\end{figure}

For our 2D calculations we have adopted a different approach,
combining mesh refinement with a ``zooming'' algorithm for the radial
coordinate.  The AMR grid is chosen such that a base-level resolution
of $48 \times 12$ zones and four levels of refinement with refinement
factors of 4 in each dimension are employed. This results in a maximum
resolution on the finest level corresponding to 3072 equidistant zones
in radius and 768 zones in angle.  Zones are flagged for refinement by
applying the same criteria as in our one-dimensional calculation for
each dimension.  Adopting transmitting boundary conditions in the
radial and reflecting boundaries in the angular direction, our
computational volume is initially set up to span the domain $(1000
\leq r/{\rm km} \leq 48\,000) \times (0 \leq \theta \leq \pi)$.  While
the grid boundaries in $\theta$ direction are kept unchanged
throughout the calculation, the zooming algorithm successively
enlarges the radial extent of the grid according to
Table~\ref{tab:2D_refinement}, when the shock is approaching the outer
grid boundary, $r_{\rm out}$. Whenever it is time to regrid, we also
move the inner grid boundary, $r_{\rm in}$, away from the central
remnant. This approach allows us to concentrate the computational
effort in the post-shock and the Rayleigh-Taylor unstable regions
while avoiding overly restrictive Courant time steps due to
fallback. The times $t_{\rm i}$ and $t_{\rm f}$ over which the radial
grid boundaries are kept fixed are given in columns one and two of
Table~\ref{tab:2D_refinement}.  The ``zooming algorithm'' allows us to
temporally achieve a radial resolution of $\Delta r = 15$\,km in the
unstable layers, that is equivalent to covering the entire star with
an effective resolution of $N^{\rm eff}_{r} = 2\,599\,933$ equidistant
zones.  Even with this combination of mesh refinement, grid
enlargement and inner boundary movement, however, the computational
load is still significant.  The two-dimensional AMR calculation that
we present in Sect.~\ref{sect:AMR_models}, and which follows the
evolution of the mixing from 820\,ms to 5.5 hrs after core bounce,
requires nearly $2\times10^{15}$ floating point operations (this
number has to be multiplied by about a factor of 5 if no adaptive mesh
is used).

\begin{figure*}[t]
\centering
\includegraphics[width=17cm]{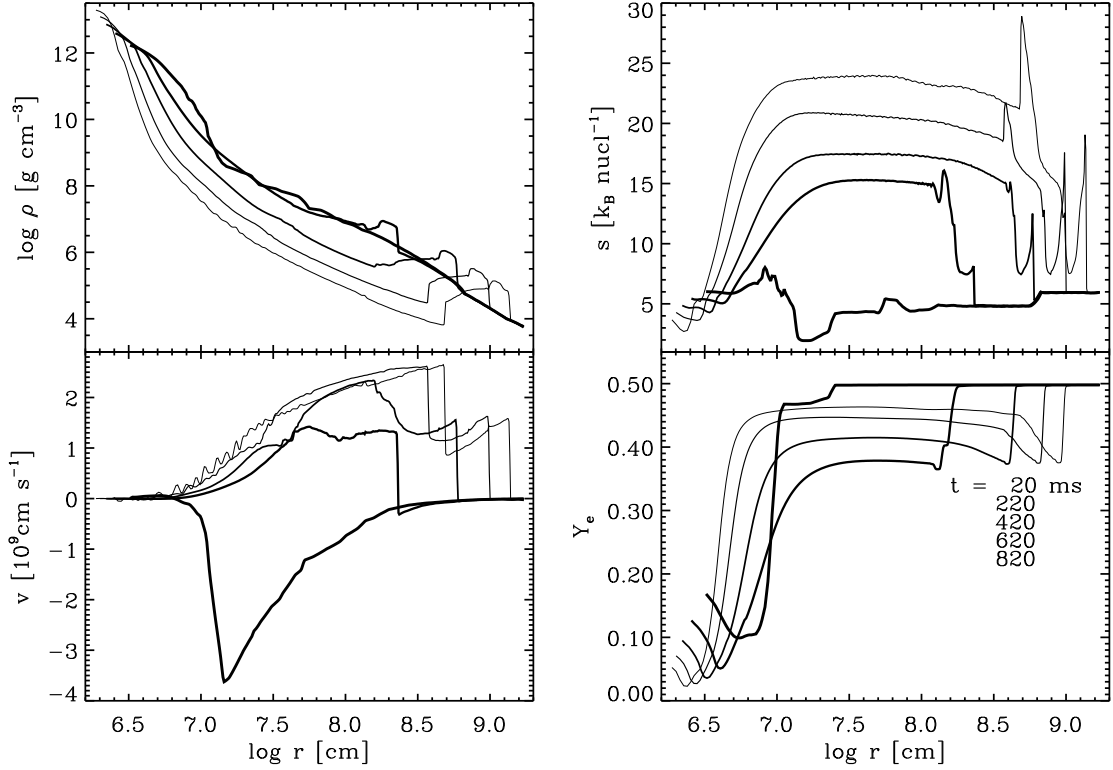}
\caption{Snapshots of the density, entropy, velocity and electron
         fraction distribution in Model O310, 20\,ms (heaviest line),
         220\,ms, 420\,ms, 620\,ms and 820\,ms after core bounce.}
\label{fig:evol_O310}
\end{figure*}

\begin{figure*}
\centering
\includegraphics[width=17cm]{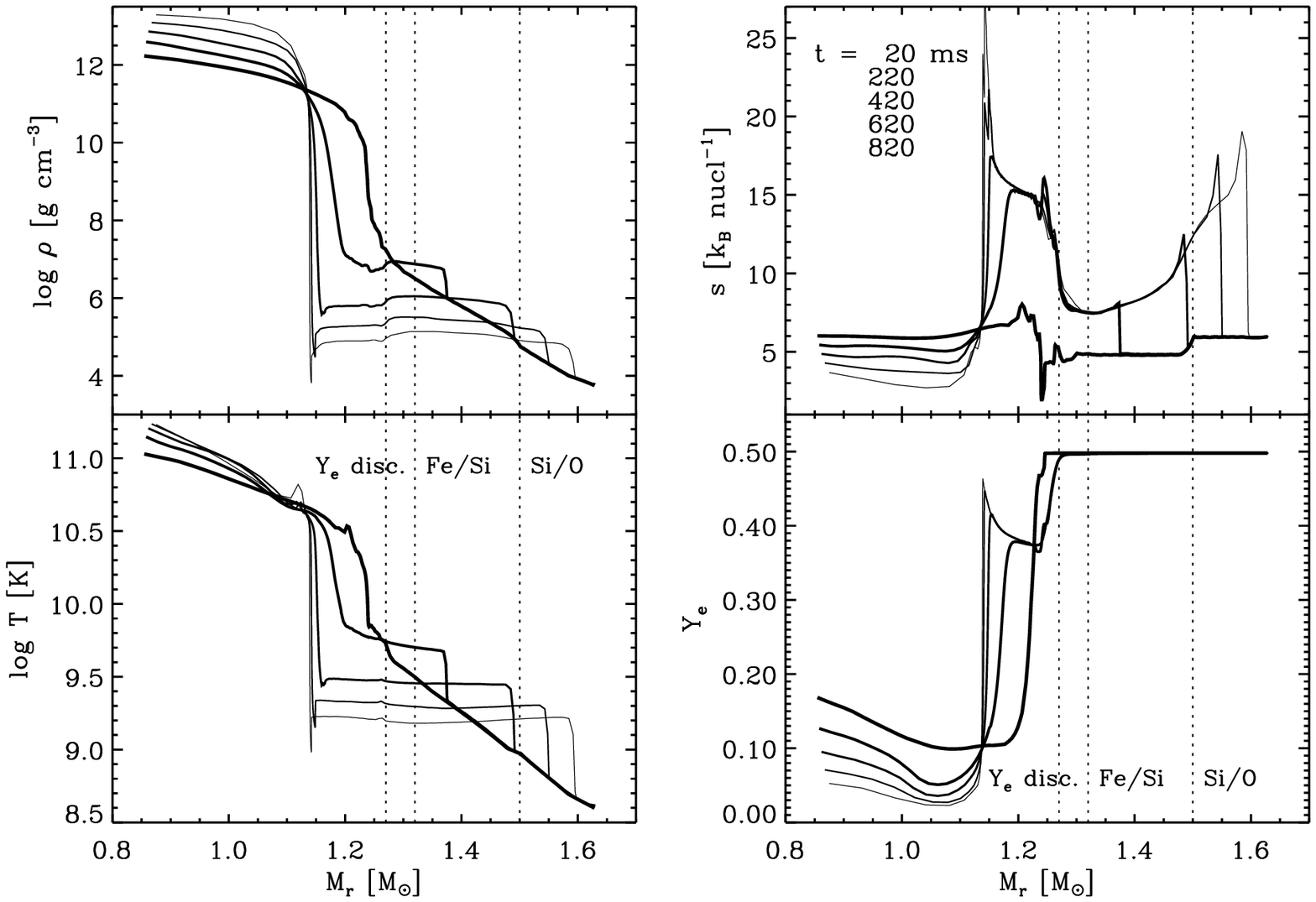}
\caption{Evolution of density, temperature, entropy and electron
         fraction as functions of the enclosed mass for Model O310
         between $t = 20$\,ms and $t = 820$\,ms post-bounce. The
         positions of the Fe/Si and Si/O interfaces, and the $Y_{\rm e}$
         discontinuity (see text) are indicated by dotted vertical
         lines.}
\label{fig:evol_O310_lagr}
\end{figure*}

Finally, we note that the choice of spherical coordinates leads to
numerical artifacts along the symmetry axis of the grid. The excessive
outflows that we observe for $\theta \approx 0$ and $\theta \approx
\pi$ for the models in this paper are much weaker than those in
\cite{Kifonidis+00}. Nevertheless, in order to avoid spurious effects,
we have excluded a cone with an opening angle of $15^{\circ}$ around
the polar axis for evaluating our simulations for the velocity
distributions of nucleosynthesis products and the angle-averaged
extent of the mixing of chemical elements.

\section{The first second}
\label{sect:first_sec}

\subsection{Evolution in one dimension}
\label{sect:first_sec_1D}

In Table~\ref{tab:explosion_models} we summarize the basic parameters
of our one-dimensional explosion, Model O310, along with its
two-dimensional counterpart, Model T310, that will be discussed in
Sect.~\ref{sect:first_sec_2D}. As described in detail in JM96 the
models are characterized by the initial values of the electron and
heavy lepton neutrino luminosities (in units of $\rm 10^{52}\,erg/s$),
$L_{\nu_e,52}^0$ and $L_{\nu_x,52}^0$, and the energy loss (in units
of ${\rm M_{\odot}} c^2$) and lepton loss parameters of the inner iron
core, $\Delta \varepsilon$ and $\Delta Y_{\rm L}$, respectively. The
latter parameter constrains the luminosity of electron antineutrinos
while $\Delta \varepsilon$ determines the characteristic time scale
for the decay of the luminosities that we assume to follow the simple
exponential law
\begin{equation}
  L_{\nu_i} = L^{0}_{\nu_i} {\rm e}^{-t/t_L},
             (\nu_i \equiv \nu_{\rm e}, \bar \nu_{\rm e}, \nu_x)
\label{eq:Lumnu}
\end{equation}
where $t_L$ is of the order of $700$\,ms. The neutrino spectra are
prescribed in the same way as in JM96.

Table~\ref{tab:explosion_models} also lists the explosion time scale,
$t_{\rm exp}$, measured in ms and the explosion energy at the end of
the shock-revival simulation, $E_{\rm exp,51}$, in units of $\rm
10^{51}\,erg$ (note that the binding energy of the envelope has not
yet been subtracted from the latter).  We define these quantities as
in JM96. The explosion energy is given by the sum of the
gravitational, kinetic and internal energy of all zones of the grid
where this sum of energies is positive. The explosion time scale is
defined as the time after the start of the simulation when the
explosion energy exceeds $\rm 10^{48}\,erg$.

\begin{table*}
\begin{center}
\caption{Parameters of Models O310 and T310. The (baryonic) mass of
the inner core, $M_{\rm core}$, that has been excluded from the
simulation, and the final neutron star mass, $M_{\rm ns}$, are given
in solar masses. The initial and final radius of the inner core,
$R_{\rm core}^{0}$ and $R_{\rm core}^{\infty}$, respectively, are
given in km, and its initial velocity of contraction, $\dot R_{\rm
core}^{0}$, in km/s.  The initial neutrino luminosities of different
flavours, $L_{\nu_{\rm e},52}^0$, $L_{\bar \nu_{\rm e},52}^0$, and
$L_{\nu_x,52}^0$ (see also the main text), are given in units of $\rm
10^{52}\,erg/s$, and the energy loss of the inner core, $\Delta
\varepsilon$, in units of ${\rm M_{\odot}} c^2$. The explosion energy,
$E_{\rm exp,51}$, and the explosion time scale, $t_{\rm exp}$, are
given in $\rm 10^{51}\,erg$ and in ms, respectively.}
\label{tab:explosion_models}
\begin{tabular}{lccccccccccccc}
\hline
\hline
  Model                     &  Resolution                           &
  $M_{\rm core}$            &  $R_{\rm core}^{0}$                   &
  $\dot R_{\rm core}^{0}$   &  $R_{\rm core}^{\infty}$              &
  $L_{\nu_{\rm e},52}^0$    &  $L_{\bar \nu_{\rm e},52}^0$          & 
  $L_{\nu_x,52}^0$          & 
  $\Delta Y_{\rm L}$        &  $\Delta \varepsilon$                 &  
  $E_{\rm exp,51}$          &  $t_{\rm exp}$                        &
  $M_{\rm ns}$             \\ 
\hline
  O310 & 400 & 0.848 & 31.7 & 50 & 18.1 & 3.094 & 3.684 & 
         2.613  & 0.0963 & 0.0688 & 1.59 & 70 & $ 1.12$ \\
  T310 & $400\times 192$ & 0.848 & 31.7 & 50 & 18.1 & 3.094  & 3.684 &
         2.613 & 0.0963 & 0.0688 & 1.77 & 62 &  $1.09$ \\
\hline
\end{tabular}
\end{center}
\end{table*}

We adopt values for $L_{\nu_e,52}^0$ and $L_{\nu_x,52}^0$ that give
rise to energetic explosions. This choice was motivated by the claim
of \cite{HB92} that premixing of the $\rm{^{56}Ni}$ (be it artificial
or due to neutrino-driven convection) in models with high explosion
energies of about $2\times10^{51}$\,ergs leads to a resolution of the
``nickel discrepancy problem''. In their simulations of
Rayleigh-Taylor instabilities in the stellar envelope, they obtained
final maximum $\rm{^{56}Ni}$ velocities of up to 3000\,km/s if they
premixed the $\rm{^{56}Ni}$ in their 1D initial models out to mass
coordinates of $1.5~{\rm M_{\odot}}$ above the Fe/Si interface
\citep[i.e. throughout 75\% of the metal core of the progenitor model
of][]{Nomoto+88}.

With the parameters listed in Table~\ref{tab:explosion_models}, shock
revival by neutrino heating is almost instant in our simulations. The
shock shows neither long stagnation times nor phases of progression
and subsequent recession. Instead, it moves out of the iron core
without noticeable delay. To illustrate matters, we show a spacetime
plot of the evolution of the density with time for Model O310 in
Fig.~\ref{fig:dens_cont}, from which one can infer a mean shock
velocity of about $19\,000$\,km/s after 0.2\,s.  Additionally, in
Fig.~\ref{fig:evol_O310} we depict the evolution of the most important
hydrodynamic and thermodynamic quantities as a function of radius. The
initial post-collapse profiles are plotted with thick lines, and the
Si/O interface can be discerned by the associated entropy step at
$\log r~ \mbox{[cm]} = 8.8$. One can easily recognize the transition
of the accretion shock to an outward propagating shock with high
post-shock velocities, the rarefaction generated by the explosion and
the deleptonization and contraction of the outer layers of the
proto-neutron star in response to the contracting inner boundary.
Figure~\ref{fig:evol_O310} also shows the neutrino-driven wind, that
is blown off the surface of the neutron star for times later than
$\sim 200$\,ms and that has also been found in the calculations of
\cite{BHF95} and \cite{JM96}. Furthermore, a reverse shock is visible,
that forms about 500\,ms after core bounce (compare
Fig.~\ref{fig:dens_cont}) when the main shock enters the oxygen core
of the star and propagates into layers with a somewhat flatter density
gradient, thereby decelerating slightly. Both shocks mark the
boundaries of a dense region that contains most of the ejecta, while
the reverse shock separates the ejecta from the neutrino-driven wind.
Shortly after the supernova shock has crossed the Fe/Si interface at
120\,ms (at this time the latter has fallen in from its initial radius
of $r = 1.38\times10^8$\,cm to $8\times10^7$\,cm), a high-density
shell forms immediately behind the shock. It becomes more pronounced
after $t = 500$\,ms when the shock has passed the Si/O interface. The
trajectory of the shell's inner boundary can be discerned in
Fig.~\ref{fig:dens_cont} in between the supernova and reverse
shocks. This shell is one of the most interesting outcomes of our
calculations. Three important physical processes take place in this
part of the ejecta, whose mutual interaction has been neglected in
previous attempts to model supernova explosions. These processes are
$\rm{^{56}Ni}$-synthesis, neutrino-driven convection, and the growth
of Rayleigh-Taylor instabilities at the Si/O interface of the
progenitor's metal core. We address the physical conditions for their
occurrence in turn, focusing first on $\rm{^{56}Ni}$-synthesis.

\begin{figure}[t]
\resizebox{\hsize}{!}{\includegraphics{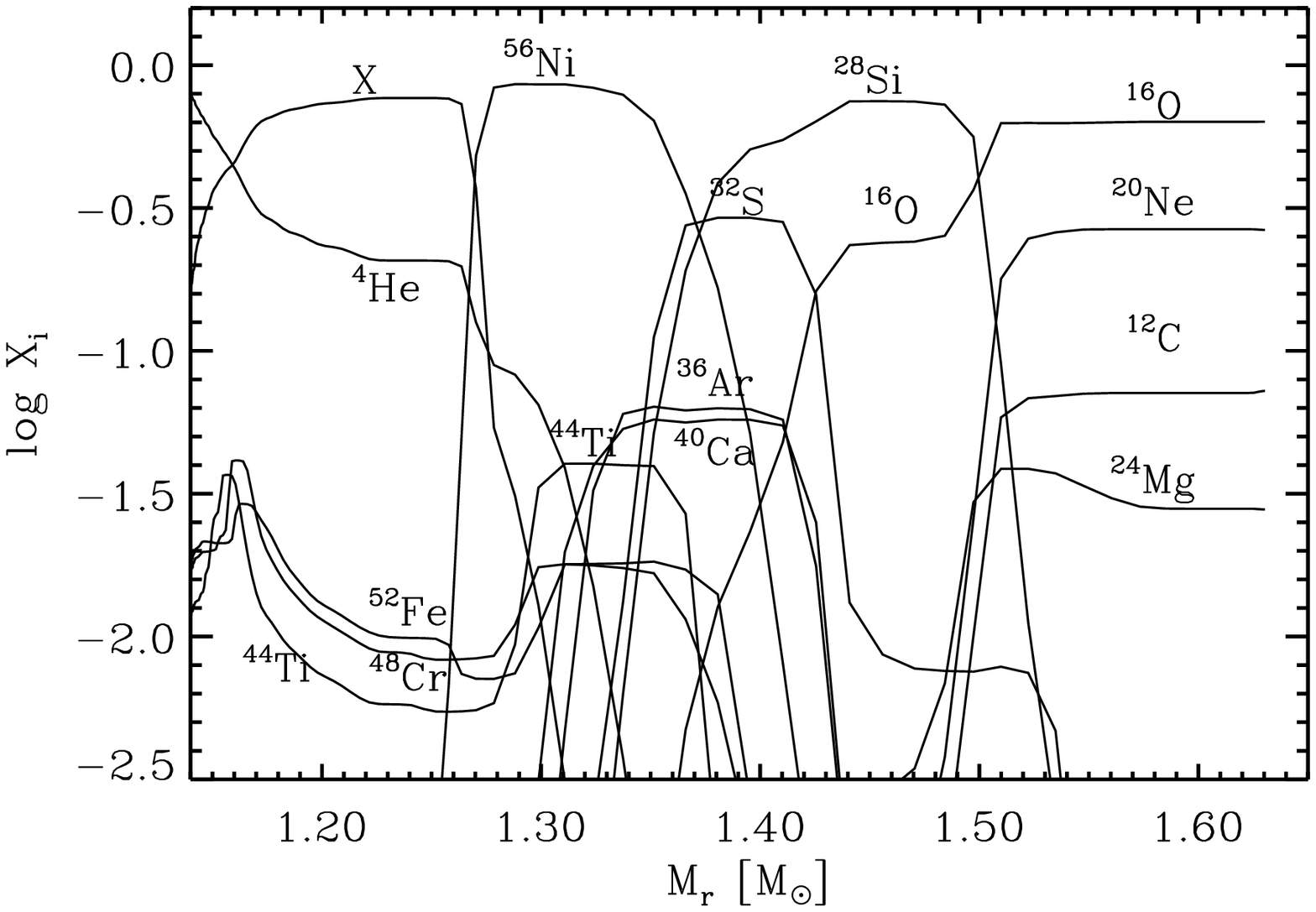}}
\caption{Chemical composition of Model O310 versus mass at $t =
         820$\,ms. X denotes the neutronization tracer nucleus.}
\label{fig:composition_O310}
\end{figure}

The left panels of Fig.~\ref{fig:evol_O310_lagr} show the evolution of
the density and temperature distribution for Model O310 as a function
of the enclosed (baryonic) mass. The inner edge of the high-density
shell is stationary at a mass coordinate of $M_r \approx 1.27\,{\rm
M_{\odot}}$.  This mass shell marks the border between low and
high-$Y_{\rm e}$ material in the expanding ejecta for times $t \geq
120$\,ms (right lower panel of Fig.~\ref{fig:evol_O310_lagr}),
i.e. for the epoch of $\rm{^{56}Ni}$ formation. Since $\rm{^{56}Ni}$
synthesis requires temperatures in the range of $\sim (5-7) \times
10^9$\,K and electron fractions $Y_{\rm e} > 0.49$, we can expect that
in Model O310 $\rm{^{56}Ni}$ does not form in large amounts in regions
with mass coordinates $< 1.27\,{\rm M_{\odot}}$, where $Y_{\rm e} \ll
0.49$. This is confirmed by Fig.~\ref{fig:composition_O310}, which
shows the composition 820\,ms after bounce, when the post-shock
temperature has already dropped below $2.1\times10^{9}$\,K (compare
Fig.~\ref{fig:evol_O310_lagr}) and nuclear reactions have frozen out.
Large $\rm{^{56}Ni}$ mass fractions ($\geq 20\%$) are indeed only
found for $1.27\,{\rm M_{\odot}} \leq M_r \leq 1.37\,{\rm M_{\odot}}$,
i.e. they are confined to the dense shell. In total, 0.08\,${\rm
M_{\odot}}$ of $\rm{^{56}Ni}$ are synthesized in Model O310. Of these,
0.04\,${\rm M_{\odot}}$ are produced by shock-induced Si-burning in
the presupernova's silicon shell (whose inner boundary was initially
located at $1.32\,{\rm M_{\odot}}$). The remaining 0.04\,${\rm
M_{\odot}}$ stem from the recombination of photo-disintegrated iron
core matter with high electron fraction at mass coordinates between
$1.27\,{\rm M_{\odot}}$ and $1.32\,{\rm M_{\odot}}$. Between the
neutron star surface and the inner boundary of the dense shell,
i.e. for $1.14\,{\rm M_{\odot}} \leq M_r \leq 1.27\,{\rm M_{\odot}}$,
$0.13\,{\rm M_{\odot}}$ of material freezes out with $Y_{\rm e} <
0.49$. About 0.087\,${\rm M_{\odot}}$ of this matter end up in the
neutronization tracer nucleus (denoted by X in
Fig.~\ref{fig:composition_O310}). The rest, which contains also a
small contribution of $\rm{^{56}Ni}$, is mainly made up of
$\alpha$-particles, $\rm{^{44}Ti}$, $\rm{^{48}Cr}$, and
$\rm{^{52}Fe}$, i.e. the products of $\alpha$-rich freezeout in
high-entropy (low-density) material.

Table~\ref{tab:yields} gives an overview of the nuclear yields for our
explosion models (not including effects of eventual fallback). For a
discussion of their accuracy regarding numerical diffusion, see
\cite{KPM01}, and \cite{PM99} who analyzed the effect in detail,
especially for the case of $\rm{^{44}Ti}$. With the spatial resolution
used in the explosion models presented in this paper, the bulk of
$\rm{^{44}Ti}$ is synthesized between about 400\,ms and 700\,ms after
core bounce at $ 1.29 \leq M_r \leq 1.36\,{\rm M_{\odot}}$, in a layer
where the abundance of $\alpha$-particles decreases and $\rm{^{40}Ca}$
comes up (Fig.~\ref{fig:composition_O310}). However, if the spatial
resolution is increased, and thus numerical diffusion at the
$\rm{^{4}He/^{40}Ca}$ boundary is reduced, the width of the
layer of $\rm{^{44}Ti}$ production via the reaction $\rm
^{40}Ca(\alpha,\gamma)^{44}Ti$ is narrowed. This decreases the
$\rm{^{44}Ti}$ yield by up to a factor of 3 \citep{KPM01}.

\begin{table}[t]
\begin{center}
\caption{Final total elemental yields (in ${\rm M_{\odot}}$) for
         Models O310 and T310. Exponents are given in brackets.
         \label{tab:yields}}
\begin{tabular}{lccccccc}
\hline
\hline
  & He  & ${\rm ^{12}C} $ & ${\rm ^{16}O}$ & ${\rm ^{20}Ne}$ & ${\rm ^{24}Mg}$
  \\
\hline
 O310  & 5.3(+0) & 1.2(-1) & 1.9(-1) & 4.5(-2) & 5.9(-3) \\  
 T310  & 5.3(+0) & 1.2(-1) & 1.9(-1) & 4.5(-2) & 6.1(-3) \\
\hline
 & ${\rm ^{28}Si}$ & ${\rm ^{32}S} $  & ${\rm ^{36}Ar}$ &  ${\rm ^{40}Ca}$ & ${\rm ^{44}Ti}$ \\
\hline
 O310  & 8.6(-2) & 2.2(-2)  & 6.2(-3) & 6.2(-3) & 4.6(-3) \\
 T310  & 8.5(-2) & 2.3(-2)  & 6.6(-3) & 6.5(-3) & 5.0(-3) \\
\hline
 & ${\rm ^{48}Cr}$ &  ${\rm ^{52}Fe}$  & ${\rm ^{56}Ni}$ & ${\rm X}$ \\
\hline
 O310  & 3.6(-3) & 3.7(-3) & 8.0(-2) & 8.7(-2) \\
 T310  & 3.9(-3) & 4.0(-3) & 7.5(-2) & 1.0(-1) \\
\hline
\end{tabular}
\end{center}
\end{table}

\begin{figure}[h]
\resizebox{\hsize}{!}{\includegraphics{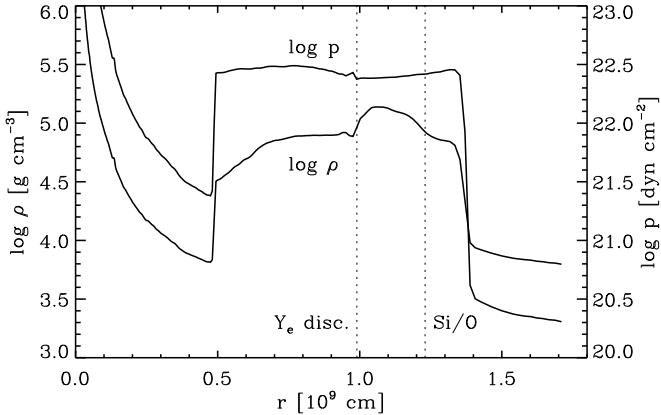}}
\caption{Density and pressure profiles of Model O310 at $t =
         820$\,ms. The position of the $Y_{\rm e}$ discontinuity and
         of the Si/O interface are indicated by dotted vertical lines.
         Note the dense shell at $r = 1.1 \times 10^{9}$\,cm, and the
         density and pressure gradients of opposite signs for $
         1.05\times10^{9}\,{\rm cm} \leq r \leq 1.35 \times 10^{9}\,
         {\rm cm}$.}
\label{fig:hump}
\end{figure}

\begin{figure*}[t]
\centering
\includegraphics[width=17cm]{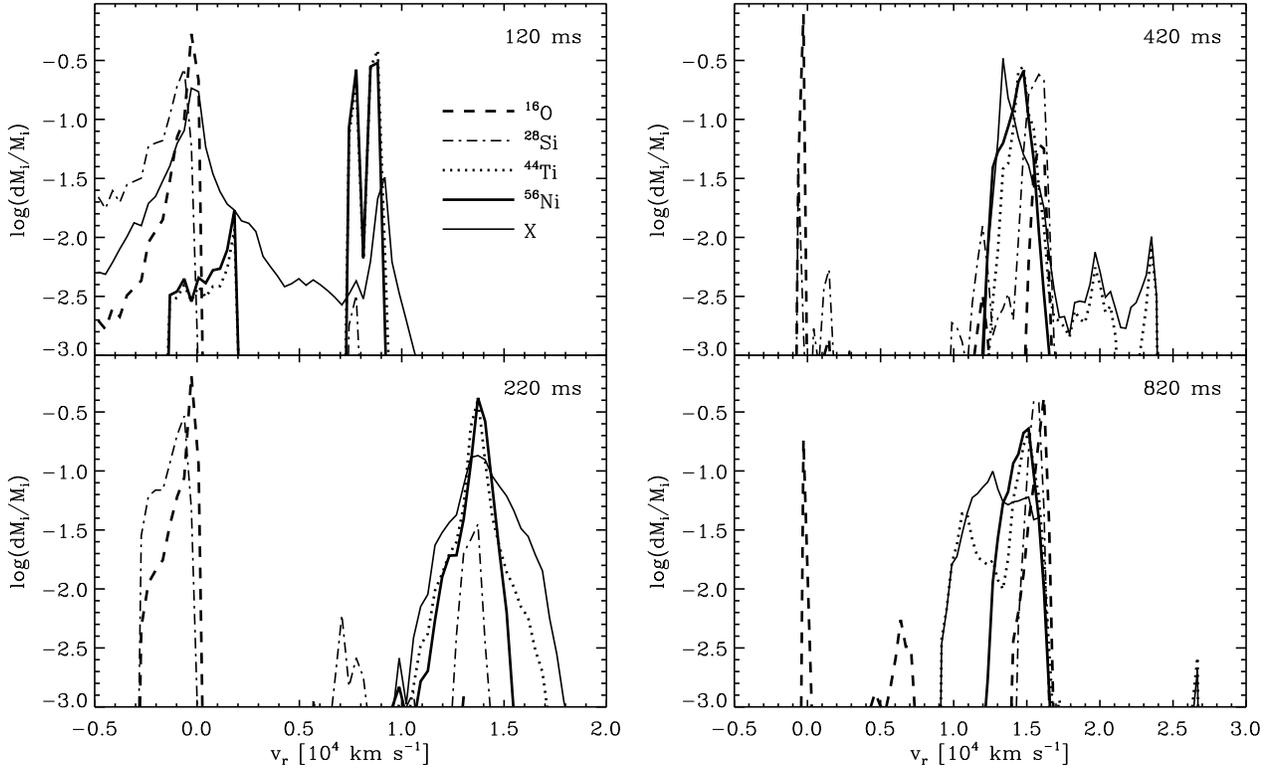}
\caption{Fractional mass of different elements contained within the
         velocity interval $[v_r,v_r+{\rm d}v_r]$ as a function of the
         radial velocity $v_r$ in Model T310.  The resolution is ${\rm
         d}v_r = 350\,{\rm km\,s^{-1}}$. }
\label{fig:massvelo_T310} 
\end{figure*}

The yield of $\rm{^{56}Ni}$ is much less sensitive to numerical
diffusion. However, it cannot be determined to much higher accuracy
than that of $\rm{^{44}Ti}$ because the contribution of the region
with $Y_{\rm e} < 0.49$ to the $\rm{^{56}Ni}$ production depends
sensitively on the ratio of the $\nu_{\rm e}$ and $\bar \nu_{\rm e}$
luminosities, $L_{\nu_{\rm e}}/L_{\bar \nu_{\rm e}}$, and its
evolution with time. This ratio, which determines the electron
fraction in regions close to the gain radius through neutrino-matter
interactions \citep[][and references therein]{JM96}, must be regarded
as uncertain in calculations like ours, that do not use accurate
neutrino transport. With the parameters adopted for Model O310 we
observe that $Y_{\rm e} \ll 0.5$ for $M_r < 1.27\,{\rm
M_{\odot}}$. However, given a higher $L_{\nu_{\rm e}}/L_{\bar \nu_{\rm
e}}$ the opposite may occur, with $Y_{\rm e}$ being increased in the
neutrino-heated layers to values up to and even above 0.5, as seen in
recent calculations including Boltzmann neutrino transport
\citep{Janka+02}. If this were the case, the composition would be
dominated by $\rm{^{56}Ni}$ also for $1.16\,{\rm M_{\odot}} \leq M_r
\leq 1.27\,{\rm M_{\odot}}$, while only insignificant amounts of
neutron-rich nuclei would be synthesized in these layers. Hence, the
$\rm{^{56}Ni}$ yield might be increased by a factor of up to about 2
compared to Model O310.

Just interior to the dense shell (and the $\rm{^{56}Ni}$-rich zone in
Model O310), we must expect substantial convective activity to occur,
because neutrino-matter interactions sustain a negative entropy
gradient in the neutrino-heated material throughout the entire
evolution up to 820\,ms. As Fig.~\ref{fig:evol_O310_lagr} shows, the
entropy starts to rise steeply when one moves inward from $1.27\,{\rm
M_{\odot}}$ and reaches a maximum near the gain radius (and at times
later than 0.5\,s at a position just downstream of the reverse
shock). This negative entropy gradient will give rise to convective
motions that, in contrast to the one-dimensional situation, will mix
high-entropy, low-$Y_{\rm e}$, matter from the deeper layers of the
iron-core with lower-entropy, high-$Y_{\rm e}$, material of the
outermost iron core and the silicon shell. We discuss the implications
of this in more detail in the next section.

Finally, we wish to draw attention to the outer boundary of the dense
shell. Figures~\ref{fig:evol_O310_lagr} and \ref{fig:hump} indicate
that after about 500\,ms this boundary coincides with the star's Si/O
interface. A pronounced negative density gradient forms at this
interface which is accompanied by a positive pressure gradient
(Fig.~\ref{fig:hump}). Such a configuration is Rayleigh-Taylor
unstable even in the absence of gravity \citep{Chevalier76}. Its
rather close spatial proximity to the convectively unstable layers at
the inner edge of the dense shell is of decisive importance for the
further evolution of the models.

\subsection{Evolution in two dimensions}
\label{sect:first_sec_2D}

Repeating the calculation of the one-dimensional model O310 in two
dimensions we obtained Model T310 that develops convective activity
already 60\,ms after core bounce. Fingers of buoyant, high-entropy
material with $Y_{\rm e} \approx 0.4$, and an initial angular width of
about $5^{\circ}-10^{\circ}$ form quickly from the imposed seed
perturbations in the neutrino heated layer near the gain radius and
start to rise toward the shock into the surrounding gas of lower
entropy.  While doing so they acquire a mushroom-capped shape and
display a tendency to merge into bubbles of about
$10^{\circ}-20^{\circ}$ lateral width
(Fig.~\ref{fig:odd-even}). Between the bubbles, low-entropy material
with $Y_{\rm e} \approx 0.498$ sinks inward in tube-like flows.
Compared to the one dimensional case, the convective rise of the
high-entropy gas leads to a somewhat more efficient energy transport,
and a reduced energy loss by the reemission of neutrinos from heated
matter.  Therefore Model T310 exploded with an energy of $E_{\rm exp}
= 1.77\times10^{51}$\,erg, which is 11\% higher than in Model
O310. The mean shock expansion velocity is, however, almost the same
as in Model O310.

While the shock propagates through the Si-rich layers of the star,
$\rm{^{56}Ni}$ is synthesized in the high-density shell
(Sect.~\ref{sect:first_sec_1D}), which is distorted at its base by the
rising bubbles of high-entropy material.  As in the one-dimensional
case, freezeout of nuclear reactions in the deleptonized bubbles leads
to high abundances of the neutronization tracer and to smaller yields
for nuclei produced by $\alpha$-rich freezeout. However, in between
the bubbles, i.e. in the down-flows mentioned above, electron
fractions are sufficiently high that $\rm{^{56}Ni}$ is synthesized.
Hence, the inner boundary of the $\rm{^{56}Ni}$ shell possesses an
aspherical shape closely tracing the inner edge of the low-entropy
(high-density) material in the convective region, while the outer
$\rm{^{56}Ni}$-boundary coincides with the (spherical) shock wave as
long as the post-shock temperatures stay above
$5\times10^{9}$\,K. After complete silicon burning has ceased at $t
\approx 270$\,ms, an anisotropic nickel shell is left behind, while
the spherical shock continues moving outward (see
Fig.~\ref{fig:odd-even}).

The differences in the spatial nickel distribution between Model T310
and its 1D analogue, Model O310, are relatively small, however. This
is caused by the high initial neutrino luminosities and their rapid,
exponential decline (Eq.~\ref{eq:Lumnu}) adopted for our models. Both
assumptions favour small explosion time scales which prevents
convective bubbles to merge to large-scale structures, as in cases
where the phase of convective overturn lasts for several turn-over
times.  Lowering the neutrino luminosities (and the explosion
energies), we obtain stronger convection that largely distorts the
shock wave by developing big bubbles of neutrino-heated material (see
\citealt{JM96}, \citealt{Kifonidis+00}, \citealt{KPM01}, and
\citealt{JKR01} for examples).  Adopting constant core luminosities in
contrast to the exponential law of Eq.~(\ref{eq:Lumnu}), we can
produce models that exhibit the vigorous boiling behaviour reported by
\cite{BHF95}. Such cases can finally develop global anisotropies,
showing a dominance of the $m=0$, $l=1$ mode of convection (see
\citealt{Janka+02}; Scheck et al., in preparation). As a consequence,
convection can lead to huge deviations of the spatial distribution of
iron group nuclei from spherical symmetry, much stronger than those
visible in Fig.~\ref{fig:odd-even}.  Convection can also change the
nucleosynthetic yields of iron group nuclei, because it enhances
neutronization by cycling high-$Y_{\rm e}$ material from outer layers
through the regions near the gain radius where the gas loses lepton
number \citep{KPM01,JKR01}.  Due to our adopted $\nu_{\rm e}$ and
$\bar \nu_{\rm e}$ luminosities the gas cannot regain high $Y_{\rm e}$
values when it moves outward again. This leads to $\rm{^{56}Ni}$
yields that are about 6\% smaller in Model T310 as compared to the 1D
case, Model O310, while the final yield of the neutronization tracer
shows the opposite behaviour and is 15\% larger in Model T310 than in
Model O310. These effects are comparatively small in Model T310, again
mainly due to the short explosion time scale of this model.

The distribution of the elements in velocity space is of significant
importance for an understanding of the evolution of the supernova
beyond the first second. Since Models O310 and T310 barely differ in
this respect we discuss only the two-dimensional case.  In
Fig.~\ref{fig:massvelo_T310} we plot as functions of the radial
velocity $v_r$ and time, the fractions of the total mass of
$\rm{^{16}O}$, $\rm{^{28}Si}$, $\rm{^{44}Ti}$ and $\rm{^{56}Ni}$ that
are contained within the velocity intervals $[v_r, v_r + {\rm d}v_r]$
with ${\rm d}v_r = 350\,{\rm km\,s^{-1}}$. At $t = 120$\,ms, when the
shock is travelling through the outermost layers of the iron core,
substantial amounts of material from the silicon and oxygen shells are
either still near hydrostatic equilibrium or falling towards the
shock. Consequently the peak and wings of the mass distributions of
the respective elements have zero and negative velocities. As the
post-shock temperature begins to drop below $7\times10^9$\,K,
reassembly of $\alpha$-particles to heavier nuclei starts in the
immediate post-shock region and leads to initially small abundances of
$\rm{^{44}Ti}$ and $\rm{^{56}Ni}$, the bulk of which has velocities
around $8\times10^{3}$\,km/s.
\begin{figure*}
\begin{tabular}{ll}
\includegraphics[width=7.5cm]{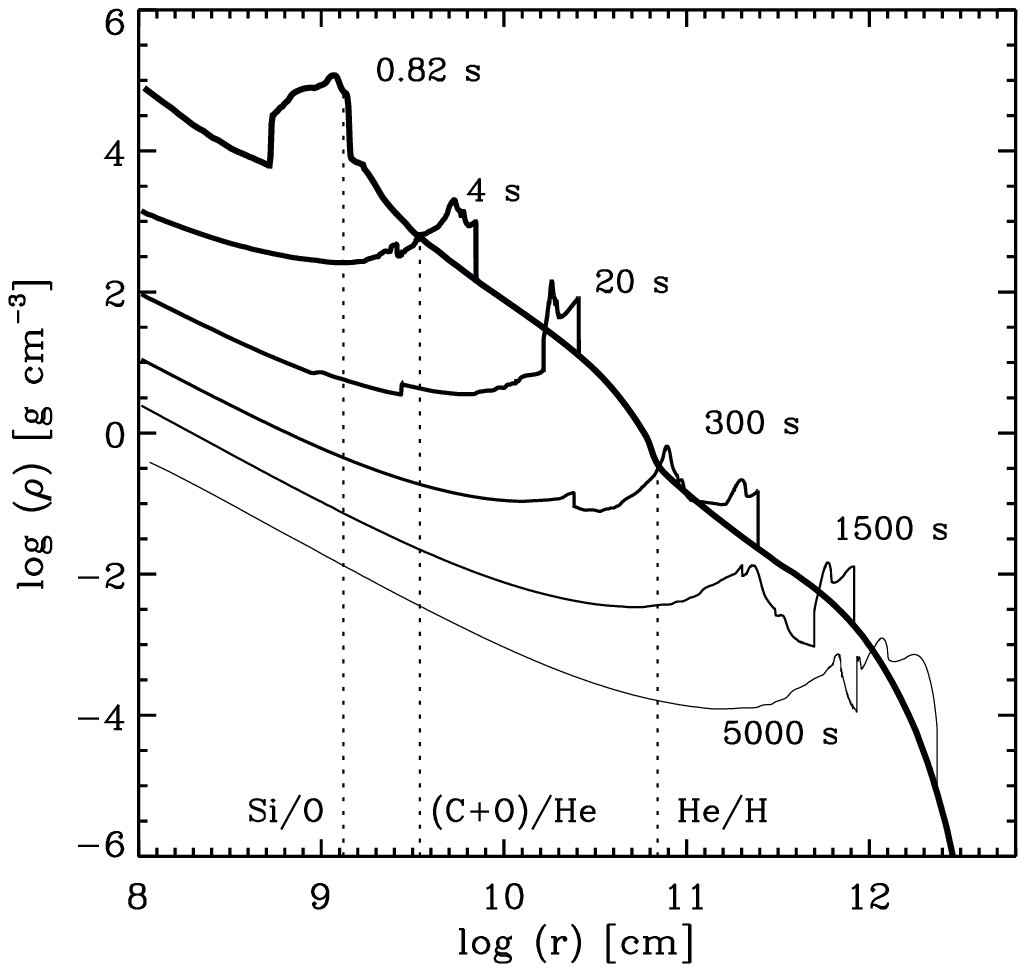} &
\includegraphics[width=10.cm]{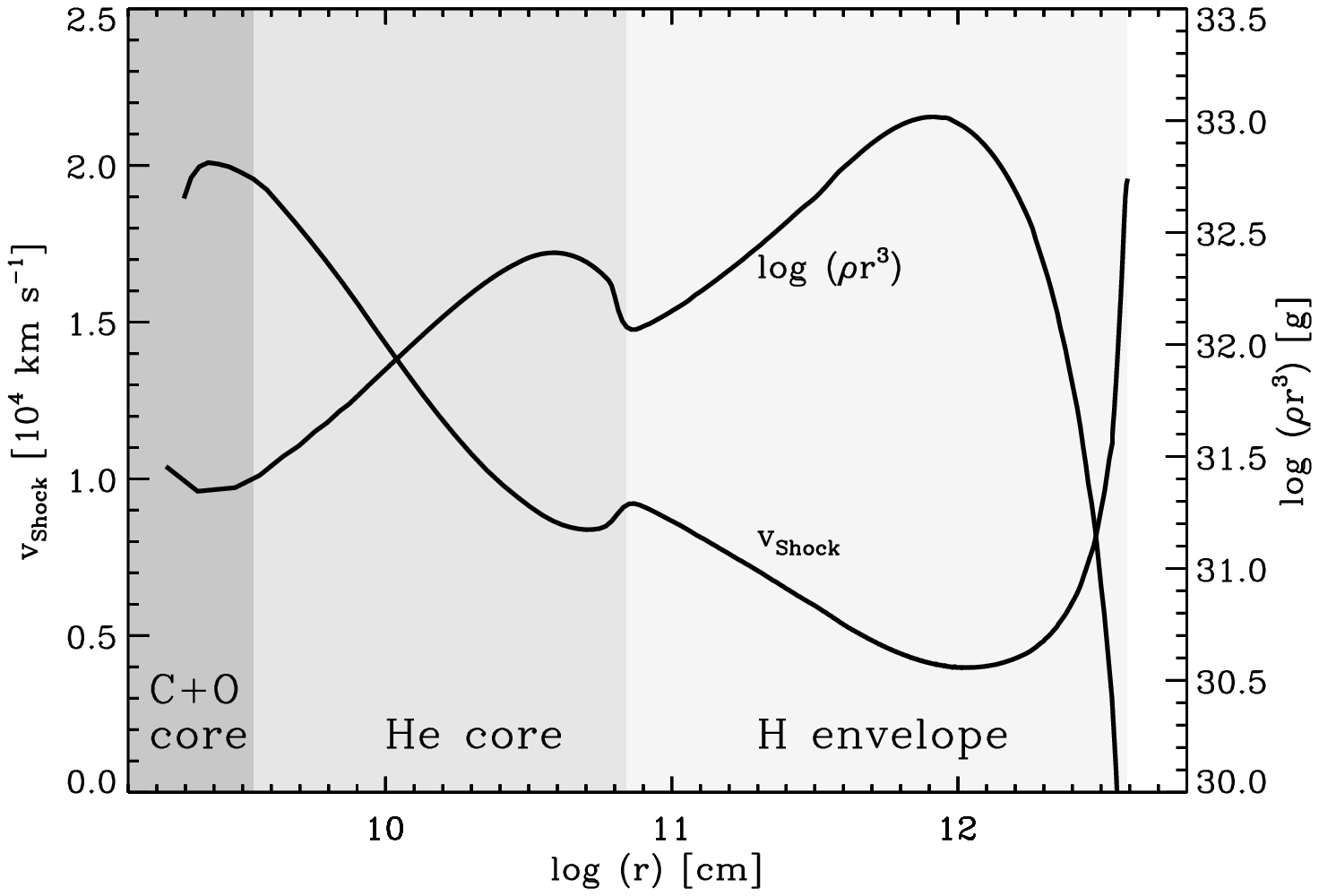}  
\end{tabular}
\caption{a) Evolution of the density in our one-dimensional Type~II
         supernova simulation Model $\overline{\rm T310}$ (left).  The
         dotted lines indicate the positions of the Si/O, (C+O)/He and
         He/H interfaces 0.82\,s after core bounce.  b) Shock velocity
         as a function of shock radius in Model $\overline{\rm T310}$,
         and radial dependence of $\rho r^3$ for our progenitor model
         (right).  Note the deceleration of the shock after it has
         crossed the (C+O)/He interface at $\log (r/{\rm cm}) = 9.55$
         and the He/H interface at $\log (r/{\rm cm}) = 10.85$. The
         increase of the shock velocity for $\log (r/{\rm cm}) \geq
         12.2$ is due to the steep drop of the density in the
         atmosphere of the star.}
\label{fig:amra_OAMR}
\end{figure*}

Only 100\,ms later ($t = 220$\,ms) the explosion gains momentum, and
the shock propagates through the silicon shell and encounters
substantially smaller maximum infall velocities of only
3000\,km/s. This leads to the cutoff of the high velocity wings of
silicon and oxygen at $-3\times10^{3}$\,km/s.  Downstream of the
shock, $\rm{^{56}Ni}$ is synthesized as the dominant nucleus in
regions of explosive silicon burning, while also $\rm{^{44}Ti}$ is
built up in smaller amounts. The spatial neighborhood of these two
nuclei is reflected in their velocity distributions, which both peak
around $1.35\times10^{4}$\,km/s.  However, in contrast to
$\rm{^{56}Ni}$, some $\rm{^{44}Ti}$ also forms by $\alpha$-rich
freezeout in the rising convective blobs that at this stage show the
highest velocities on the grid. This causes the broad wing in the
velocity distribution of $\rm{^{44}Ti}$ up to $1.7\times10^{4}$\,km/s,
while maximum $\rm{^{56}Ni}$ velocities reach
$1.55\times10^{4}$\,km/s. The velocity distribution for the
neutronization tracer shows a behaviour similar to $\rm{^{44}Ti}$.

At 420\,ms the supernova's explosion energy is still increasing and
has led to maximum $\rm{^{56}Ni}$ velocities as high as
$1.65\times10^{4}$\,km/s.  Substantial amounts of silicon have also
been accelerated outward. A stratification of these elements in
velocity space is beginning to emerge because the positive velocity
gradient behind the shock (see Fig.~\ref{fig:evol_O310}) results in
higher velocities in the post-shock region as compared to gas in the
deeper layers of the ejecta. Meanwhile, the neutrino-driven wind has
begun to blow off material from the proto-neutron star surface with
velocities in excess of $2.0\times10^{4}$\,km/s
(Fig.~\ref{fig:evol_O310}). Traces of $\rm{^{44}Ti}$ in the wind, that
has a composition dominated by $\alpha$-particles and neutron-rich
nuclei, can be seen at these velocities. The distribution of
$\rm{^{44}Ti}$ in velocity space is somewhat uncertain.  We have noted
in Sect.~\ref{sect:first_sec_1D} that the bulk of $\rm{^{44}Ti}$ is
synthesized at the boundary of layers with high $\rm{^{4}He}$ and
$\rm{^{40}Ca}$ abundances. The location of this region coincides with
the maximum of the $\rm{^{56}Ni}$ abundance
(Fig.~\ref{fig:composition_O310}). Thus, the peaks of both elements
are found at comparable velocities ($\sim 1.4\times10^{4}$\,km/s;
Fig.~\ref{fig:massvelo_T310}).  We note again, however, that the
spatial distribution and the yield of $\rm{^{44}Ti}$ are severely
affected by numerical diffusion \citep{PM99}.  The location of the
$\rm{^{44}Ti}$ peak at $t = 420$\,ms (Fig.~\ref{fig:massvelo_T310})
might actually occur at much lower velocities, if numerical diffusion
(and thereby $\rm{^{44}Ti}$ synthesis at the
$\rm{^{4}He/^{40}Ca}$ boundary) is substantially reduced. In
this case, the contribution from the slower, deeper layers of the
ejecta, just downstream of the reverse shock, might dominate. Indeed
this material causes the second, smaller $\rm{^{44}Ti}$ peak near
$1.0\times10^{4}$\,km/s at $t = 820$\,ms (lower right panel of
Fig.~\ref{fig:massvelo_T310}).

The velocities of the fastest regions that contain $\rm{^{56}Ni}$
reach a maximum of $1.7\times10^{4}$\,km/s around $t = 620$\,ms and
decrease slightly to $1.65\times10^{4}$\,km/s at $t = 820$\,ms. It is
obvious, that these $\rm{^{56}Ni}$ velocities are much larger than
those observed in SN 1987\,A and that some slow-down of this material
must have taken place later. In fact, we will show in
Sect.~\ref{sect:AMR_2D} that excessive deceleration of this material
during the subsequent evolution is the main problem for an explanation
of the observed $\rm{^{56}Ni}$ velocities in SN 1987\,A.

Summarizing our one and two-dimensional results for the evolution
within the first second, we note that a Rayleigh-Taylor unstable
density inversion can form at the Si/O interface of a blue supergiant.
The quickly expanding $\rm{^{56}Ni}$ is situated just interior to this
interface and is distributed anisotropically in an inhomogeneous
layer. This results in a seed perturbation for the Rayleigh-Taylor
unstable regions. We will show in the following sections that the
magnitude of this perturbation is sufficient to induce significant
outward mixing of $\rm{^{56}Ni}$.

\section{Beyond the first second: A Type~II model}
\label{sect:AMR_models}

\subsection{One-dimensional evolution}
\label{sect:AMR_1D}

\begin{figure*}
\centering
\includegraphics[width=17cm]{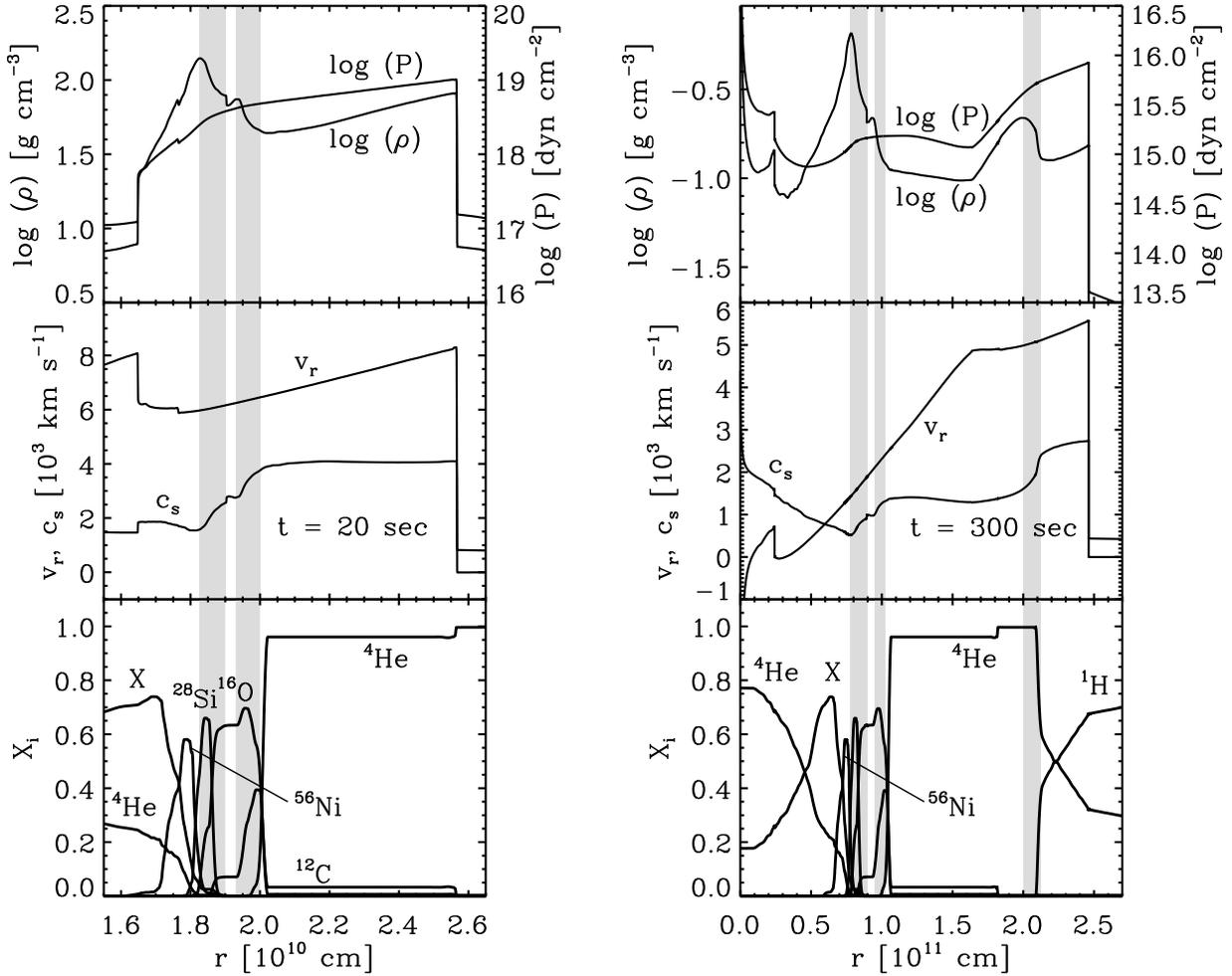}
\caption{Structure of Model $\overline{\rm T310}$, 
         a) 20\,s (left), and b) 300\,s
         after bounce (right). Shown are (from top) the density,
         pressure, velocity, sound speed, and mass fractions as
         functions of radius. Note the unstable regions (shaded in
         grey) near the Si/O, (C+O)/He and He/H composition
         interfaces, where the density and pressure gradients have
         opposite signs. At $t=300$\,s, the density gradient at the
         inner edge of the dense shell that has formed below the He/H
         interface is in the process of steepening into a reverse
         shock (compare also Fig.~\ref{fig:amra_OAMR}a). Note also the
         enormous drop of the velocity in the layers that are enriched
         in $\rm{^{56}Ni}$, from 6000\,km/s at 20\,s to 1300\,km/s at 300\,s.}
\label{fig:amra_late_evol_1d}
\end{figure*}

Once the shock wave has been launched by neutrino heating and the
explosion energy has saturated, the further evolution of the supernova
depends strongly on the density profile of the progenitor star.
According to the analytic blast wave solutions of \cite{Sedov}, the
shock decelerates whenever it encounters a density profile that falls
off with a flatter slope than $\propto r^{-3}$ (i.e. for increasing
$\rho r^{3}$), while for density profiles steeper than $\propto
r^{-3}$ (i.e. for decreasing $\rho r^{3}$) it accelerates.  Since
models of supernova progenitors \citep[see e.g.][]{Woosley_Weaver95}
do not show a density structure that can be described by a single
power-law, an unsteady shock propagation results that in turn gives
rise to Rayleigh-Taylor unstable pressure and density gradients at the
composition interfaces of the star.  In this section we follow shock
propagation beyond the first second with our AMR code in one spatial
dimension to illustrate these effects.  In order to obtain the closest
possible approximation to the energetics of the two-dimensional case,
with which we will compare our results in the following section, we
start this 1D calculation from an angularly averaged version of our 2D
explosion, Model T310, (instead of the one-dimensional model O310).
We will henceforth refer to the one-dimensional results of this
section as Model $\overline{\rm T310}$.

\begin{figure*}
\centering
\includegraphics[width=15cm]{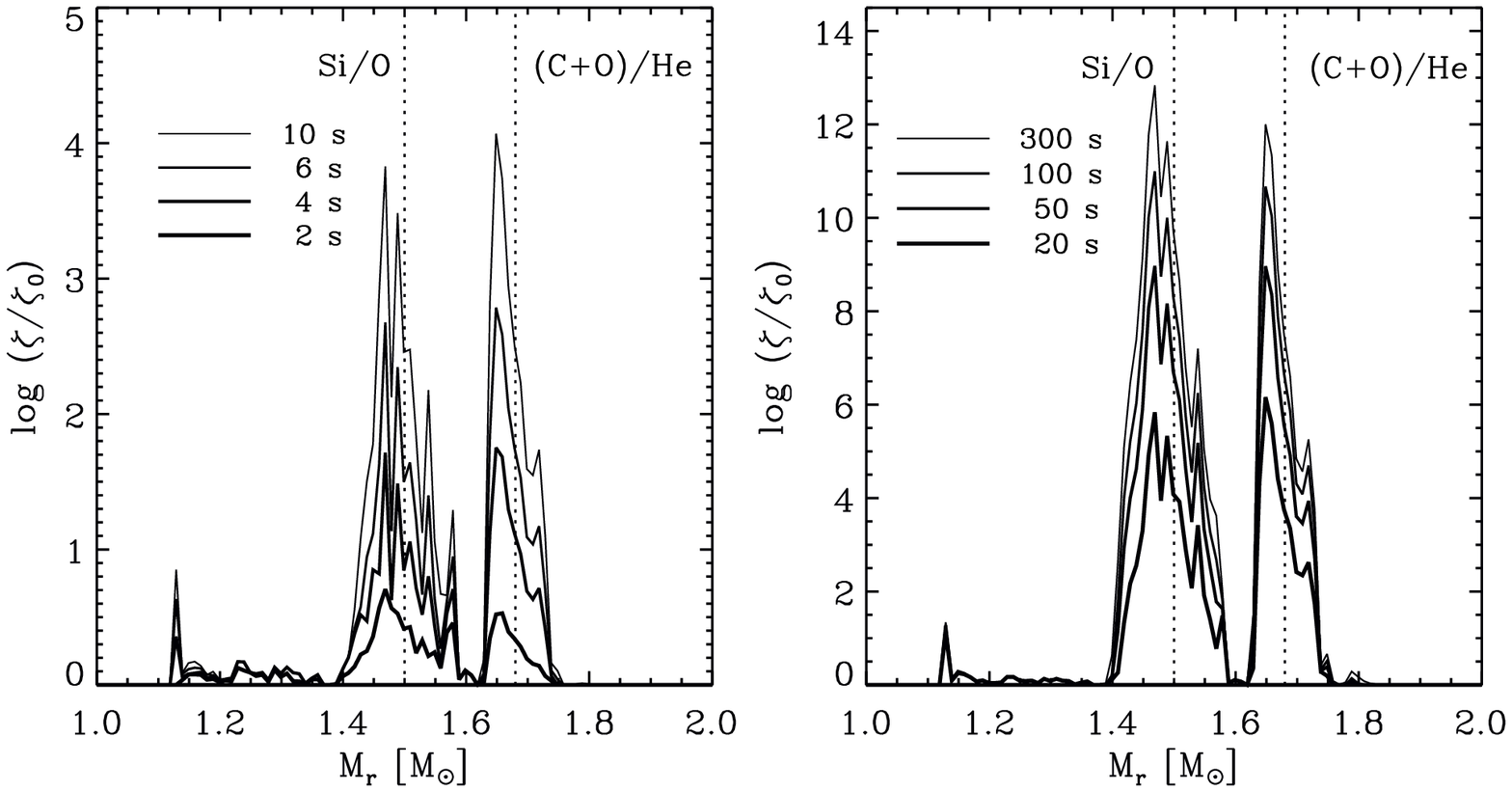}
\caption{Logarithm of the total, time-integrated, \emph{linear} growth
         rate in the unstable layers near the Si/O and (C+O)/He
         composition interfaces of Model $\overline{\rm T310}$ at
         different times.  The linear growth rates increase with time
         and reach amplification factors $\geq 10^{12}$ within the
         first 300 seconds (i.e. five minutes) of the explosion.}
\label{fig:amra_growth_a}
\end{figure*}

Figure~\ref{fig:amra_OAMR}a displays the evolution of the density for
this model.  The locations of the Si/O, (C+O)/He, and He/H interfaces
at the start of the calculation (i.e. 0.82\,s after bounce) are also
indicated.  In Fig.~\ref{fig:amra_OAMR}b we show the shock velocity as
a function of shock radius for times later than about one second after
core bounce, along with the radial variation of $\rho r^3$ for our
progenitor model.  When the shock wave crosses the (C+O)/He interface
at $\log (r/{\rm cm}) = 9.55$, 1.8\,s after bounce, its propagation
speed is as high as $\sim 20\,000$\,km/s.  Thereafter, a rapid
decrease of the shock velocity can be seen, since $\rho r^3$
increases, i.e. the density profile in the He core falls off with a
more shallow slope than $\propto r^{-3}$. The deceleration lasts until
the blast wave has reached the outer layers of the He core and enters
a narrow region of decreasing $\rho r^3$ around $\log (r/\rm{cm}) =
10.7$. This region is characterized by a rapid drop of the density
just below the He/H interface (see Fig.~\ref{fig:amra_OAMR}a). Here
the shock temporarily accelerates to about 9000\,km/s
(Fig.~\ref{fig:amra_OAMR}b). Once it has passed the He/H interface
around 80\,s after core bounce, the evolution resembles that after the
crossing of the (C+O)/He interface. A gradual deceleration of the
shock to 4000\,km/s occurs, until the blast wave enters the atmosphere
of the star and finally accelerates to more than 20\,000\,km/s, while
propagating off the numerical grid.

Every time the shock decelerates, it leaves behind a positive pressure
gradient which slows down the post-shock layers. The shocked material
thus piles up and forms a high-density post-shock shell.  Note that
layers which are sufficiently far downstream of the shock may be out
of sonic contact with it, i.e. on the time scale of the supernova
expansion these layers may not be reached by sound waves originating
in the immediate post-shock region. In this case, the only way in
which information can be mediated to the deeper layers is through a
supersonic wave. Therefore their slow-down proceeds via a reverse
shock, that forms at the inner boundary of the high-density shell of
decelerated matter and starts to propagate inwards in mass. The
reverse shocks as well as the dense shells that form after the main
shock has passed the Si/O, (C+O)/He and He/H interfaces, can be
clearly seen in Fig.~\ref{fig:amra_OAMR}a in the density structures
for 0.82\,s, 20\,s and 1500\,s, respectively. It is also apparent that
the density profiles for the latter two times show a striking
similarity in the region that is bounded by the forward and the
reverse shocks.

In a one-dimensional calculation, where hydrodynamic instabilities are
absent, dense shells, that have formed during earlier phases of the
evolution, are preserved provided the numerical diffusivity of the
employed hydrodynamic scheme is sufficiently small.  This is
demonstrated by the left panel of Fig.~\ref{fig:amra_late_evol_1d}
which shows some flow quantities 20\,s after core bounce. Note that
two high-density peaks are visible between the forward and reverse
shock at that time. The inner is the one that formed at the Si/O
interface (Sect.~\ref{sect:first_sec_1D}). It contains mainly silicon
and oxygen. The outer one has formed at the (C+O)/He interface and
contains carbon and oxygen. In the layers of these shells that are
shaded in gray in Fig.~\ref{fig:amra_late_evol_1d}, negative density
and positive pressure gradients exist. Thus, already at this early
time, Rayleigh-Taylor instabilities may grow in these regions
\citep{Chevalier76}. This is confirmed quantitatively by performing a
linear stability analysis \citep[see e.g.][and the references therein;
note that our calculations differ from these previous simulations
because we start from a model that has a consistent explosion history,
instead of having the explosion energy artificially deposited in the
stellar core]{MFA91,Iwamoto+97}.

\begin{figure}[t]
\resizebox{\hsize}{!}{\includegraphics{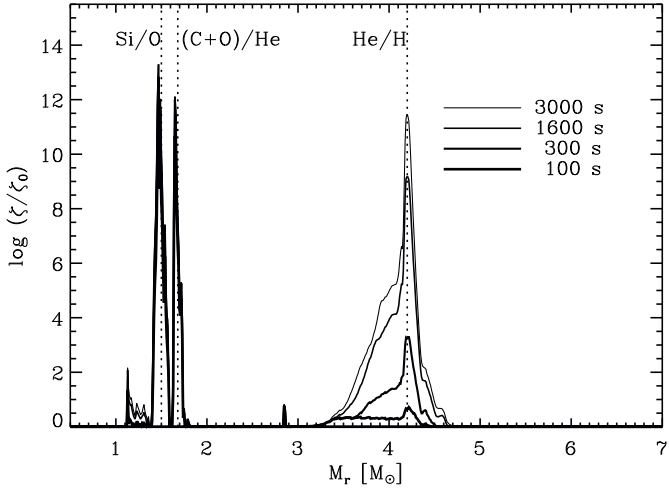}}
\caption{Logarithm of the total, time-integrated, \emph{linear} growth
         rate in the unstable layers near the Si/O, (C+O)/He and He/H
         composition interfaces of Model $\overline{\rm T310}$ at
         different times.
         Comparing the amplitudes of the growth rates at the Si/O and
         (C+O)/He interfaces with those given in
         Fig.~\ref{fig:amra_growth_a} one can see that these have
         saturated after 300\,s (in fact the curves for 300\,s,
         1600\,s, and 3000\,s are not visible individually because
         they are identical within plotting accuracy).  Note, however,
         that the rate for the instability at the He/H interface is
         still increasing between 300\,s and 3000\,s.}
\label{fig:amra_growth_b}
\end{figure}

In Fig.~\ref{fig:amra_growth_a} we plot for different times the total,
time-integrated growth rate
\begin{equation}
{\zeta \over \zeta_0} = \exp \left( {\int_0^t \sigma dt' } \right),
\end{equation}
i.e. the factor by which a (small) perturbation with an initial
amplitude $\zeta_0$ would be magnified at time $t$ according to linear
perturbations theory. The growth rate of the instability, $\sigma$, is
given by
\begin{equation}
\sigma = \sqrt{- {P \over \rho} {\partial \ln P \over \partial r}
                                {\partial \ln \rho \over \partial r}},
\end{equation}
\citep[e.g.][]{MFA91}, where we have assumed the fluid to be
incompressible for simplicity. Within only 20\,s after bounce, the
amplification factor grows to about $10^6$ in both unstable regions,
which are separated by a narrow stable layer at $1.6\,{\rm
M_{\odot}}$. Both growth rates are increasing until 300 seconds after
bounce and saturate thereafter. The right panel of
Fig.~\ref{fig:amra_late_evol_1d} shows that at $t=300$\,s the metal
core of the star has piled up to a high-density zone that includes the
two unstable interfaces.  At this time, the shock has already passed
the He/H interface and a third unstable region has formed at the outer
edge of the dense shell that the shock has left behind at the He/H
interface.  The evolution of the growth rate in this region is shown
in Fig.~\ref{fig:amra_growth_b}. At $t=300$\,s it is much smaller than
the rates for the instabilities at the Si/O and (C+O)/He
interfaces. However, while the latter reach their maxima at this
epoch, the growth rate at the He/H interface still increases.  It
reaches about the same level as the growth rates at the two inner
interfaces at 3000 seconds after core bounce.

This behaviour is opposite to that found in the calculations of
\cite{MFA91}, where the instability initially grows faster at the He/H
interface and only after some time is surpassed by the growth at the
(C+O)/He interface. It is very likely, that this result of
\cite{MFA91} is caused by the fact that they started their stability
analysis as well as their 2D calculations 300\,s after core bounce.
\emph{Reliable calculations of the Rayleigh-Taylor growth, however,
require consideration of the very early moments of the
explosion}. While the results of the linear stability analysis are
certainly invalid quantitatively once non-linear growth sets in, the
above conclusion is substantiated by our two-dimensional
calculations. It is also supported by results of \cite{Iwamoto+97} who
found a similar behaviour for the growth of the instability at the
(C+O)/He and He/H interfaces of their 2D models for SN~1993\,J.

To our knowledge, large growth rates for Rayleigh-Taylor instabilities
at the Si/O interface of a massive star have never been reported
before. We can only speculate about the reason for the absence of this
unstable layer in previous simulations.  Either it was caused by the
fact that the explosion was initiated artificially by depositing
energy in an ad hoc way, or it was caused by insufficient numerical
resolution, or by differences in the structure of the progenitor stars
that were employed in these studies. The latter is obviously the case
for the simulations of \cite{AFM89}, \cite{FMA91} and \cite{MFA91} who
used a stellar model at the end of carbon burning for their
calculations. Also a combination of the effects listed above is
possible. High-resolution studies with our hydrodynamic code applied
to different progenitor models may be required to clarify this issue.
We stress this point because the two-dimensional models to be
discussed in the next section demonstrated that the instability at the
Si/O interface {\em is the most crucial one\/} for determining the
subsequent evolution.

\subsection{Two-dimensional evolution}
\label{sect:AMR_2D}

\subsubsection{Clump formation and mixing}
\label{sect:clump_formation}

In contrast to the one-dimensional case, in two (and three) spatial
dimensions the dense shells that form at the composition interfaces
can fragment into a set of smaller clumps that may decouple from the
flow and move ballistically through the ejecta.  Such a fragmentation
actually occurs in our models due to the spatial proximity of the
Rayleigh-Taylor unstable zones near the Si/O and (C+O)/He interfaces
and the convective layers. To demonstrate these effects, we present a
two-dimensional Type~II supernova calculation that was started from
the shock-revival model T310. In what follows we will refer to this
simulation as Model T310a. To illustrate its hydrodynamic evolution we
show plots\footnote{Movies from this simulation are available under
\tt{http://www.mpa-garching.mpg.de/\mbox{$\sim$}kok/MPG}} of the mass
density and the partial densities of the nuclei $\rm{^{16}O}$,
$\rm{^{28}Si}$ and $\rm{^{56}Ni}$.

\begin{figure*}
\centering
\begin{tabular}{ccc}
\includegraphics[width=5.7cm]{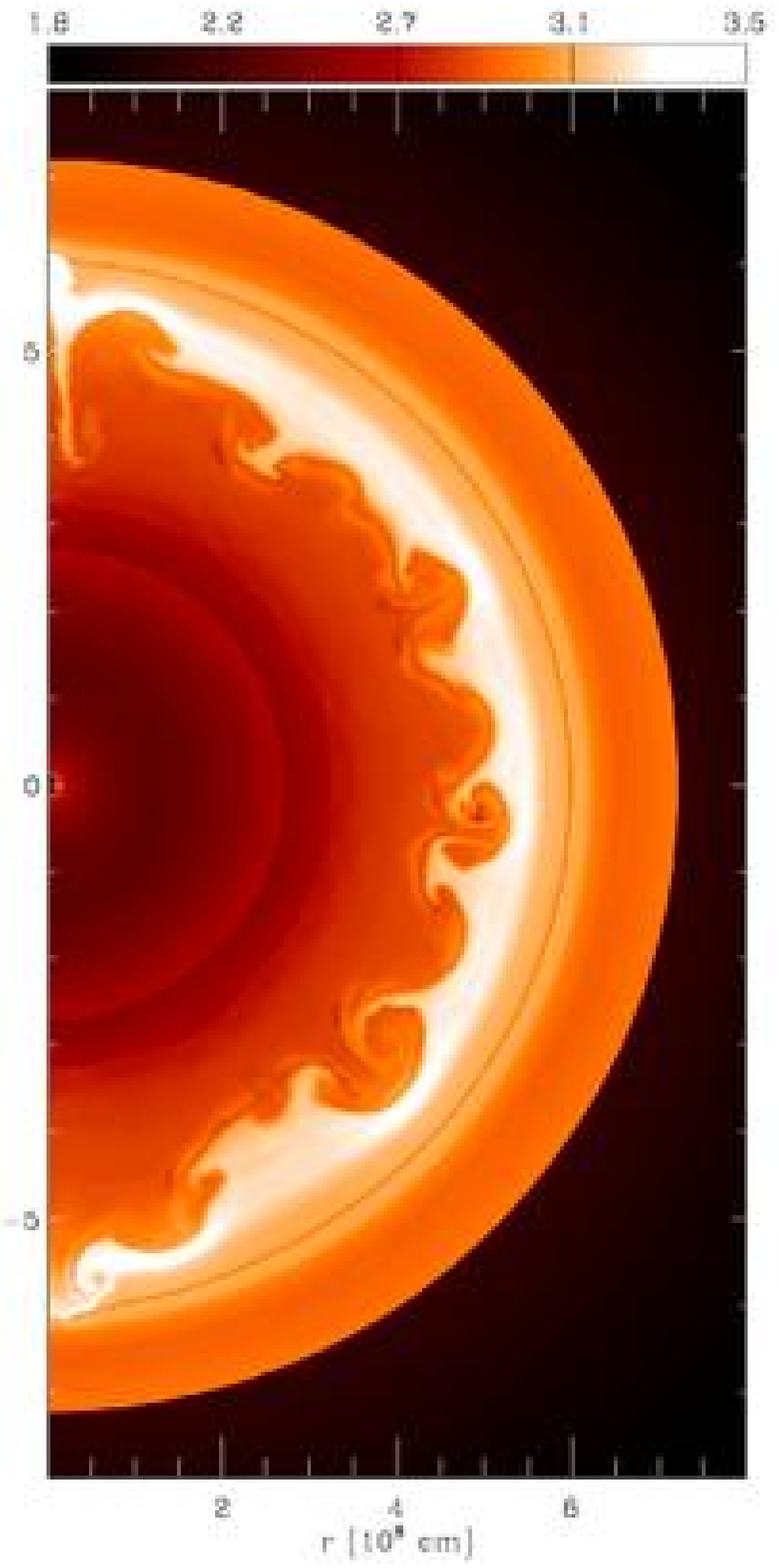}  &
\includegraphics[width=5.7cm]{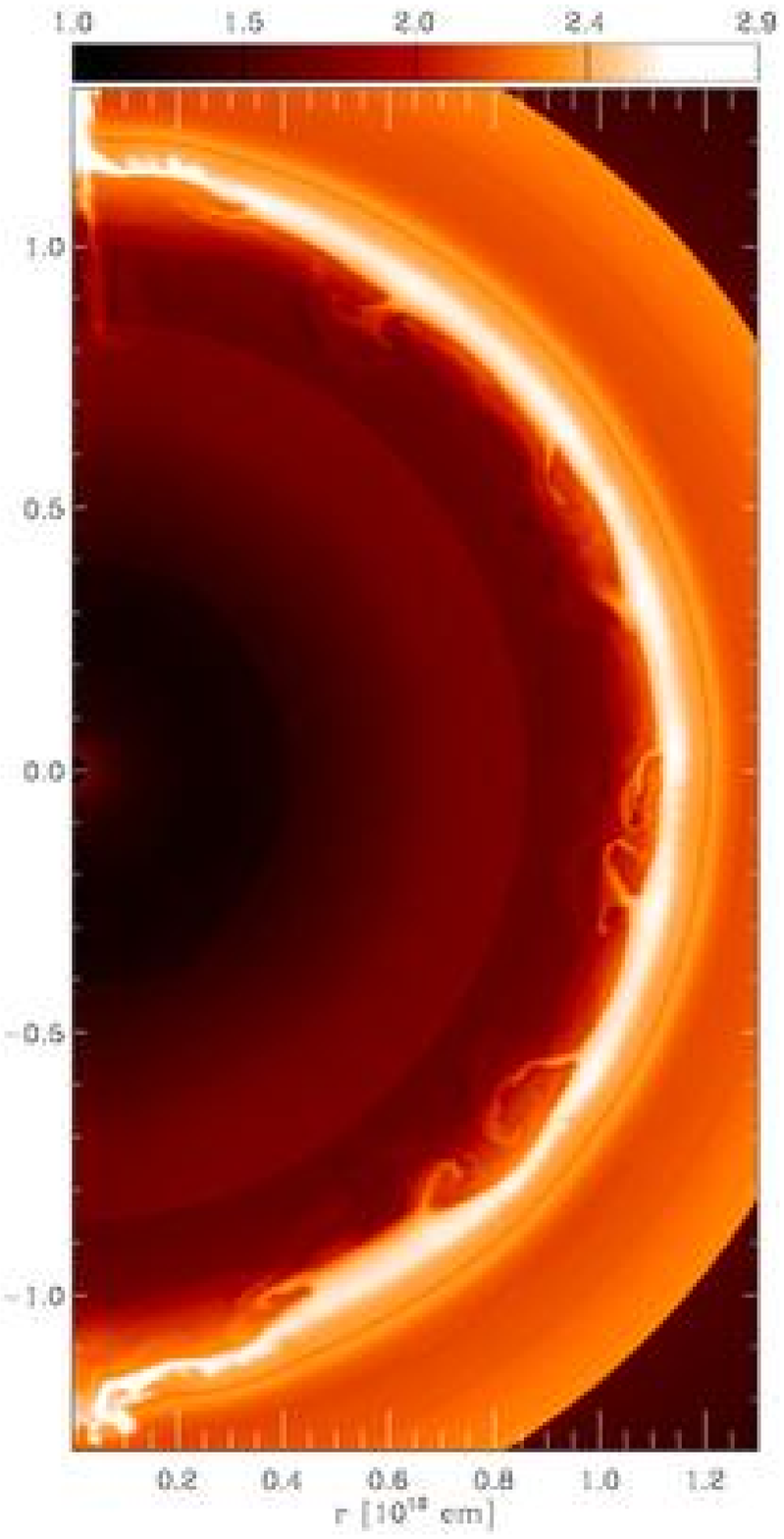} &
\includegraphics[width=5.7cm]{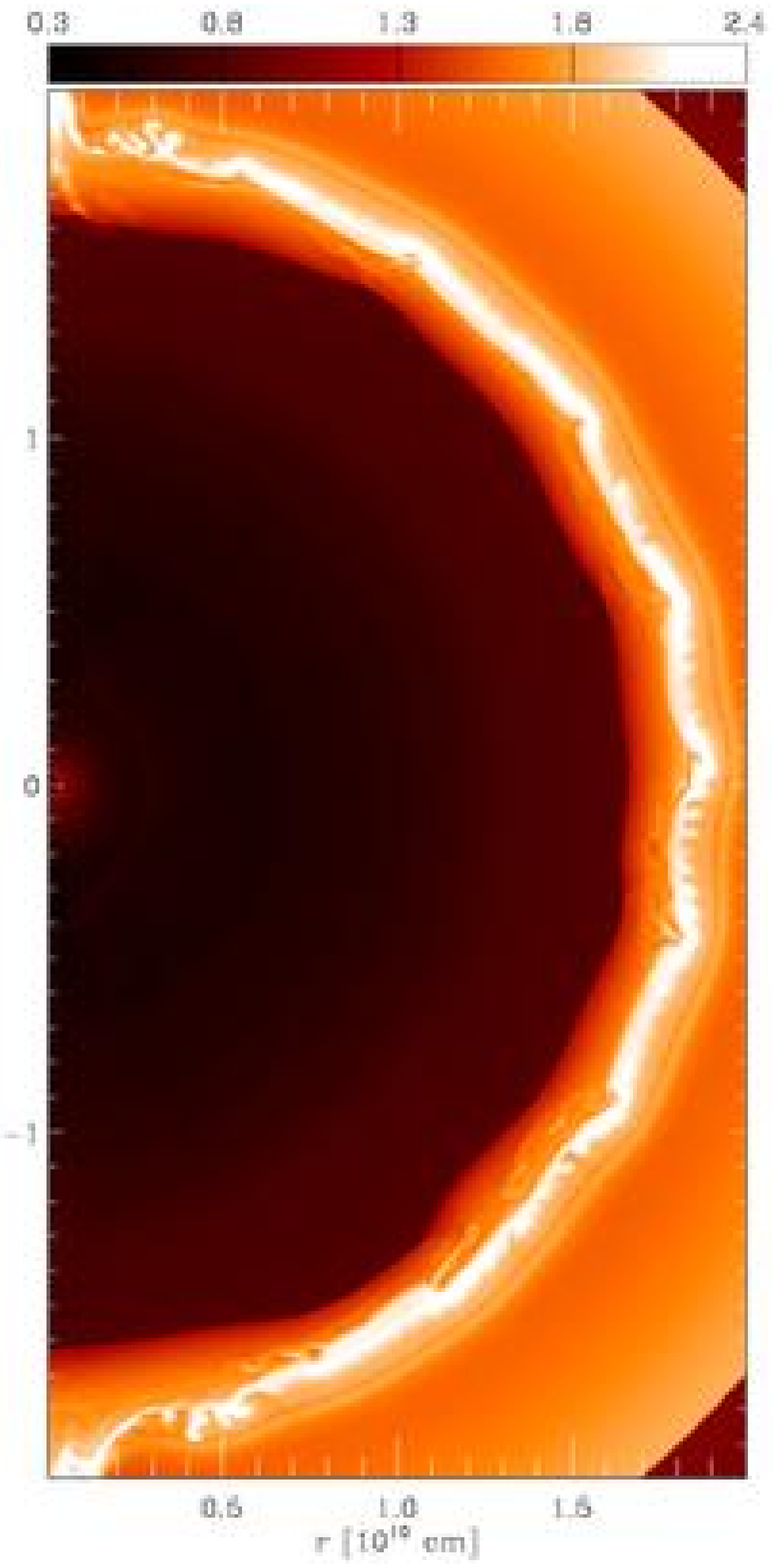} \\
\includegraphics[width=5.7cm]{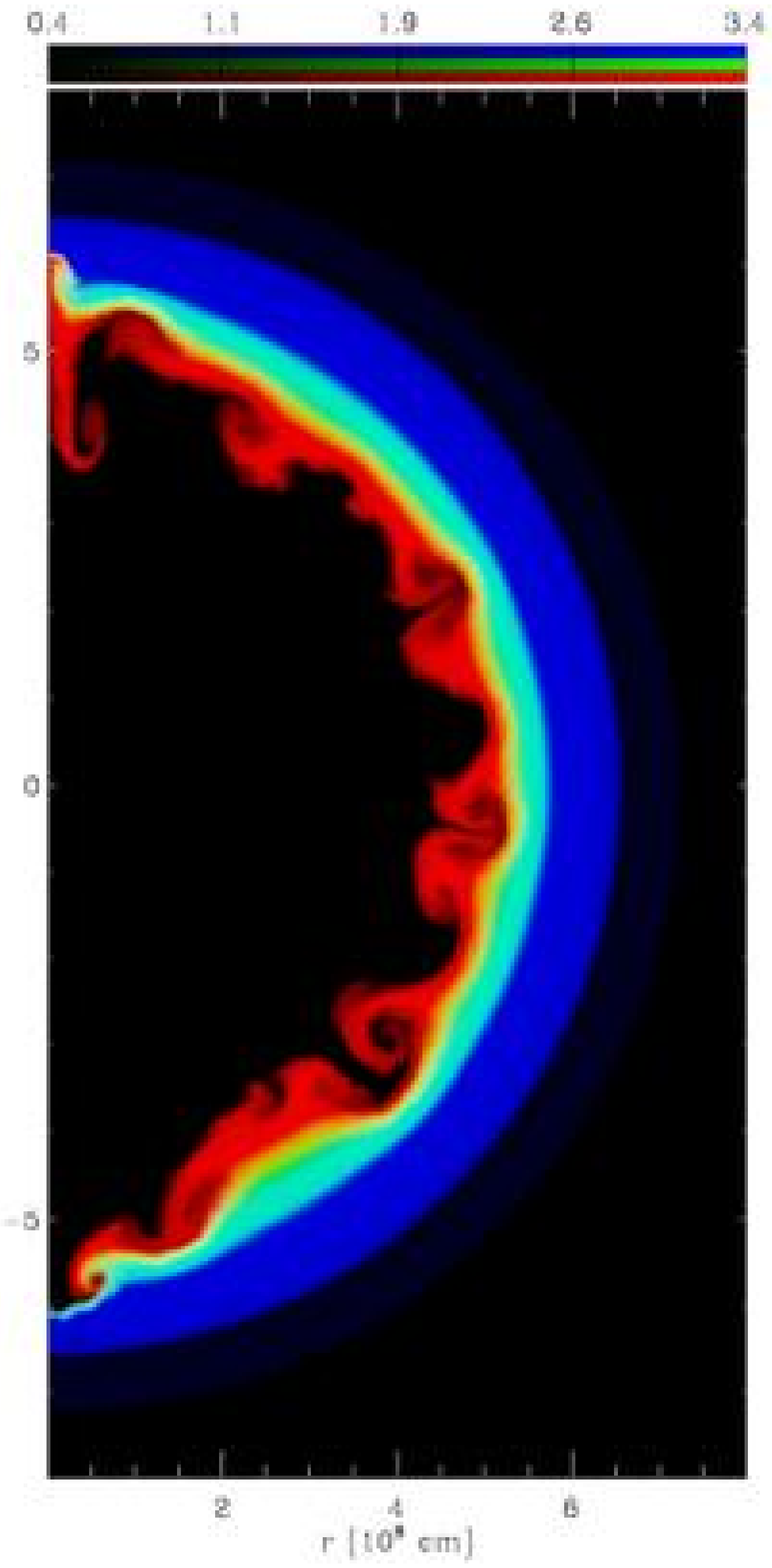}  &
\includegraphics[width=5.7cm]{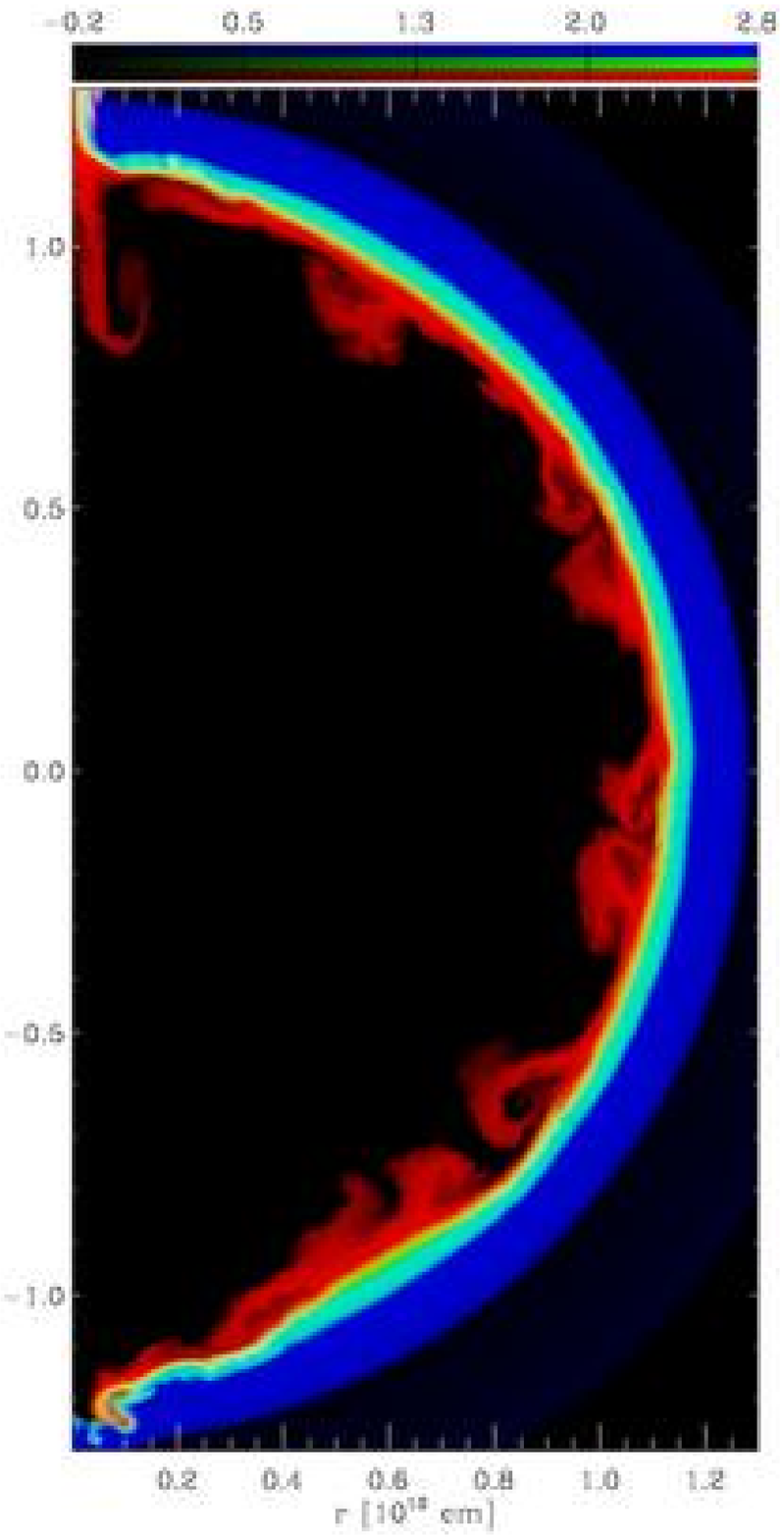} &
\includegraphics[width=5.7cm]{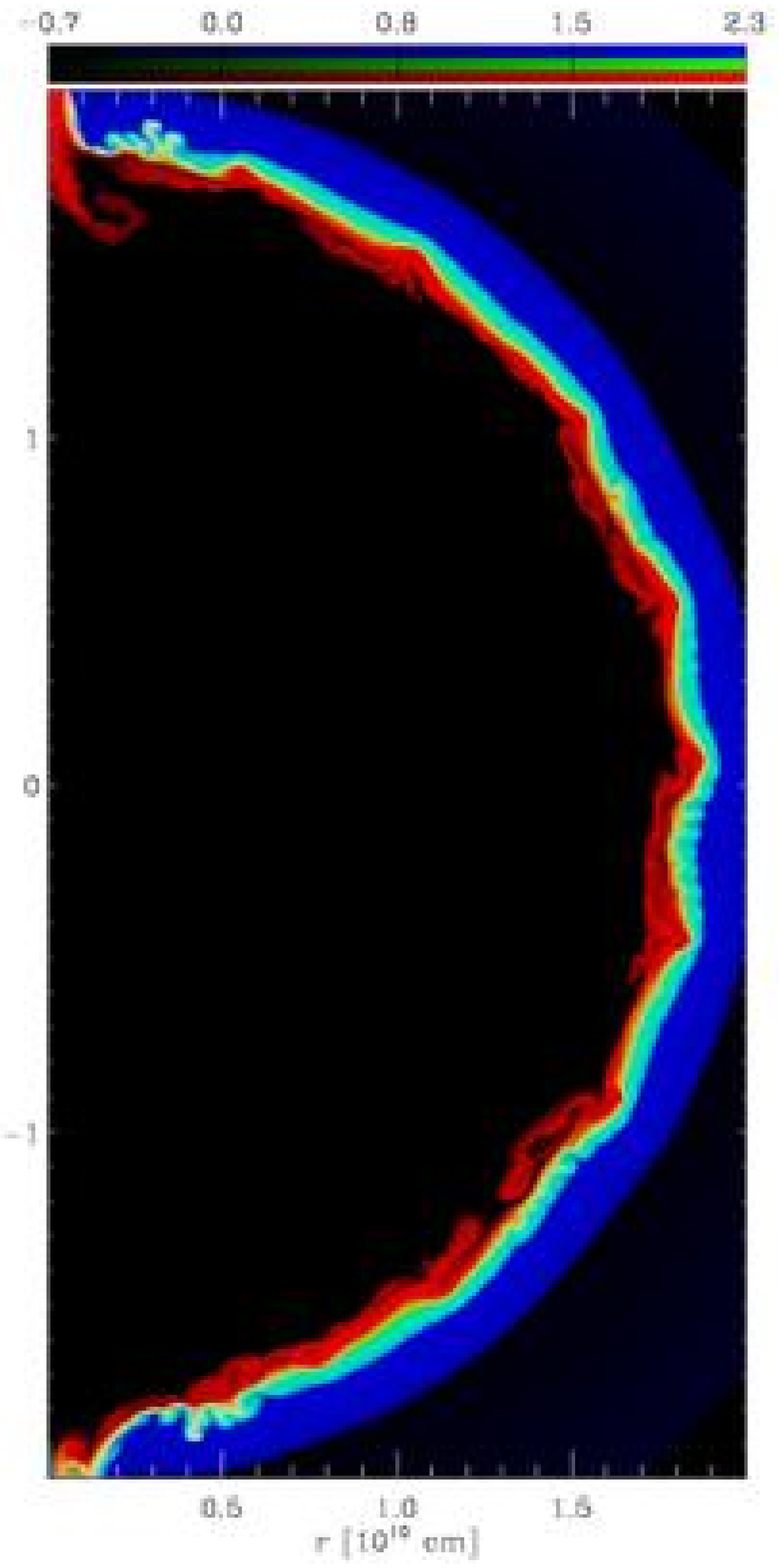}
\end{tabular}
\caption{Logarithm of the density (top) and of the partial densities
         (bottom) of $\rm{^{16}O}$ (blue), $\rm{^{28}Si}$ (green), 
         and $\rm{^{56}Ni}$ (red) in
         Model T310a at selected early epochs. Data values are coded
         according to the color bars given for each frame. In case of
         the partial densities, colors other than red, green and blue
         (resulting from the superposition of these color channels)
         indicate mixing of the composition.  From left to right a) $t
         = 4$\,s, b) $t = 10$\,s, c) $t = 20$\,s. Note the change of
         the radial scale. The supernova shock is visible as the
         outermost circular discontinuity in the density plots.}
\label{fig:dens_elem_4-20}
\end{figure*}

\begin{figure*}
\centering
\begin{tabular}{ccc}
\includegraphics[width=5.7cm]{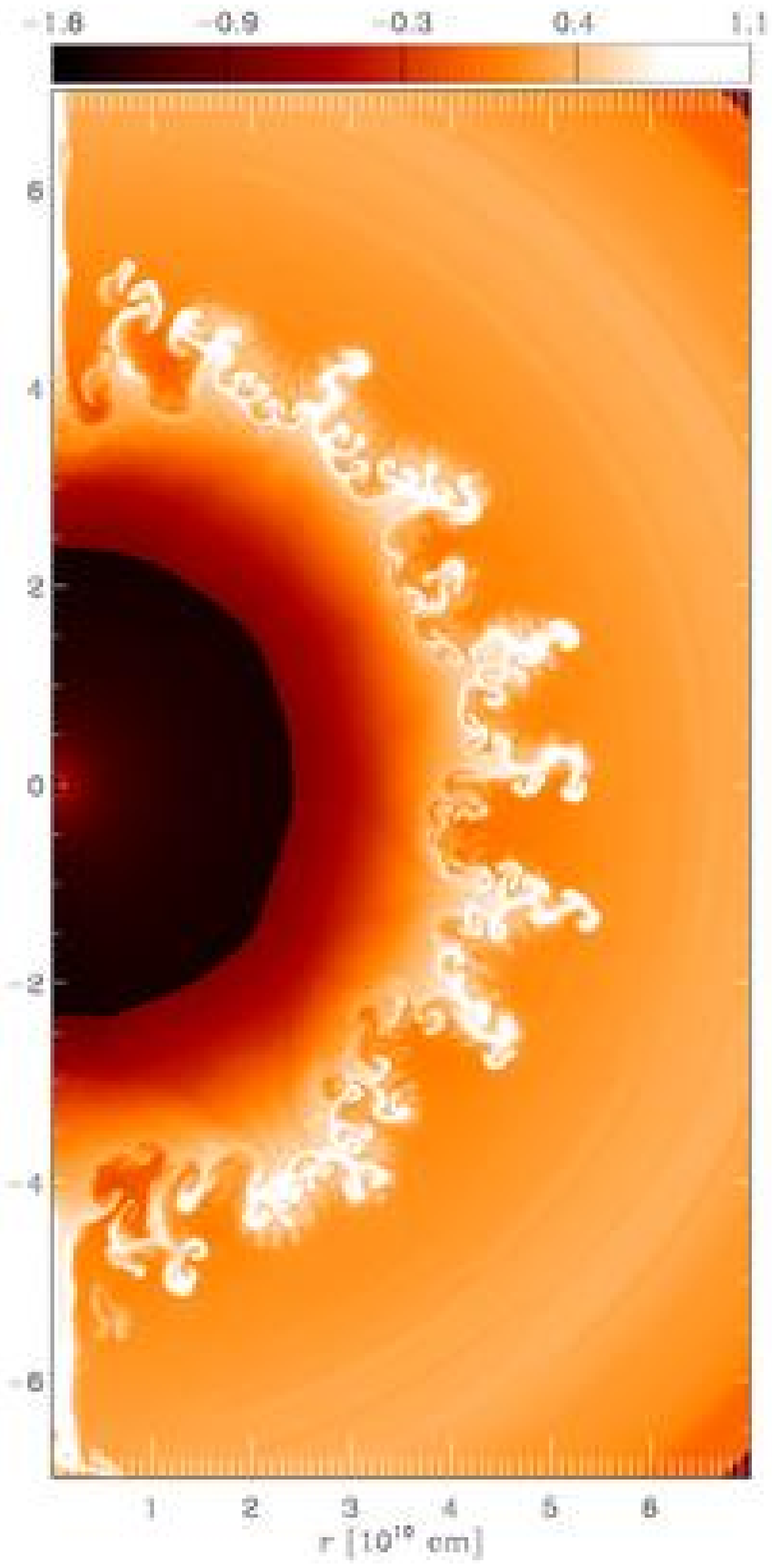}  &
\includegraphics[width=5.7cm]{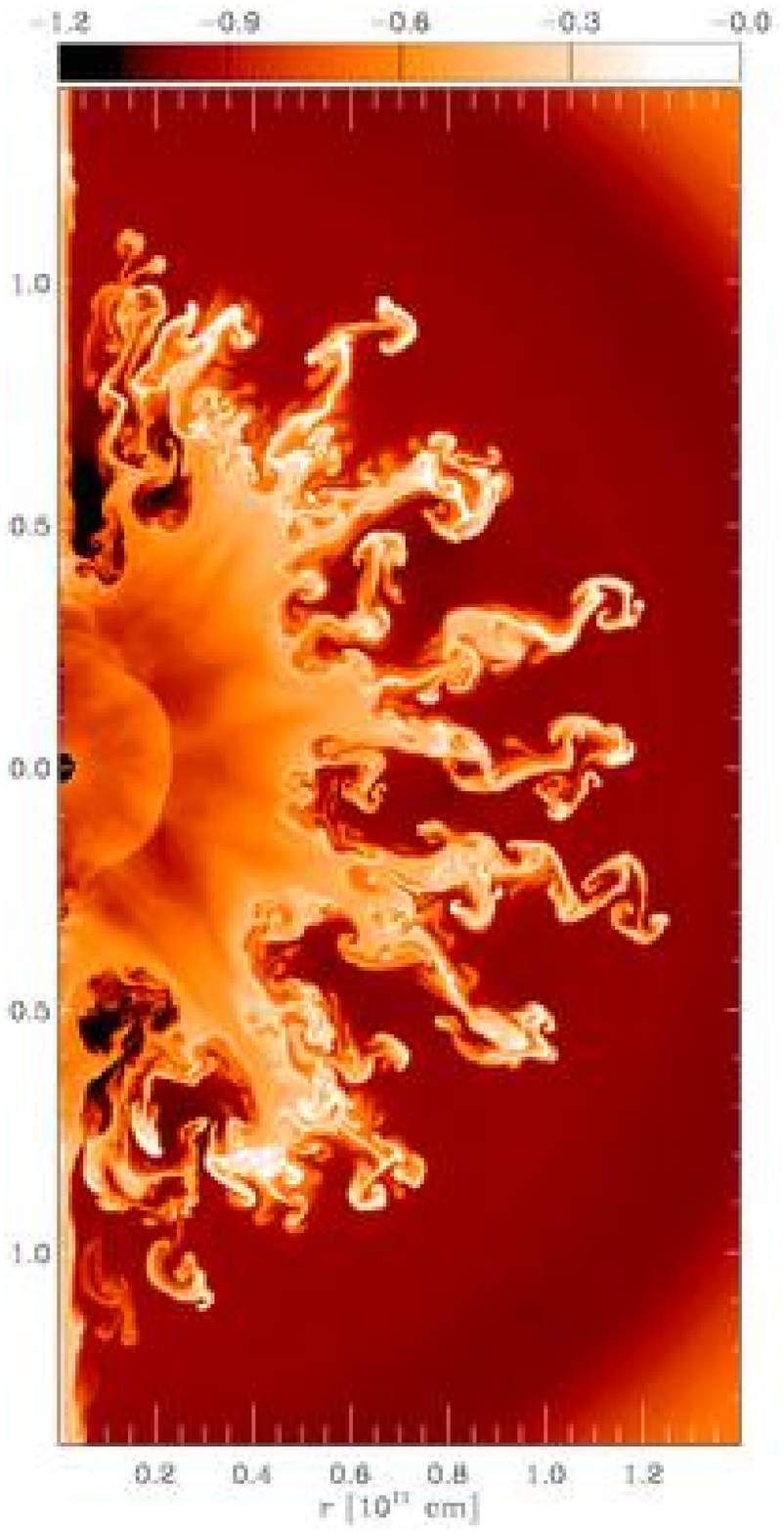}  &
\includegraphics[width=5.7cm]{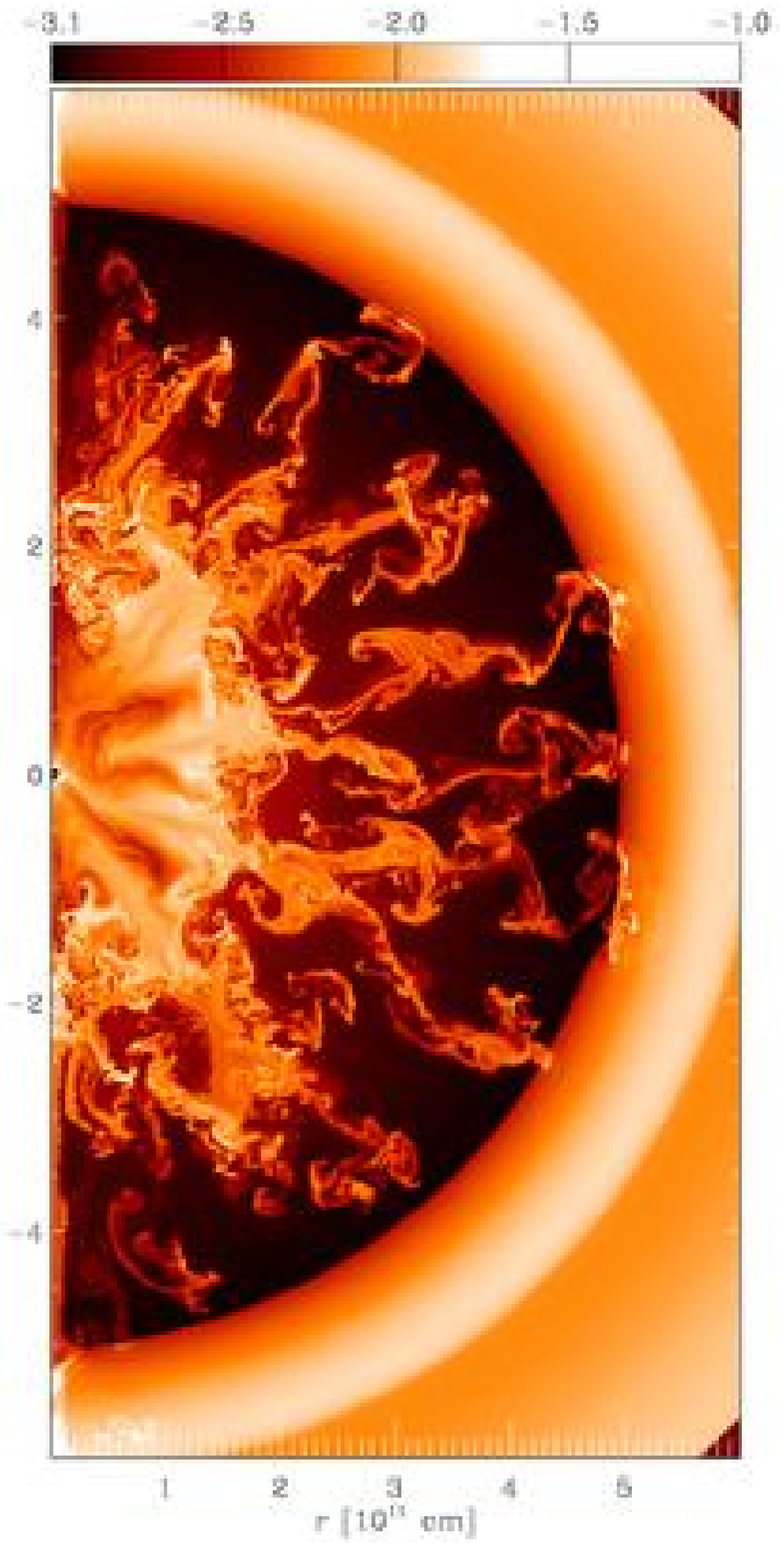} \\
\includegraphics[width=5.7cm]{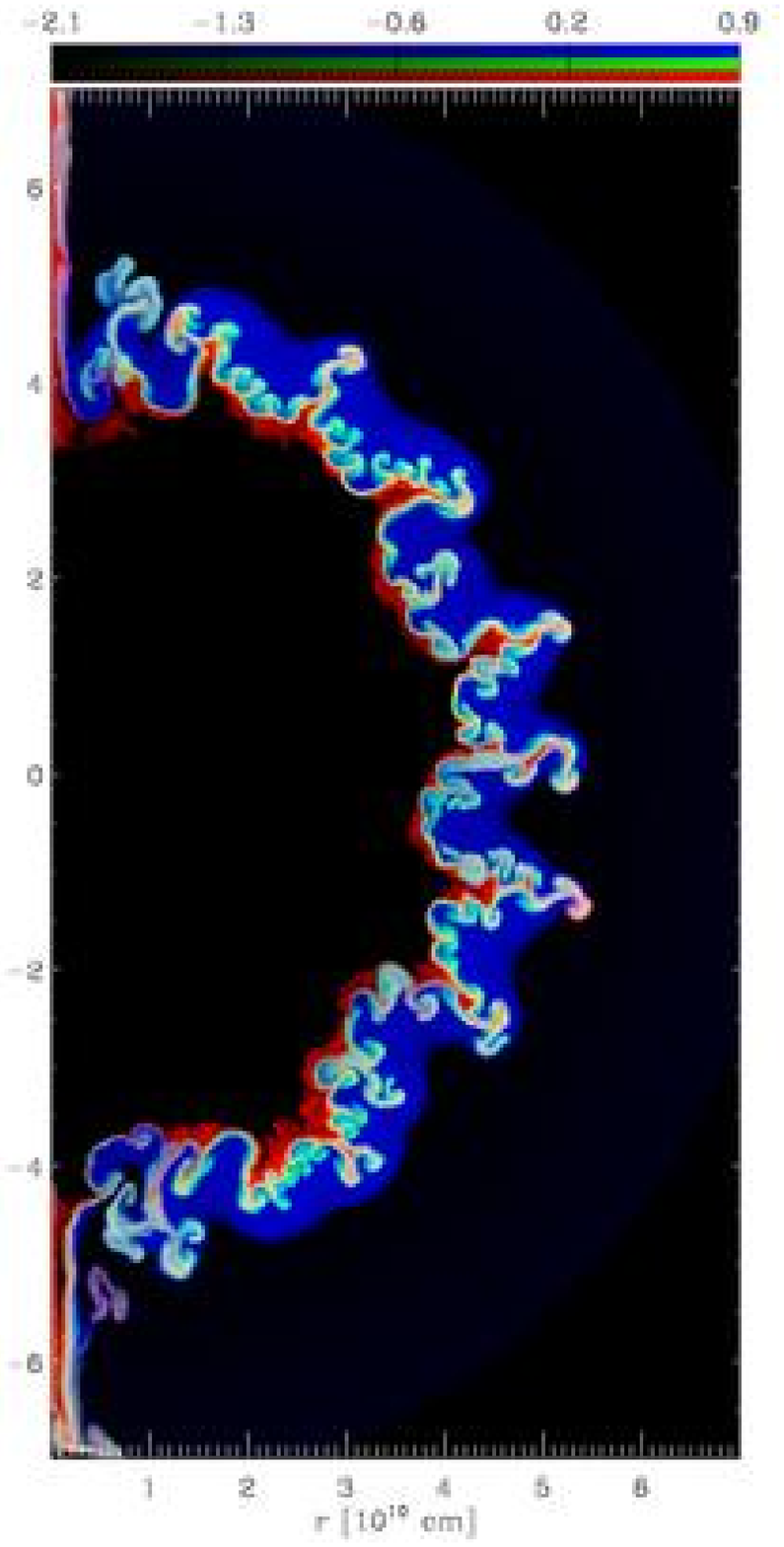}  &
\includegraphics[width=5.7cm]{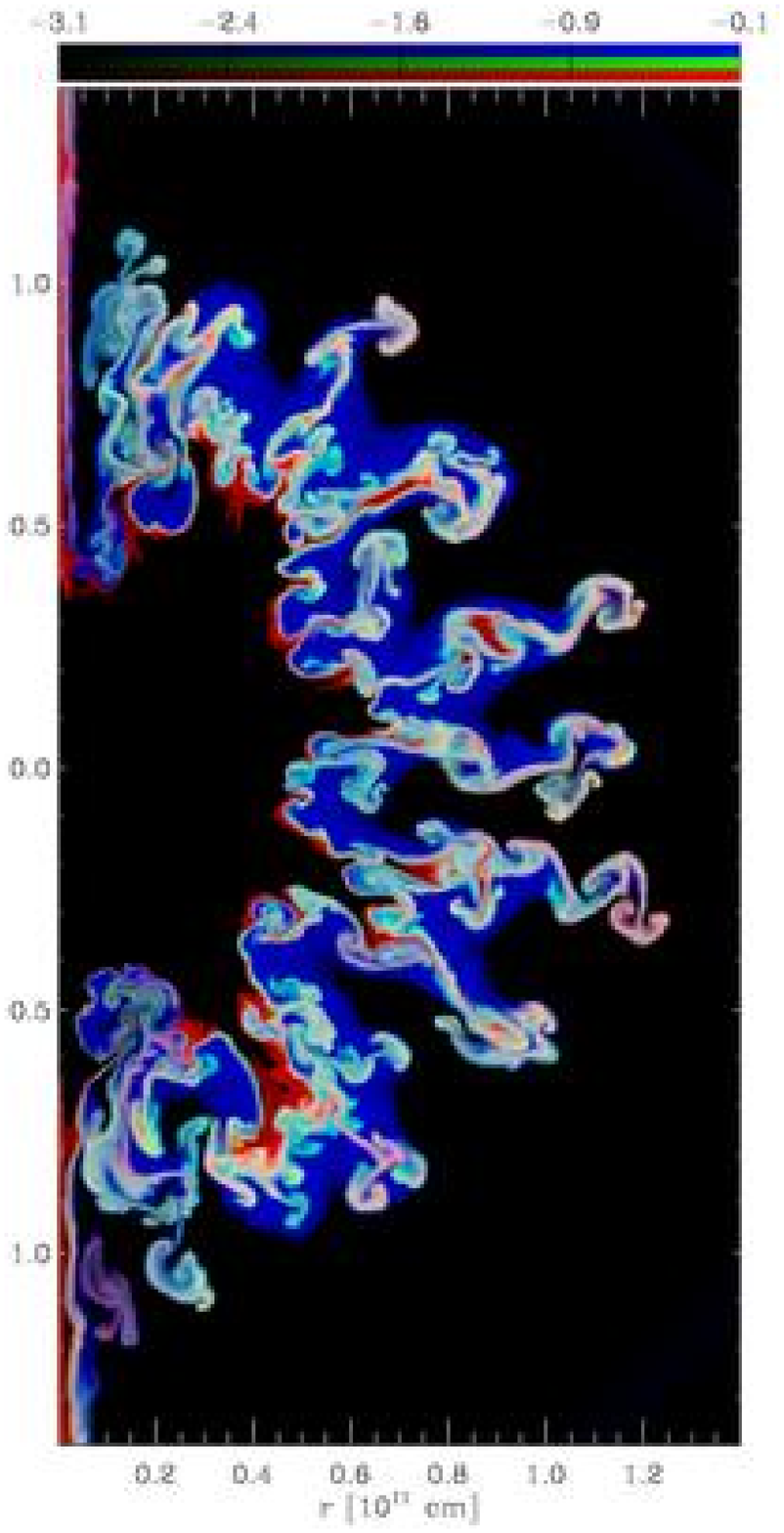}  &
\includegraphics[width=5.7cm]{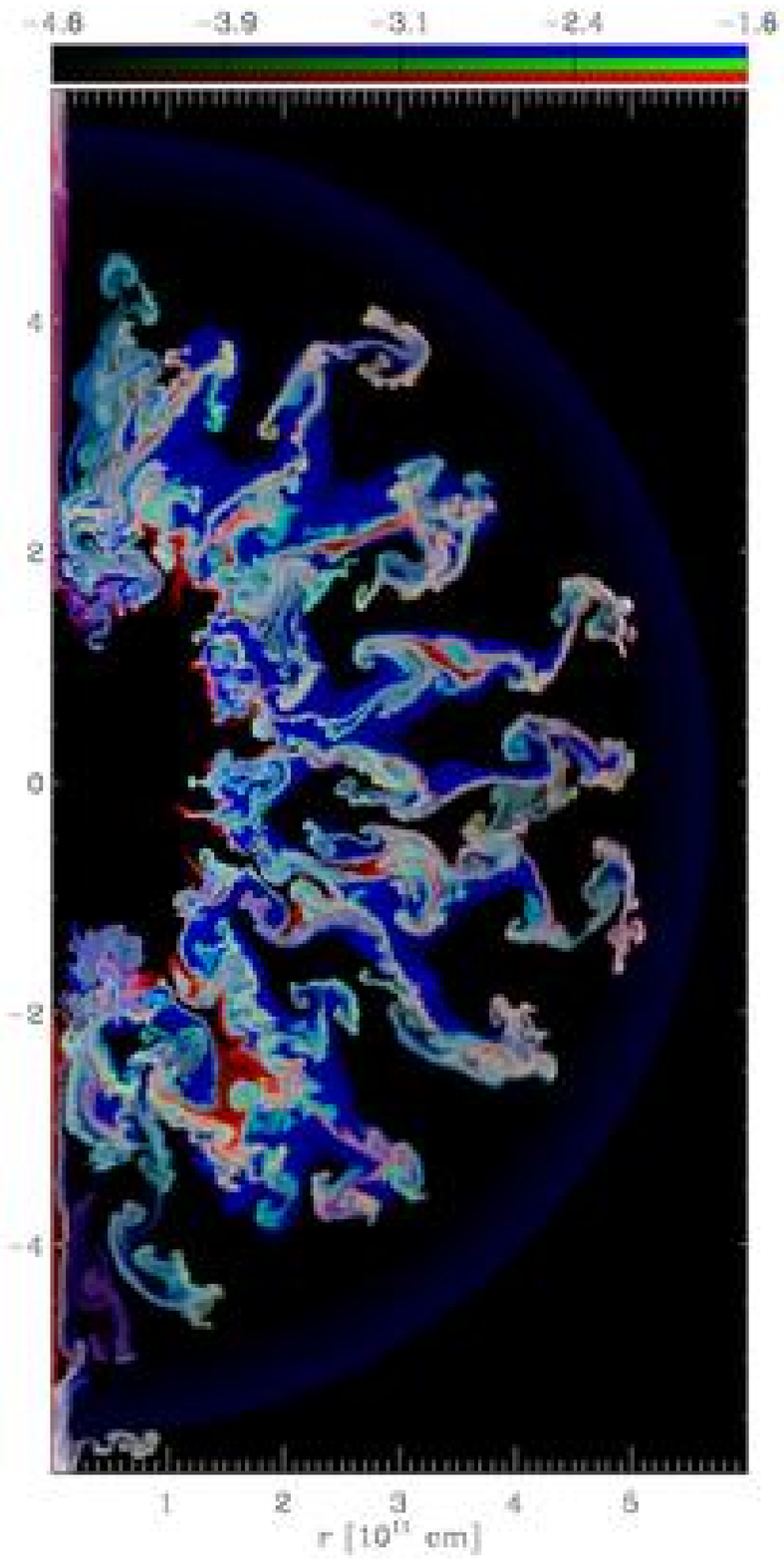}
\end{tabular}
\caption{Same as Fig.~\ref{fig:dens_elem_4-20}. a) $t = 100$\,s
         (left), b) $t = 300$\,s (middle), c) $t = 1500$\,s
         (right). Note the change of the radial scale.}
\label{fig:dens_elem_100-1500}
\end{figure*}

\begin{figure*}
\centering
\begin{tabular}{ccc}
\includegraphics[width=5.7cm]{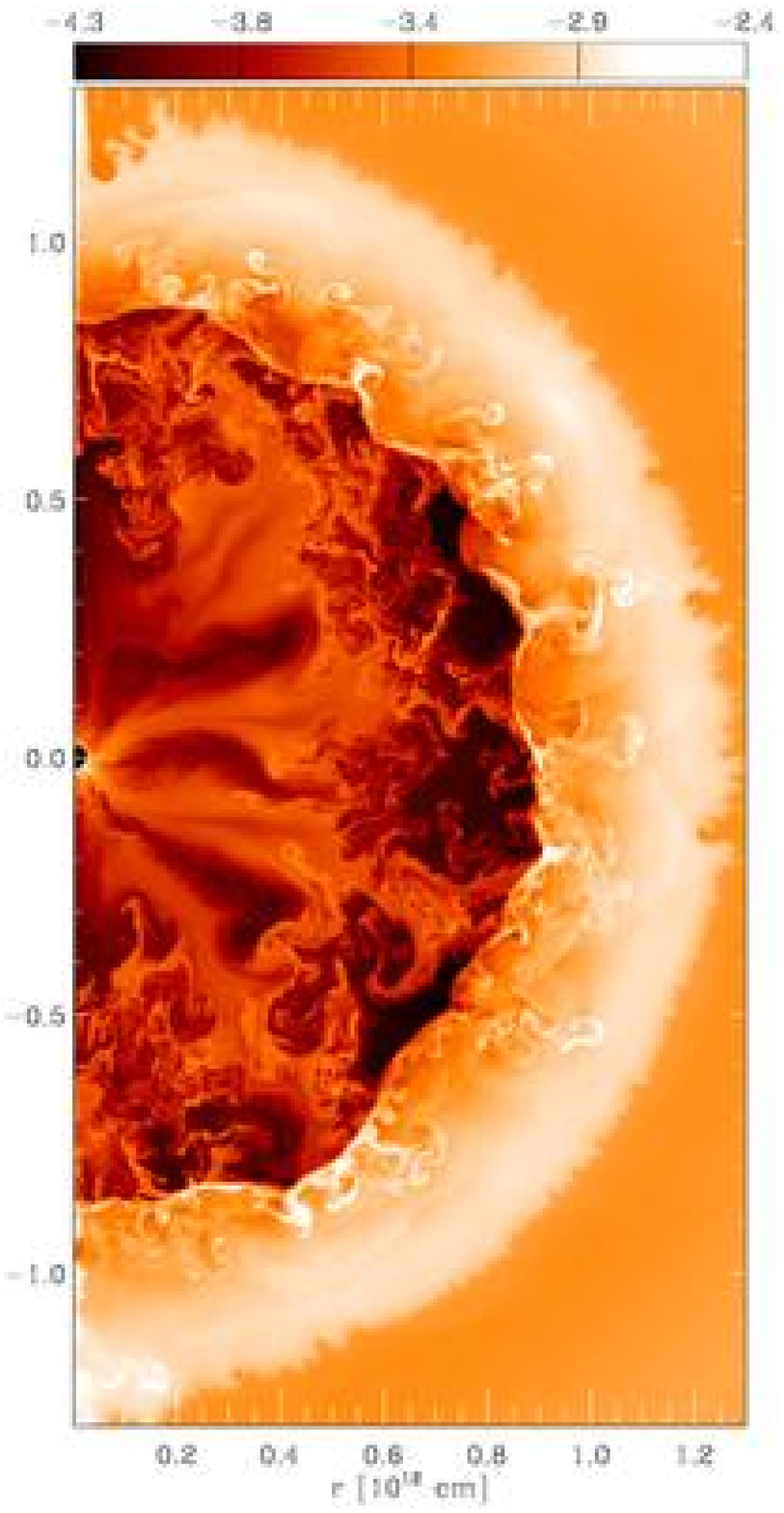}  &
\includegraphics[width=5.7cm]{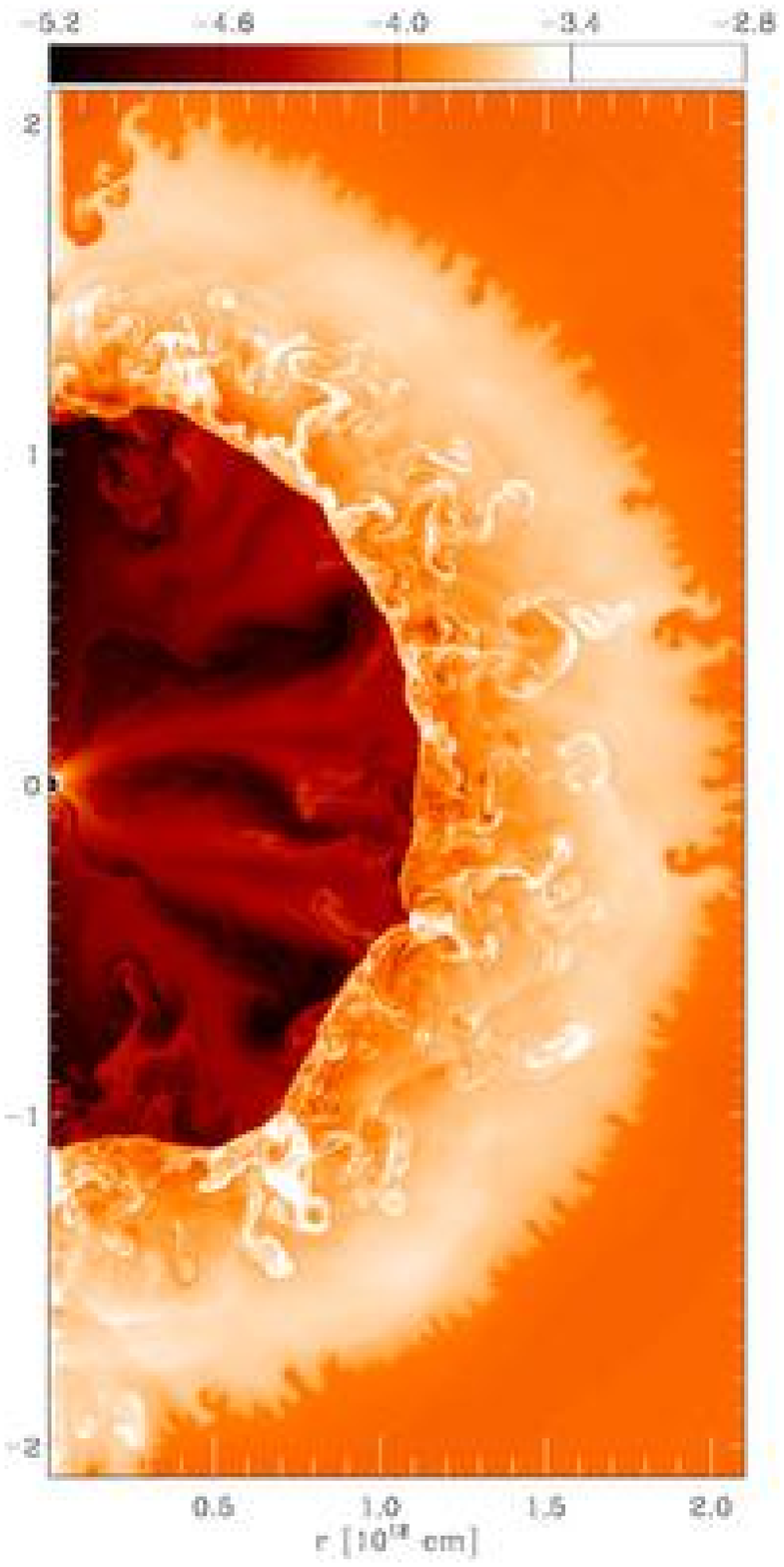}   &
\includegraphics[width=5.7cm]{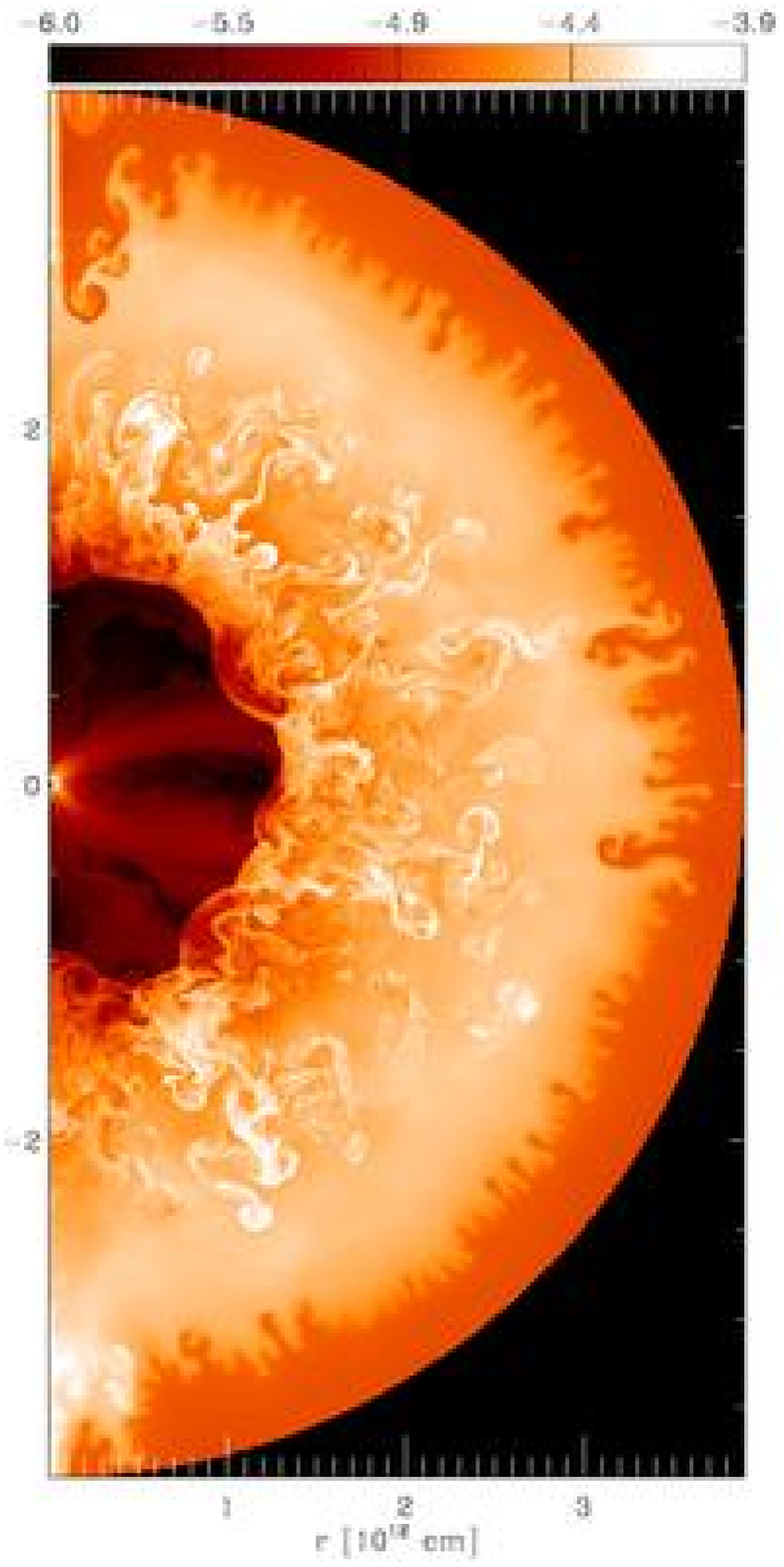}   \\
\includegraphics[width=5.7cm]{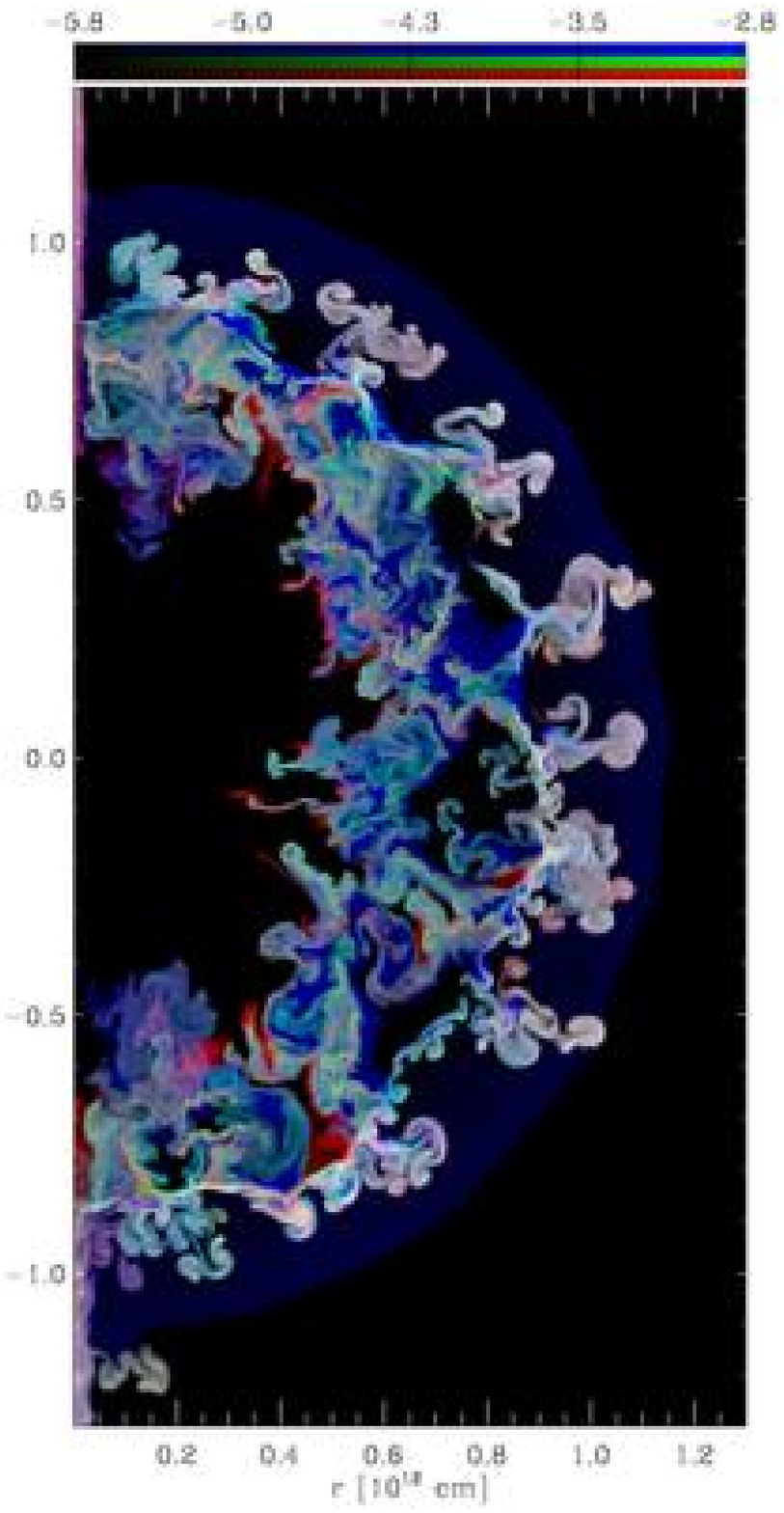}  &
\includegraphics[width=5.7cm]{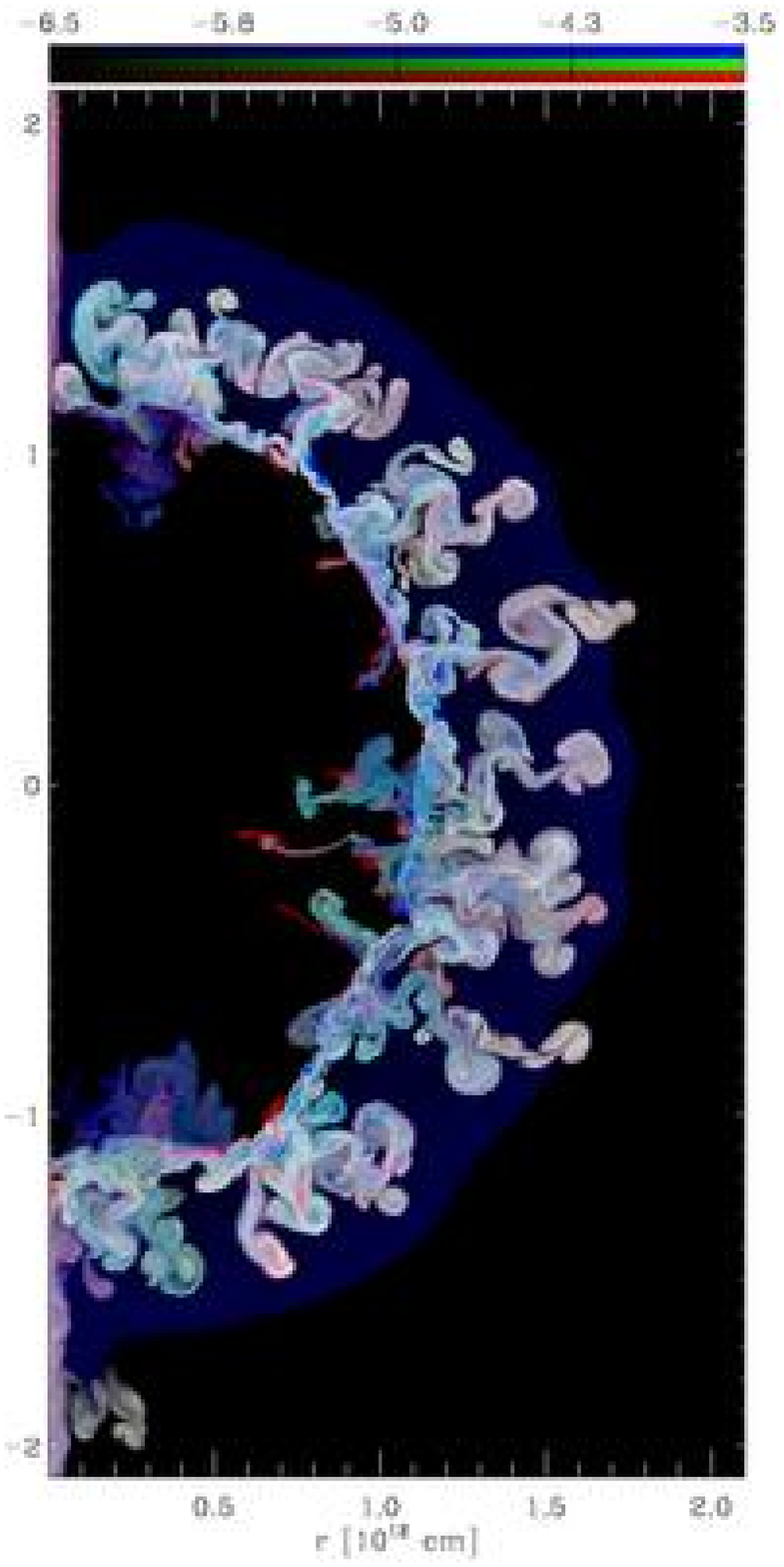}   &
\includegraphics[width=5.7cm]{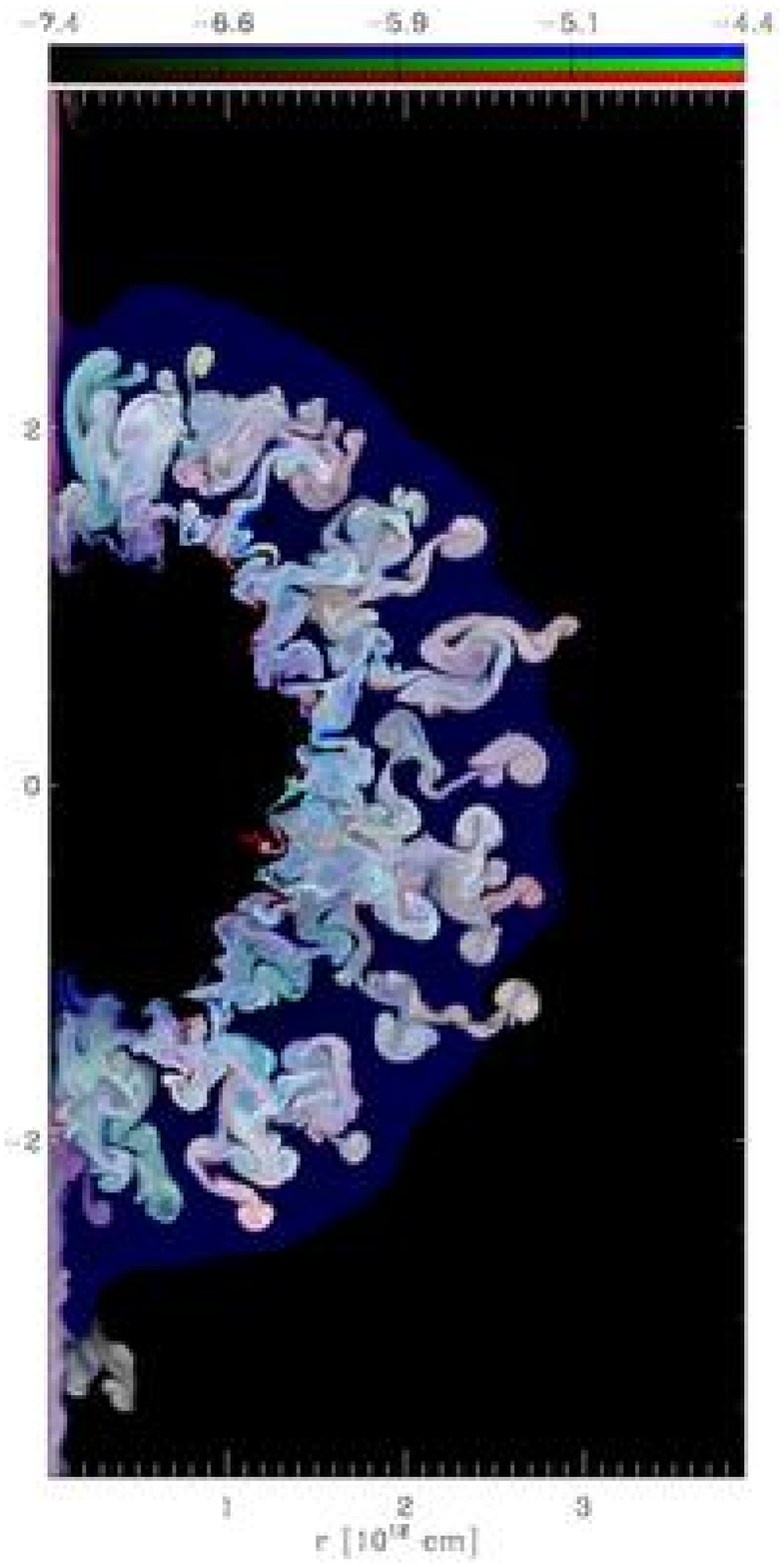}
\end{tabular}
\caption{Same as Fig.~\ref{fig:dens_elem_4-20}.  a) $t = 5000$\,s
         (left), b) $t = 10\,000$\,s (middle), c) $t = 20\,000$\,s
         (right). Note the change of the radial scale.  The circular
         orange/black boundary in the upper right panel corresponds to
         the outer edge of the computational domain, which the
         supernova shock has crossed long before.}
\label{fig:dens_elem_5e3-2e4}
\end{figure*}

Figure~\ref{fig:dens_elem_4-20}a displays the situation 4\,s after
core bounce, when the supernova shock has already crossed the (C+O)/He
interface of the star. The anisotropic structures that can be seen in
this figure are strikingly similar to those visible 420\,ms after
bounce (Fig.~\ref{fig:odd-even}b). This indicates that the low-density
(high-entropy) layers of the ejecta, that were heated by neutrinos,
expand essentially self-similarly during the first few seconds of the
explosion. However, outside of the high-entropy mushrooms, and exactly
as in the one-dimensional case, material of the Si and C+O layers of
the star piles up in two dense shells. Ten seconds after core bounce
(Fig.~\ref{fig:dens_elem_4-20}b) also the high-entropy gas is
affected, and the low-density mushrooms have been compressed to flat
structures.  Only 10\,s later (Fig.~\ref{fig:dens_elem_4-20}c) the
compression has become so strong that the density distribution no
longer resembles the flow-structures of the shock revival phase. A
second reverse shock has formed as a consequence of the deceleration
of the main shock in the He core (Sect.~\ref{sect:AMR_1D}). It is
visible as the dark discontinuity at $r = 1.6 \times 10^{10}$\,cm in
Fig.~\ref{fig:dens_elem_4-20}c.  The initial mushrooms have imprinted
a strong long-wavelength perturbation both on the dense shell behind
the Si/O interface and at the (C+O)/He interface. Superposed upon this
perturbation, small-scale disturbances start to grow along the entire
Si/O interface where also about 10 cusps (not counting the features
near the symmetry axis) begin to develop that are located about
$20^{\circ}$ apart. Interestingly, the positions of these cusps do
\emph{not} coincide with those of the former mushrooms that are
located in the regions \emph{between} the cusps. The cusps themselves
seem to be pushed by the denser material that constituted the former
down-flows between the mushrooms. It is exactly in these dense
regions, where $\rm{^{56}Ni}$ had formed during the first 250\,ms of
the explosion. Although most of the $\rm{^{56}Ni}$ is originally
located in a layer just interior to the unstable zone at the Si/O
interface (see Fig.~\ref{fig:dens_elem_4-20}a and compare also
Fig.~\ref{fig:amra_late_evol_1d} for the one-dimensional model), the
dense regions with their larger momentum cannot be slowed down as
strongly as their neighbouring matter so that they penetrate outward
much farther. This becomes evident in the plots for $t = 100$\,s
(Fig.~\ref{fig:dens_elem_100-1500}a). The cusps have already grown
into separate fingers that start to rise through the (C+O)/He
interface and impose a long-wavelength perturbation onto these
layers. Concurrently, the smallest scale perturbations that are
resolved on our grid have grown to small mushrooms at the Si/O
interface and start to mix $\rm{^{28}Si}$ outwards while $\rm{^{16}O}$
is mixed inwards in between them. The growth of Rayleigh-Taylor
fingers from the former dense Ni-``pockets'' has important
consequences for the interpretation of observational data and will be
addressed in more detail in Sect.~\ref{sect:SN_Ib_model}.

At $t=100$\,s the supernova shock has already passed the He/H
interface.  Furthermore, the reverse shock that had formed due to the
slow-down of the main shock in the He core starts to propagate inward
in radius. This reverse shock decelerates the innermost layers of the
ejecta until it is reflected at the inner boundary and moves outward
again, compressing the inner metal core of the star
(Figs.~\ref{fig:dens_elem_100-1500}b and c). Five minutes after
bounce, the metal core itself is totally shredded by the instability.
Showing a density contrast to the ambient material of up to a factor
of 5, the fingers have grown into the typical mushroom shape, which is
caused by Kelvin-Helmholtz instabilities
(Fig.~\ref{fig:dens_elem_100-1500}b). Almost the entire metal core has
thereby been carried through a substantial fraction of the He core,
while in turn helium-rich gas was mixed into the metal core in
``pockets'' between the fingers. These low-density helium tongues can
be discerned as the black regions around the blueish oxygen plumes in
the abundance plots of Fig.~\ref{fig:dens_elem_100-1500}.  Note that
there is a second helium-rich region closer to the center of the
models, interior to the red, green, and blue-colored zones.  This
region contains material that experienced a high-entropy freezeout of
nuclear reactions, and thus shows high abundances of
$\alpha$-particles and nuclei like $\rm{^{44}Ti}$. In between this
$\alpha$-rich layer and the regions of $\rm{^{56}Ni}$ dominance (the
patches of dense material colored in red and pink in the abundance
panel of Fig.~\ref{fig:dens_elem_100-1500}; see also
Fig.~\ref{fig:amra_late_evol_1d} for the one-dimensional case) one
encounters neutron-rich nuclei. These nuclear species participate in
the mixing.  The closer they are located to the unstable interfaces,
the stronger their spatial distribution is affected.

At 1500 seconds after bounce, the flow in the mixing region has become
very complex because of the interaction of the instabilities at the
former Si/O and (C+O)/He interfaces and the action of Kelvin-Helmholtz
instabilities in the shear flows along the fingers
(Fig.~\ref{fig:dens_elem_100-1500}c). During this phase, and depending
on the spatial resolution, our simulations show a tendency for the
edges of the fingers to fragment into smaller structures. This
fragmentation appears to be more pronounced for models with small
explosion energies. In this case, the time-scale for the fingers to
cross the He core may increase sufficiently to allow for substantial
growth of fluid-dynamic instabilities at the boundaries of the
fingers.

\begin{figure*}
\includegraphics[width=17cm]{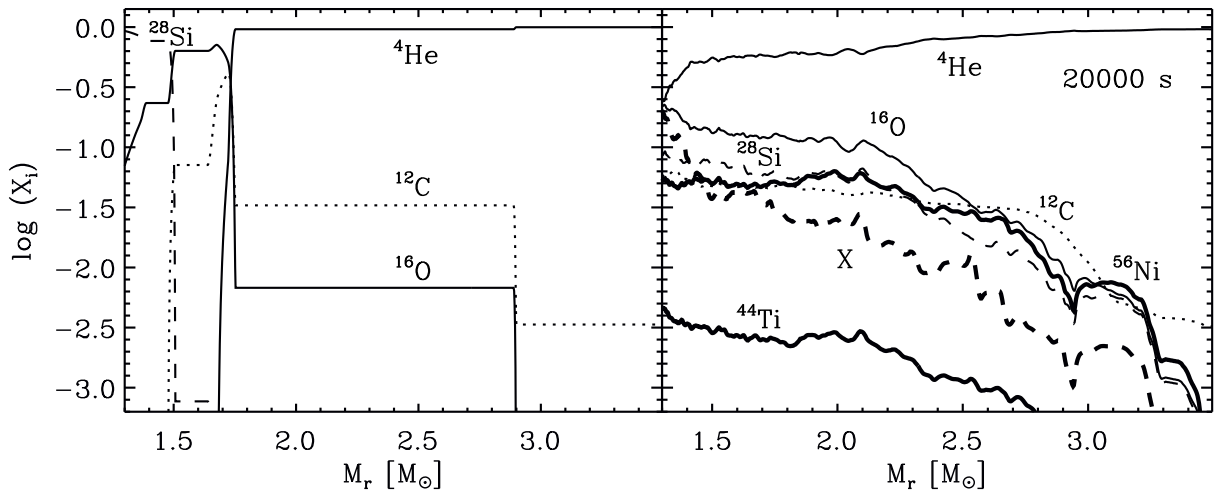}
\caption{Evolution of the extent of mixing in Model T310a. Left:
         Initial composition. Right: Composition 20\,000\,s after core
         bounce. The neutronization tracer is denoted with ``X''.
         Note that all heavy elements are confined to the helium core
         (i.e. to the innermost $4.2\,{\rm M_{\odot}}$ of the star) 
         even as late as 20\,000\,s after core bounce.}
\label{fig:amra_mixing_310}
\end{figure*}  

While the central regions of the star are turned inside out, a dense
shell can be seen 1500\,s after bounce just below the border between
helium core and hydrogen envelope
(Fig.~\ref{fig:dens_elem_100-1500}c). It is this prominent shell that
we also found in the one-dimensional calculation
(Sect.~\ref{sect:AMR_1D}) as a result of the deceleration of the main
shock in the hydrogen envelope. Note that like in the one-dimensional
case, the density gradient at the inner edge of the shell steepens
into a reverse shock. In the one-dimensional calculation, the layers
of the metal core are well behind this reverse shock until $t =
2600$\,s. Only then has the reverse shock propagated sufficiently deep
inward in mass to reach the (C+O)/He interface (see also
\citealt{Kane+00} and \citealt{MFA91}).  In our two-dimensional
simulation, however, the fastest mushrooms already start to penetrate
through the reverse shock at 1500\,s after bounce
(Fig.~\ref{fig:dens_elem_100-1500}c). This interaction of the
metal-enriched clumps with the reverse shock has not been pointed out
in any previous study of Rayleigh-Taylor instabilities in Type~II
supernova progenitors.  It has important consequences for the velocity
distribution of the elements because it leads to a strong deceleration
of the clumps. We will address this issue in more detail below. For
the moment we only note that after penetrating through the reverse
shock and entering the high-density He-shell, the clumps move
supersonically relative to the ambient medium. As a result they are
dissipating a large fraction of their kinetic energy in bow shocks and
strong acoustic waves. The wave fronts can be seen in
Figs.~\ref{fig:dens_elem_5e3-2e4}a to c which show the interaction of
the clumps with the shell between 5000\,s and 20\,000\,s after
bounce. Figure~\ref{fig:dens_elem_5e3-2e4} also suggests that during
this interaction the spatial distribution of the heavy elements inside
the clumps is almost entirely homogenized. This causes the light blue
and whitish regions in the abundance plots (lower panels of
Figs.~\ref{fig:dens_elem_5e3-2e4}a to c) due to the superposition of
the single colors that were assigned to the different elements. The
interaction with the reverse shock also leads to mixing of ambient
helium into the clumps, with the helium mass fraction becoming
comparable to that of the heavy elements. Before, the clumps contained
only small amounts of helium that was admixed to them during the first
$\sim 300$\,s from the central zone of $\alpha$-rich freezeout by the
development of the Rayleigh-Taylor instabilities.

To facilitate a comparison of Model T310a with one-dimensional work,
upon which most attempts to reproduce observations of nucleosynthetic
yields and their distribution are based, we summarize the extent of
mixing as a function of the enclosed mass in
Fig.~\ref{fig:amra_mixing_310}. The left panel shows the distribution
of the mass fractions for the original presupernova model.  At a time
of 20\,000\,s after bounce (right panel), elements like $\rm{^{16}O}$
and $\rm{^{28}Si}$, that made up the original metal core have been
mixed almost homogeneously throughout the inner $3.4\,{\rm
M_{\odot}}$, along with the newly synthesized $\rm{^{56}Ni}$.  The
only nuclear species that are not mixed this far out in mass are the
neutronization tracer and, to an even lesser extent,
$\rm{^{44}Ti}$. These nuclei were synthesized in the innermost layers
of the neutrino-heated ejecta very close to the collapsed core.

Note that the mixing is confined to the former He core of the star
(i.e. the inner $4.2\,{\rm M_{\odot}}$). This can be understood from
Fig.~\ref{fig:dens_elem_5e3-2e4}: The dense He-shell, that forms just
below the He/H interface, acts like an impenetrable wall for the metal
clumps and shields the hydrogen envelope from becoming enriched with
heavy elements. This result may appear somewhat surprising because, as
we have shown by the linear stability analysis
(Sect.~\ref{sect:AMR_1D}), the He/H interface of the \cite{WPE88} star
is Rayleigh-Taylor unstable.  However, since we did not impose any
seed perturbations in our AMR simulations and since the supernova
shock is almost perfectly spherically symmetric when it crosses the
He/H interface, the evolution of these layers proceeds initially
one-dimensionally. This turns out to be true even in models where much
more vigorous neutrino-driven convection imposes noticeable
asphericities on the shock during its revival. Such asphericities,
however, are smoothed out during the subsequent propagation of the
shock through the helium core. The only deviations from spherical
symmetry that perturb the He/H interface in Model T310a are the waves
that are excited when the metal-enriched clumps hit the inner edge of
the dense shell below the interface at times $t \geq
1500$\,s. However, these perturbations encounter only a moderately
unstable He/H interface at these late times, in accordance with the
linear stability analysis (Fig.~\ref{fig:amra_growth_b}) which shows
that after $t \sim 1500$\,s the integrated growth rate, $\sigma$,
increases only slightly. The mixing at the He/H interface is therefore
rather weak. The instability at this interface has apparently evolved
into the non-linear regime by $t \approx 10\,000$\,s, showing a
multitude of small and somewhat bent fingers. However, these do not
grow appreciably up to 20\,000\,s after bounce, when we stopped our 2D
simulation. At this time the flow near the interface expands nearly
self-similarly, i.e. the small fingers move with the same velocity as
the medium between them. Note that the perturbations of the He/H
interface that result from the accoustic waves are quite different
from those used to initiate the instability in all previous studies.
In the latter, seed perturbations of the order of 10\% of the radial
velocity were imposed on the entire star about 300\,s after shock
formation. With perturbations of this magnitude it is indeed possible
to obtain strong mixing at the He/H interface of our progenitor.  We
have found such extended mixing in the late evolution of the
hydrodynamic model discussed in \cite{Kifonidis+00}, which suffered
strongly from odd-even decoupling, and was therefore perturbed in the
radial velocity by $\sim 20\%$ at $t=300$\,s. However, even in this
model the nickel was mixed out only to a mass coordinate of $4.5\,{\rm
M_{\odot}}$ after 10\,000\,s.

\begin{figure*}
\centering
\includegraphics[width=15cm]{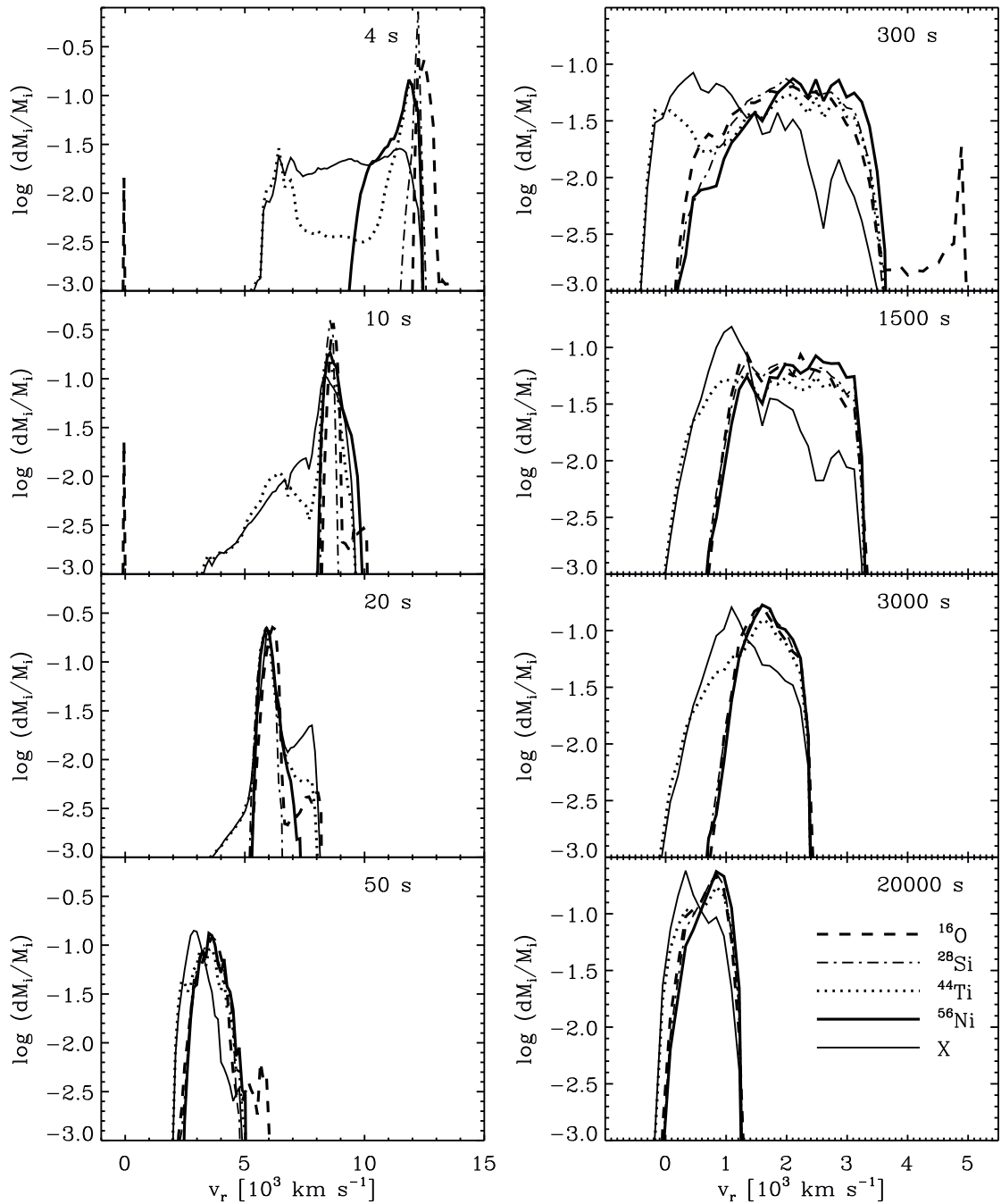}
\caption{Logarithm of the fractional mass of different elements that
         is contained within the velocity interval $[v_r,v_r+{\rm
         d}v_r]$ as a function of the radial velocity $v_r$ in Model
         T310a at various epochs. The resolution is ${\rm d}v_r
         \approx 130\,{\rm km\,s^{-1}}$. Note the different scales for
         the radial velocity in the left and right panels.}
\label{fig:amra_massvelo}
\end{figure*}

\subsubsection{Clump propagation and velocity 
               distribution of nucleosynthetic products}
\label{sect:velodist_amra}

In Fig.~\ref{fig:amra_massvelo} we show for times from 4\,s until
20\,000\,s after bounce the logarithm of the fractional masses for
several nuclei that are contained within the velocity interval
$[v_r,v_r+{\rm d}v_r]$ as a function of the radial velocity.
Obviously, there occurs a bulk deceleration of the material from
velocities as large as $\sim 14\,000$\,km/s at a time of 4\,s after
bounce to $\sim 5000$\,km/s after 50\,s. This is caused by the
enormous deceleration that the shock and the post-shock material
experience after the shock has entered the He core. The deceleration
and the associated compression also cause the clear tendency of the
mass distributions of the elements to narrow down in velocity space
(Fig.~\ref{fig:amra_massvelo}).

Shortly before the shock crosses the (C+O)/He interface (i.e. for
times $1\,{\rm s} \leq t \leq 2\,{\rm s}$) the abundance maxima of
different elements are well-separated in radius. Together with the
post-shock velocity gradient (Sect.~\ref{sect:first_sec_2D}) this
results in maxima of the velocity distributions of different nuclei
that are well-separated in velocity space. However, this separation
disappears within the first 50\,s after bounce due to the strong
compression: Being squeezed into a very dense shell, different nuclear
species are getting very close in radius and thus they reflect the
velocities that are prevailing in this shell.  The first nuclei to
become affected by the compression are $\rm{^{16}O}$, $\rm{^{28}Si}$
and $\rm{^{56}Ni}$, since these are located closest to the supernova
shock.  A narrow peak forms in the velocity distribution of these
elements already between 4\,s and 10\,s after bounce
(Fig.~\ref{fig:amra_massvelo}). Being centered around 8500\,km/s at $t
= 10$\,s, the peak gradually recedes to $\sim 3500$\,km/s at
$t=50$\,s. The maximum velocities of $\rm{^{56}Ni}$ and $\rm{^{28}Si}$
are larger than this value by about 1500\,km/s, while a small fraction
of $\rm{^{16}O}$ is found at somewhat larger velocities,
still. Elements that are located closer to the inner boundary of the
computational domain (i.e. $\rm{^{44}Ti}$ and the neutronization
tracer) show broader distributions in velocity space during the first
$\sim 10$\,s of the explosion, until the strong compression also
affects their spatial distribution and leads to a homogenization of
all profiles in Fig.~\ref{fig:amra_massvelo} around 50\,s after
bounce. This homogenization is actually fortunate for a comparison of
our results to observational data, since it makes the late-time
velocity distributions of nuclei in our models fairly robust against
uncertainties that affect the nucleosynthesis in the neutrino-heated
matter. In particular, the velocity profiles of $\rm{^{56}Ni}$ and of
the neutronization tracer nucleus are finally rather similar. Hence,
it is not very relevant for the velocity distributions of these nuclei
that nucleosynthesis in the inner ejecta depends on unknown details of
the neutrino luminosities and spectra, which determine whether
$\rm{^{56}Ni}$ or mere neutron-rich nuclei are produced.

Up to a time of 300\,s, when the Rayleigh-Taylor instabilities are
still developing, the maximum velocity of $\rm{^{56}Ni}$ is steadily
decreasing, dropping to values of $\sim 3500$\,km/s.  This maximum
velocity is found in the mushroom caps of the Rayleigh-Taylor fingers.
In the early stages of their existence these ``clumps'' have to
propagate through a rather dense medium. In addition they ``feel'' the
positive pressure gradient in the unstable layers. Hence they are
decelerated appreciably, but less than the expansion of the
surrounding matter is slowed down by this pressure gradient. Around
300\,s, however, the density contrast between the clumps and their
environment has grown substantially.  In addition, at this time the
sound speed is about 1400\,km/s in the layers of the He core through
which the clumps propagate, while the ``background'' flow is expanding
with $\sim 3000$\,km/s. Comparing this to the clump velocity of
3500\,km/s we find that at $t = 300$\,s the clumps move relative to
the He core with about 500\,km/s, i.e. their motion relative to the
background is \emph{subsonic} and the drag they experience is rather
small. As a result, the clumps ``decouple'' from the flow and start to
move ballistically through the remaining part of the helium core, with
the maximum $\rm{^{56}Ni}$ velocity remaining roughly constant for
times between 300\,s and $1500$\,s.  At approximately the latter time
the clumps penetrate through the reverse shock at the inner edge of
the dense shell that has formed below the He/H interface. After
entering this shell they are strongly decelerated, and the maximum
velocities of all elements drop to $\sim 1200$\,km/s as is visible for
a time of 20\,000\,s after core bounce in
Fig.~\ref{fig:amra_massvelo}.  These values are significantly smaller
than those observed in SN 1987\,A. Note that prior to the interaction
of the clumps with the shell, the nickel velocities of our models {\em
are in accordance\/} with the velocities observed in SN 1987\,A,
i.e. a good match to the observations is prevented by the formation of
the dense shell in the outer He core of our models. Several effects
cause the strong slow-down of the clumps in the shell. They can be
discussed by considering the expression for the drag force
\begin{equation}
     F_D = c_D \frac{\rho v^2}{2} A,
\label{eq:drag}
\end{equation}
where $A$ is the cross sectional area of a clump, $\rho$ the ambient
fluid density, $v$ the velocity of the clumps relative to the ambient
fluid, and $c_D$ the drag coefficient. Since the clumps have to pass
through a {\em reverse\/} shock they enter an environment with a {\em
smaller\/} mean expansion velocity than in the layers upstream of the
shock.  Their velocity $v$ relative to the new background is therefore
\emph{higher} than it was before they crossed the shock and amounts to
$v \approx 1300$\,km/s at $t = 1500$\,s.  In addition $\rho$ is larger
in the post-shock region.  Both effects increase the drag by
increasing the ram pressure $\rho v^2$.  The sound speed in the
post-shock layers is rather small (890\,km/s) at $t =
1500$\,s. Therefore, the clumps move \emph{supersonically} through the
dense medium, with Mach numbers $M \approx 1.5$.  The drag coefficient
$c_D$ of a projectile increases steeply in the subsonic/supersonic
transition (this is the well-known ``sound barrier''). The effect is
even more pronounced if the projectile's shape is roundish (as in the
case considered here) and not pointed.  Together with the increase in
ram pressure, the supersonic motion leads to a large rise of $F_D$,
and hence to significant energy dissipation.

The interaction of the clumps with the reverse shock resembles the
well-studied problem of a (planar) shock interacting with an
overdense, spherical, interstellar cloud that is in pressure
equilibrium with its surroundings \citep[see][and the references
therein]{KMC94}. In this case it has been shown that the interaction
can be divided into four stages. In the first phase, the blast wave
strikes the cloud driving a shock into the cloud and a reflected shock
into the (shocked) intercloud medium. The reflected shocks resulting
from this phase correspond to the bow-shocks that we observe and which
pervade the entire dense helium shell in our simulation between
3000\,s and 20\,000\,s after bounce
(Sect.~\ref{sect:clump_formation}). In the second stage the cloud is
compressed into a pancake-like structure.  Indications of this
phenomenon can be seen in Fig.~\ref{fig:dens_elem_100-1500}c.  The
third stage is the reexpansion phase in which the cloud expands
laterally.  This increases its cross section, $A$, and thus hastens
its deceleration (see Figs.~\ref{fig:dens_elem_5e3-2e4}a to c). In the
fourth and final phase the cloud is destroyed by vorticity generation
due to the development of Rayleigh-Taylor and Kelvin-Helmholtz
instabilities at its surface.  We do not observe a complete
disintegration of the metal clumps according to this last phase in our
simulations.  It is not clear, whether this is due to numerical
reasons, due to the fact that we might have simply stopped our
simulations too early for the instabilities to grow appreciably, or
due to differences in the physics of both problems.  In our case the
density contrast between a clump and the medium upstream as well as
downstream of the reverse shock (the former quantity is a constant in
\citealt{KMC94}) is time-dependent, because the overall expansion
decreases the densities.  Furthermore, our reverse shock has a Mach
number $M \sim 1$ and is thus not very strong, in contrast to the
shocks studied by \cite{KMC94}, for which $M \gg 1$. In addition, the
mushroom caps of our Rayleigh-Taylor fingers have typical radii
$r_{\rm clump} \sim 0.1 R_{\rm s}$, with $R_{\rm s}$ being the radius
of the reverse shock.  Thus they do not satisfy the small clump
criterion ($r_{\rm clump} \leq 0.01 R_{\rm s}$) that \cite{KMC94}
assumed for their study. Finally, our problem has a more complicated
geometry and equation of state.  Recent experiments of $M \sim 1$
shocks in air interacting with a cylindrical column of high-density
gas \citep{Fishbine02,Zoldi_PhD} indicate that the gas column is
strongly ablated, developing significant vorticity at its
edges. However, it is not entirely destroyed.

\cite{KMC94} report that for the shock-cloud interaction problem,
about 120 zones per cloud radius are required to reduce numerical
viscosity in a second-order Godunov-type hydrodynamic scheme to the
point that converged results can be obtained.  With $\sim 20$ zones
per clump radius our resolution is much lower. Resolution studies are
required in order to decide whether our velocity distributions are
numerically converged. The same holds for answering the question
whether the clumps will get dispersed and mixed with the material of
the ``helium wall'', or whether they can survive as individual
entities. Note also that in the end 3D calculations will be required
since the drag coefficient, $c_D$, and the cross section, $A$, in
Eq.~(\ref{eq:drag}) depend on clump shape, which can lead to
quantitative differences in three-dimensional as compared to our
two-dimensional simulations.  The Rayleigh-Taylor fingers that one
finds in a two-dimensional calculation are comparable to genuine 3D
``mushrooms'' only along the polar axis of the 2D grid. Along the
equator one actually obtains toroidal structures because of the
assumption of axial symmetry \citep[see][and the references
therein]{Kane+00}. Judging from the 2D results of \cite{KMC94} and the
3D calculations of \cite{SN92} for the shock-cloud interaction
problem, we have little doubt, however, that independent of the
dimensionality of the calculation the metals must slow down
appreciably.  We consider this to be a potentially serious problem for
obtaining high velocities for the metals in the H-envelopes of Type~II
supernova models. Efficient re-acceleration mechanisms operating at
later phases of the evolution are currently unknown.

Actually there is no physical reason to expect an acceleration of
matter (relative to the observer's frame) by the instabilities.  The
pressure gradient in the Rayleigh-Taylor unstable layers of the ejecta
is \emph{positive}, i.e. these layers are \emph{decelerated}. In
fact it is this deceleration that leads to the formation of
Rayleigh-Taylor fingers: Since overdense parcels of gas have a larger
momentum than the neighbouring material, they \emph{cannot be slowed
down as efficiently} as the surrounding matter and hence they start to
move outward relative to their ambient medium. During no time,
however, the clumps are accelerated with respect to the observer's
rest frame.  The reason why we obtain high $\rm{^{56}Ni}$ velocities
in the early evolution of our models is the fact that the nickel
clumps decouple from the (continuously decelerated) background flow
and subsequently move ballistically through the ejecta.  By decoupling
from the background, the clumps can nearly conserve their high initial
velocity beyond the time of decoupling and thus move outward in mass
until they become comoving with the outer ejecta. It must be noted,
though, that these high initial $\rm{^{56}Ni}$ velocities are only
obtained if clump formation starts sufficiently early, i.e. within the
first $\sim 10$\,s after bounce, when the velocity in the
$\rm{^{56}Ni}$-rich layers is still high (compare
Fig.~\ref{fig:amra_late_evol_1d}a).  If clump formation is delayed by
a few 100\,s, the $\rm{^{56}Ni}$ velocities have already dropped to
values $< 1500$\,km/s (Fig.~\ref{fig:amra_late_evol_1d}b) and it
becomes impossible to obtain high velocities by ``clump decoupling''
during the subsequent evolution. This is the reason why earlier
investigations of Rayleigh-Taylor mixing in Type~II supernovae, which
started from one-dimensional models 300\,s after bounce, obtained only
low $\rm{^{56}Ni}$ velocities. The only exception were some of the SPH
calculations of \cite{HB92} which showed nickel velocities of $\sim
3000$\,km/s. As stated by the authors, this result is a direct
consequence of an assumed premixing of $\rm{^{56}Ni}$ throughout 75\%
of the metal core of the 1D explosion model of a $20\,{\rm M_{\odot}}$
progenitor, from which they started their 2D calculations at
$t=300$\,s.  In contrast to their approach, in which this (early)
mixing was put in by hand, we have attempted to perform consistent and
continuous high-resolution 2D simulations of the entire shock
propagation through the star. Doing so, we find that even if one
succeeds to obtain high metal velocities during the first minutes of
the explosion, there is no guarantee that these velocities will remain
high throughout the later evolution. As we discussed earlier, a main
obstacle to overcome is the interaction of the metal clumps with the
reverse shock below the He/H interface, which dramatically reduces the
nickel velocities in our simulations. It is not clear to us why this
nickel deceleration is not present in the models of
\citeauthor{HB92}.

\cite{HB92} reported that the energy release due to the radioactive
decay of $\rm{^{56}Ni}$ and $\rm{^{56}Co}$ is inefficient to
significantly boost the nickel velocities over the first few
months. However, further studies of this effect, which should start
from models like ours and use more accurate hydrodynamic schemes and
higher resolution than the calculations of \cite{HB92} are clearly
required.  At present, we have to conclude that high final clump
velocities can only be obtained if the clumps decouple sufficiently
early from the background flow, and if the helium wall and the
associated reverse shock do not form. The latter requires that the
main shock does not decelerate in the hydrogen envelope, i.e. the
density gradient outside the He core has to be steeper than $\propto
r^{-3}$. A situation where this might hold is the case of a Type~Ib
supernova explosion. The progenitors of Type~Ib supernovae are thought
to be stripped He cores that lack a thick hydrogen envelope. In this
case the shock directly enters the tenuous atmosphere of the star once
it leaves the dense part of the He-rich layers.

\section{Beyond the first second: A Type Ib model?}
\label{sect:SN_Ib_model}

\begin{figure}[t]
\resizebox{0.8\hsize}{!}{\includegraphics{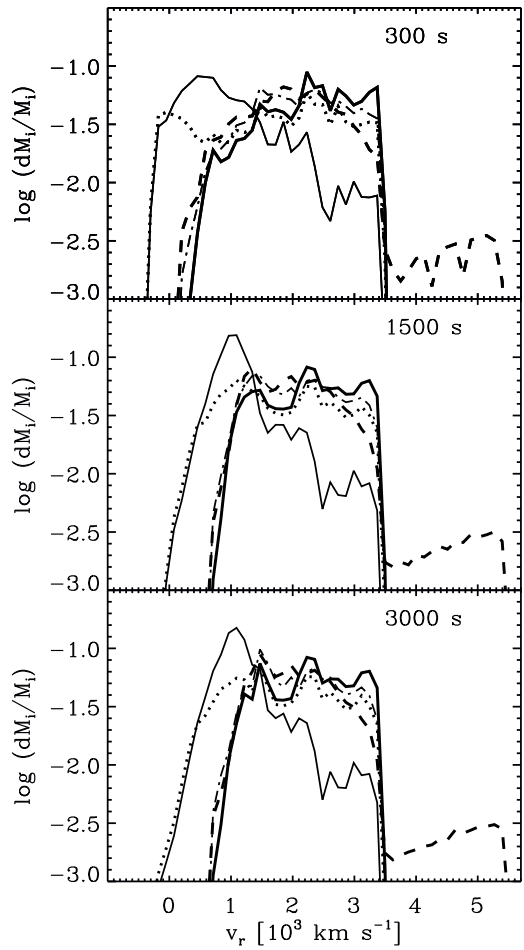}}
\caption[]{Same as Fig.~\ref{fig:amra_massvelo} but for Model T310b
           and times of 300\,s, 1500\,s and 3000\,s after bounce.}
\label{fig:amra_massvelo_Ib}
\end{figure}  

To confirm the effect of a steeper density decline upon clump
propagation we have removed the outer envelope of our progenitor model
and replaced it by power-law profiles for the density and temperature
of the form
\begin{eqnarray}
\rho(r) &=& \rho_c \frac{r_{c}^{n}}{r^{n}}, \\
   T(r) &=& T_c   \frac{r_{c}^{m}}{r^{m}}, 
                   \quad \mbox{for}~r \geq r_c.
\end{eqnarray}
Herein $r_c$ is the radius at which the original model was cut,
$\rho_c$ and $T_c$ are the density and temperature at that radius, and
$n=4.25$ and $m = 1.32$ are the adopted power-law indices for the
density and temperature profiles, respectively. We have chosen $r_c =
5.0\times 10^{10}$\,cm, i.e. we placed the cut still inside the He
core at a mass coordinate $M_r =3.89\,{\rm M_{\odot}}$, while the mass
of the new envelope amounted to $1.2\,{\rm M_{\odot}}$.
\begin{figure}[t]
\centering
\resizebox{0.75\hsize}{!}{\includegraphics{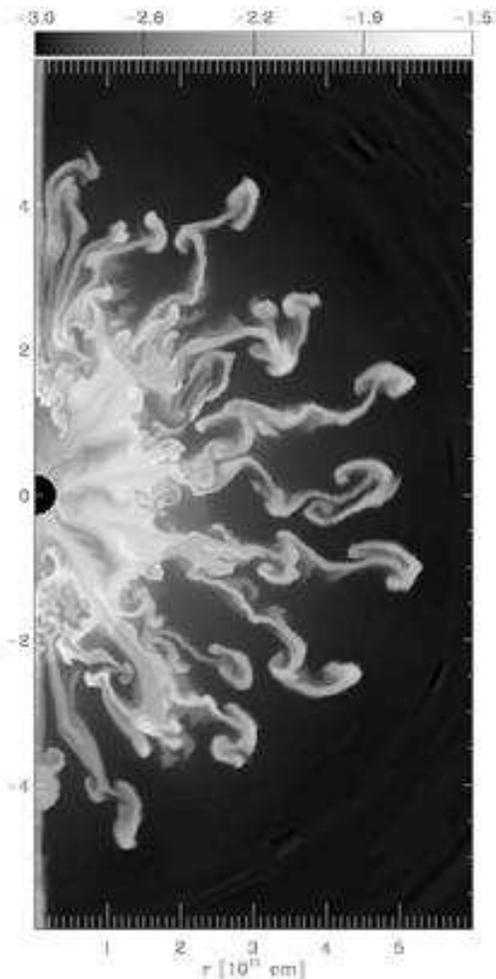}}
\caption[]{Logarithm of the density in Model T310b at a time of
           1500\,s after bounce.}
\label{fig:dens_Ib}
\end{figure}
Note that this simple setup of a ``Type~Ib progenitor'' lacks internal
consistency. It does not take into account that mass-loss due to winds
(or binary interaction) will gradually alter the entire structure of
the star during its evolution and that the density profile at the
onset of core collapse obtained from a consistent stellar evolutionary
calculation will therefore look different from our model.  For this
reason the calculations discussed below can only demonstrate effects
in principle, and the simulations should be repeated with generic
Type~Ib progenitors \citep[e.g.][]{WLW95} to discuss observationally
relevant aspects in detail. In the context of the present work we
refrained from doing so but wished to start with our explosion model
T310 as described in Sect.~\ref{sect:first_sec_2D}, and follow clump
propagation in our truncated progenitor with \textsc{AMRA} until
3000\,s after bounce. At this epoch the expansion is already
self-similar, i.e. the clumps are ``frozen in'' and move with the same
velocity as the ambient medium.

\begin{figure*}[t]
\includegraphics[width=17cm]{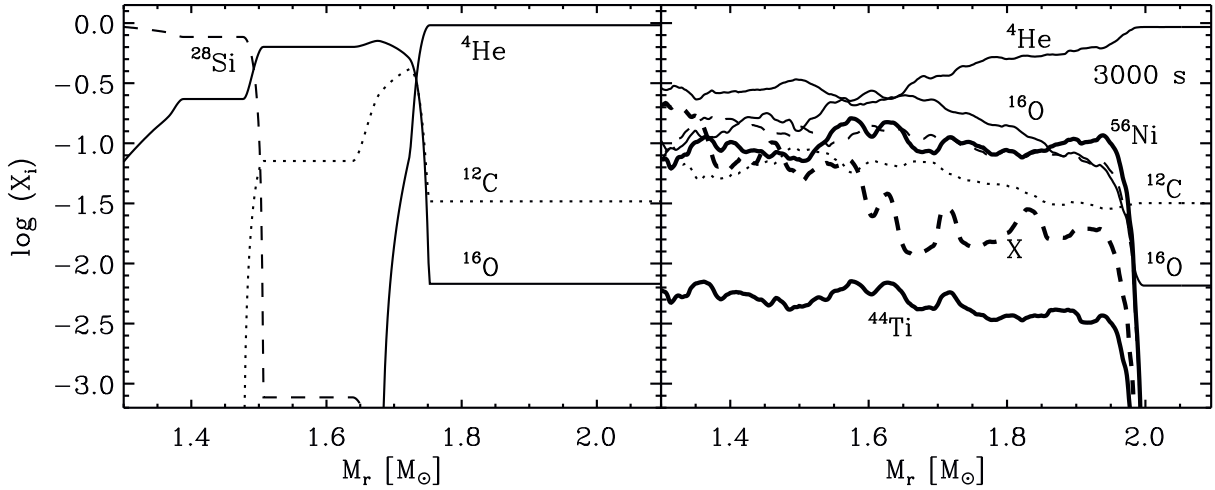}
\caption[]{Same as Fig.~\ref{fig:amra_mixing_310}, but for Model
           T310b. Left: Initial composition. Right: Composition 
           3000\,s after core bounce.}
\label{fig:amra_mixing_310b}
\end{figure*}

The simulation in this section, to which we will henceforth refer to
as Model T310b, was performed with a maximum resolution of $1536
\times 384$ zones. Otherwise we used the same computational strategy
as for Model T310a of Sect.~\ref{sect:AMR_2D}. Not surprisingly, the
evolution in Model T310b proceeded identically to the previous Type~II
supernova model, T310a, for the first $50$\,s, i.e. the time it took
the shock to reach the mass coordinate where we truncated the original
progenitor and where the density in the modified model now starts to
decline steeply. It is remarkable, however, that the velocity profiles
for all chemical elements show only minor differences between Model
T310b (Fig.~\ref{fig:amra_massvelo_Ib}) and Model T310a (compare
Fig.~\ref{fig:amra_massvelo}) even as late as 300\,s after bounce.
This also indicates that the lower resolution used in Model T310b has
only small effects on the overall dynamics. However, it suppresses
smale-scale fragmentation due to Kelvin-Helmholtz instabilities which
is visible in Fig.~\ref{fig:dens_elem_100-1500}c but not in
Fig.~\ref{fig:dens_Ib}. Significant differences between both cases
start to develop only for times later than 300\,s.  As we expected,
the formation of a reverse shock and an associated dense helium shell
is absent in the outer core of Model T310b (compare
Fig.~\ref{fig:dens_Ib} to Fig.~\ref{fig:dens_elem_100-1500}c).
Instead of suffering from a deceleration in such a dense environment
the metal clumps can now expand freely into a tenuous medium, i.e. the
velocity profiles of all chemical elements do not change any more
after 300\,s (Fig.~\ref{fig:amra_massvelo_Ib}). Maximum velocities of
3500\,km/s are obtained for $\rm{^{28}Si}$, $\rm{^{56}Ni}$,
$\rm{^{44}Ti}$, and for the neutronization tracer, while $\rm{^{16}O}$
expands with velocities as high as 5500\,km/s.

Figure~\ref{fig:amra_mixing_310b} displays the extent of the mixing
(averaged over angle) for this model at a time of 3000\,s. The newly
synthesized nuclei are mixed throughout the inner $2.0\,{\rm
M_{\odot}}$ of the star.  Figure~\ref{fig:amra_mixing_310b} should be
compared to the plots of ``Model 4B'' of \cite{WE97}, who computed
synthetic spectra from one-dimensional explosion models of a
$2.3\,{\rm M_{\odot}}$ He core (originating from a $4\,{\rm
M_{\odot}}$ He star) and demonstrated that very good agreement with
observed spectra of SN 1984\,L near maximum light could be
achieved. This was only the case, however, if they artificially mixed
$\rm{^{56}Ni}$ into the helium rich layers assuming an exponential
decline of the $\rm{^{56}Ni}$ mass fraction, $X({\rm ^{56}Ni})$, with
the enclosed mass. The value of $X({\rm ^{56}Ni})$ decreased below
$10^{-3}$ only at a mass coordinate of $\sim 2\,{\rm
M_{\odot}}$. Different from Model 4B of \cite{WE97} our Model T310b
(Fig.~\ref{fig:amra_mixing_310b}) shows a plateau-like distribution
with high nickel mass fractions ($\log X({\rm ^{56}Ni}) \approx -1.0$)
up to the outer edge of the nickel-enriched core.  The latter
coincides with that of the \citeauthor{WE97} model.  Whether this is
sufficient to cause an adequately large flux of ionizing
$\gamma$-photons in the He envelope in case of our more massive star
($5.1\,{\rm M_{\odot}}$ instead of $2.3\,{\rm M_{\odot}}$ for
\citeauthor{WE97}'s ``Model 4B'') can only be decided by calculating
detailed spectra.  Our purpose here is only to demonstrate that
multi-dimensional hydrodynamic models, with a consistent treatment of
the early phases of the explosion, can yield strong mixing of the
metal and He core that is a prerequisite to produce a good match
between calculated and observed spectra and light curves of Type~Ib
supernovae (\citealt{Shigeyama+90}; \citealt{HMNS94};
\citealt{WE97}). \cite{HMNS94} have shown that in order to obtain a
sufficient amount of mixing in multi-dimensional simulations with
artificially triggered Rayleigh-Taylor instabilities, the amplitude of
the seed perturbations must exceed 5\% of the radial expansion
velocity at the time the shock crosses the unstable interfaces in the
metal core.  Our models demonstrate that such a degree of perturbation
can be naturally obtained as a result of neutrino-driven convection
during the first second of the explosion.

A most remarkable feature of Model T310b is the fact that the number
of Rayleigh-Taylor fingers at times $t \geq 1500$\,s is correlated
with (indeed it is essentially equal to) the number of down-flows that
developed in the convective layer of neutrino heating behind the shock
(compare Fig.~\ref{fig:dens_Ib} to Fig.~\ref{fig:odd-even}b).  This
correlation, although initially present, is destroyed in Model T310a,
where the interaction of the clumps with the reverse shock and the
dense helium shell results in a stronger non-linear evolution with
substantial vorticity generation and mixing. As the structure of
down-flows and neutrino-heated bubbles characterizes the mode of the
convective pattern that prevails within the shock-revival phase, it
carries important information about the mechanism that initiates the
explosion.  It would be interesting if our finding is generic, and if
it also holds for cases where the pattern of neutrino-driven
convection develops much larger modes than those obtained in our
simulations. At present, however, one must be cautious not to
overinterpret our result, which is based on a single 2D model. A
confirmation of the existence of a tight correlation between the
patterns of the early post-shock convection and the final
instabilities in the mantle of the exploding star will probably
require well-resolved 3D calculations and parameter studies with
different explosion and progenitor models.

\section{Conclusions}
\label{sect:conclusions}

We have presented a study of hydrodynamic instabilities in a Type~II
and a Type~Ib-like supernova model that has improved upon earlier
simulations by starting the explosion by means of neutrino heating
behind the stalled supernova shock and by following the hydrodynamics
of this explosion well beyond shock-breakout from the stellar
photosphere. The consistency of the hydrodynamic evolution that was
thereby achieved allows for the following conclusions.

\begin{itemize}

\item
We find newly synthesized $\rm{^{56}Ni}$ and the products of
incomplete silicon and oxygen burning to be initially located in a
high-density (low entropy) inhomogeneous shell that forms behind the
outward sweeping shock within the first $\sim 500$\,ms of the
evolution. Depending on the explosion time scale and the strength of
the neutrino-driven convection this shell is either markedly or weakly
deformed by rising, high-entropy bubbles of neutrino-heated matter.
The angular extent of these bubbles is larger for long explosion time
scales (usually leading to smaller explosion energies of our
progenitor) and large seed perturbations.  Besides being formed by
Si-burning $\rm{^{56}Ni}$ is also produced in significant amounts in
the neutrino-heated high-entropy bubbles if their neutronization is
moderate. The latter is sensitive to the $\nu_{\rm e}$ and $\bar
\nu_{\rm e}$ luminosities and spectra imposed at the inner boundary in
our simulations.

\item
The outer boundary of the $\rm{^{56}Ni}$-rich shell is located close
to the Si/O composition interface where a negative density and a
positive pressure gradient exist. This state is unstable to the
Rayleigh-Taylor instability.

\item
For an explosion energy around $1.8\times10^{51}$\,ergs,
$\rm{^{56}Ni}$ is born with velocities $ \leq 17\,000$\,km/s. These
velocities are much larger than those typically observed for the metal
clumps in supernovae, and indicate that a substantial deceleration of
this material must occur when it penetrates through the overlying
layers of the star.

\item
Three Rayleigh-Taylor unstable regions develop in the explosion of the
$15\,{\rm M_{\odot}}$ progenitor model of \cite{WPE88}: one at the
Si/O interface, a second one at the (C+O)/He interface and a third one
at the He/H interface of the star.

\item
Seeded by the perturbations induced by neutrino-driven convection,
Rayleigh-Taylor mixing at the Si/O interface sets in already about 20
seconds after core bounce and leads to a fragmentation of the
$\rm{^{56}Ni}$-enriched shell with the formation of fingers that
distort also the (C+O)/He interface farther out.

\item
Within only about five minutes after bounce the entire metal core of
the \cite{WPE88} star is shredded.

\item
Dense ``bullets'' and clumps of $\rm{^{56}Ni}$, $\rm{^{28}Si}$ and
$\rm{^{16}O}$-rich material decouple from the flow and start to
propagate ballistically through the stellar helium core.

\item
During this phase, metal velocities as high as 3500-5500\,km/s are
observed, and the pattern of the Rayleigh-Taylor structures carries
information about the geometry of the neutrino-driven convection from
the shock-revival phase.

\end{itemize}
The effects discussed above could not be seen in previous studies of
Rayleigh-Taylor instabilities in core collapse supernovae where the
explosion was started by ad hoc energy deposition (using pistons or
thermal bombs). Especially the high $\rm{^{56}Ni}$ velocities, the
strong mixing triggered at the Si/O interface, and the very short
growth time-scale of the instabilities at the Si/O and (C+O)/He
interfaces (minutes as compared to hours) are in clear contrast to
what has been reported in previous studies, especially in those which
dealt with the mixing in Type~II supernovae.  Of all earlier works,
only the Type~Ib studies of \cite{HMNS94} and the SN 1993\,J
simulations of \cite{Iwamoto+97} agree qualitatively with the
evolution that we observe for the first minutes of the explosion. Both
of these studies found, as we do, that Rayleigh-Taylor instabilities
at the (C+O)/He interface of the progenitor stars grow within minutes
after core collapse. However, they did not observe the instability at
the Si/O interface. Moreover, they had to assume large seed
perturbations to induce the mixing, whereas in our case the
perturbations are a natural consequence of neutrino-driven convection.

The crucial difference between our Type~II model T310a and all other
Type~II supernova simulations (except for the work of
\citealt{Iwamoto+97}) is the fact that our metal core starts to
fragment within only about 20\,s after bounce, when the velocity in
the $\rm{^{56}Ni}$-rich layers is still high (see
Fig.~\ref{fig:amra_late_evol_1d}a). This leads to an early decoupling
of the clumps from the background flow, enabling the clumps to
conserve the high velocity that the flow possesses at early times.
Not allowing for early clump formation and decoupling, as it has been
done in nearly all previous multi-dimensional simulations of Type~II
supernovae by starting the calculations hundreds of seconds after core
bounce from 1D initial models, has the effect that the velocities of
the $\rm{^{56}Ni}$-rich layers have dropped to values $\leq
1500$\,km/s at the start of the multi-dimensional calculation
(Fig.~\ref{fig:amra_late_evol_1d}b). It therefore becomes impossible
to obtain high velocities by the Rayleigh-Taylor instabilities during
the subsequent evolution.

In fact, the situation becomes even worse at later times.  We find
that our Model T310a is initially successful in yielding high
$\rm{^{56}Ni}$ velocities: Once the $\rm{^{56}Ni}$ clumps in this
model have decoupled from the flow, they propagate ballistically and
subsonically through the helium core. Due to their subsonic motion,
they do not dissipate much of their kinetic energy and the maximum
$\rm{^{56}Ni}$ velocities remain high at $3500$\,km/s. However, this
changes when the clumps reach the outer helium core and encounter the
reverse shock and the dense (Rayleigh-Taylor unstable) ``wall'' that
are left behind by the main shock below the He/H interface. When the
clumps penetrate through the reverse shock into this wall, 1500\,s
after core bounce, they are decelerated due to supersonic drag to
velocities $< 1500$\,km/s.  An interaction of the clumps with the
reverse shock is also present in data of previous Type~II supernova
simulations \citep{MFA91,MFA2Ds_91}. However, its importance for clump
propagation was not recognized because
\begin{itemize}
\item
due to the late phase when previous multi-dimensional calculations
were started, clump formation at the (C+O)/He interface was only about
to set in at $t \approx 2000$\,s, when the reverse shock had
propagated sufficiently far inward in mass to reach this interface,
\item
the initial clump velocities and the densities in the shell downstream
of the reverse shock were low, leading to less pronounced effects
during the interaction,
\item
and the employed numerical resolution was insufficient to resolve the
non-linear waves that characterize this interaction.
\end{itemize}

A final difference between Model T310a and previous Type~II supernova
simulations is the fact that we do not obtain strong mixing at the
He/H interface. Unless we add artificial perturbations to these
layers, as it was done in all previous studies, the mixing remains
very weak.  This might indicate that neutrino-driven convection is not
able to provide perturbations that are sufficiently strong to cause
the large-scale mixing that was observed in SN 1987\,A. However, final
conclusions regarding this point are not possible before
Rayleigh-Taylor simulations have been performed which start from a
globally aspherical situation, as seen in recent calculations of
neutrino-driven explosions (\citealt{Janka+02}; Scheck et al., in
preparation).

Taken together, the effects revealed by our simulations demonstrate
that the evolution proceeds multi-dimensionally from the earliest
moments. Convective instabilities during the shock-revival phase
cannot be neglected. They lead to very early growth of Rayleigh-Taylor
instabilities and thus to high initial clump velocities what in turn
determines the character of the late-time hydrodynamic evolution. By
starting from 1D initial models hundreds of seconds after core bounce,
all earlier investigations of Rayleigh-Taylor mixing in Type~II
supernovae have therefore missed essential parts of the physics of the
problem.

Supernova 1987\,A still remains an enigma, though. Given the (late)
deceleration of the clumps below the He/H interface in Model T310a, we
wonder whether the conflicting maximum $\rm{^{56}Ni}$ velocities
between our models and the observations point to deficiencies of the
``standard'' blue supergiant progenitors, or to missing physics in our
shock-revival simulations.  There are indications that the progenitor
of SN 1987\,A might have been the result of a merger of two smaller
stars \citep{Podsiadlowski92} and that therefore its structure might
not be accounted for reliably by ``standard'' stellar evolutionary
calculations. Numerical studies of such mergers have recently been
performed by \cite{IPS02}. Although their models show density profiles
in the hydrogen envelope that look similar to the one of our
progenitor (N. Ivanova, private communication), it remains to be seen
in future simulations whether for these models the evolution differs
from what we have reported here.

Sources of uncertainty in our calculations that may affect the
velocities of chemical elements are the still limited numerical
resolution and especially the differences between two-dimensional
versus three-dimensional hydrodynamics. The smaller drag that clumps
experience in three dimensions \citep{Kane+00} may lead to higher
initial clump velocities after the fragmentation of the metal core and
thus to a farther penetration of the clumps through the helium
``wall''. This might result in higher final $\rm{^{56}Ni}$ velocities
than in two dimensions. An extension of our calculations to three
dimensions with a similar resolution as in the 2D case appears hardly
feasible in view of present computer resources.  Further insight
might, however, be gained by laser experiments \citep[][and the
references therein]{Robey+01,Kane+01,Drake+02,Klein+03}, and
hydrodynamic code validation experiments as those described by
\cite{Fishbine02} and \cite{Zoldi_PhD} before well-resolved 3D
hydrodynamic simulations will become available.

There has been much speculation about large-scale anisotropies caused
by jets in the explosion of SN 1987\,A
\citep{Khokhlov+99,Wheeler+02,Wang+02}. However, convincing and
consistent (MHD) calculations of the formation and propagation of such
jets and an associated explosion have not yet been performed.  It is
unclear whether in this scenario nickel velocities can be obtained
that are in agreement with those observed in SN 1987\,A. An initial
anisotropy of the explosion, that might lead to high initial nickel
velocities, would have to work against the same effects in the
envelope that we find to be important in our simulations. The
interaction with the reverse shock and the dense helium shell might
not only slow down fast metal clumps, it might also reduce an initial
anisotropy of the ejecta.  Judging from the experience gained from our
calculations, we doubt that the spatial resolution in simulations of
anisotropic explosions that have been performed to date
\citep{NHSY97,NSS98,Khokhlov+99} was sufficient to study these effects
reliably.  The origin of the prolate deformation of the ejecta of SN
1987\,A \citep{Wang+02} must therefore be regarded to be unknown.  In
fact, there is currently neither observational nor theoretical
evidence that unambiguously demands the assumption of a ``jet-driven''
explosion. The deformation could as well be linked to an anisotropic
initiation of the explosion due to neutrino heating
(\citealt{Janka+02}; Scheck et al., in preparation). On the other hand
it might also be caused by stellar rotation as a result of a merger
history of the progenitor.

While clearly substantial work is required in case of SN 1987\,A, our
``Type~Ib'' Model T310b is in reasonably good agreement with
observations of extragalactic Type~Ib supernovae.  Due to the absence
of dense shell and reverse shock formation in the He core of this
model, the metal-rich clumps are not decelerated once they start to
propagate ballistically through the ambient gas and the final metal
velocities of $3500-5500$\,km/s are sufficiently high.  In addition,
the extent of the mixing in this model is comparable to what
\cite{WE97} had to assume in their ``Model 4B'' to model the spectrum
of SN 1984\,L. This result is very encouraging. It indicates that the
interaction of neutrino-driven convection with the Rayleigh-Taylor
instabilities in the stellar mantle is able to account for important
aspects of the mixing of different elements in Type~Ib supernovae
which are known from spectral and light curve calculations for more
than a decade.

Another interesting aspect of Model T310b is the fact that without the
strong non-linear interaction of the clumps with the reverse shock,
perturbations originating from the shock-revival phase are reflected
in the final flow structures of the Rayleigh-Taylor instabilities.
This may enable one to deduce information about the pattern of the
neutrino-driven convection from observations of the distribution of
metals in the ejecta of Type~Ib supernovae. The possibility to obtain
such a unique piece of evidence from ejecta properties is exciting.
However, before such conclusions can be drawn, our results need to be
ascertained by making use of a larger number of shock-revival
simulations with different properties and a greater range of (Type~Ib
and Type~Ic) progenitors. We believe that in the end very
well-resolved 3D simulations will be required to prove the existence
of such a correlation.  If confirmed it might provide us with a means
to probe the physics linked to the explosion mechanism even for
extragalactic supernovae.

Finally, we emphasize that clump deceleration due to the interaction
with the reverse shock might be crucial for a correct interpretation
of observations of the different supernova types.  It can cause
different final metal clump velocities even if the initiation of the
explosion proceeds similarly in different stars. The latter must
actually be expected for the neutrino-heating mechanism, because the
post bounce models of supernova cores are rather similar, regardless
of whether they originate from Type~II or Type~Ib progenitors (Rampp
et al., in preparation). On the other hand, clump deceleration depends
on the structure of the envelope of the progenitor and a sequence is
conceivable where its importance varies with the type of the
supernova, being strongest in (some fraction of?)  Type~II events and
weaker or absent in Type~Ib and Ic supernovae. Unfortunately,
observational data of metal clump velocities are currently sparse and
hence they do not allow one to test this hypothesis.  Photospheric
velocities, as those published recently by \cite{Hamuy03}, are not
very helpful in this respect because they only probe the outer, faster
expanding layers of the ejecta. To enlarge the data base recourse must
be made to measurements of velocities in supernova remnants.  Clumps
moving with up to $\sim 6000$\,km/s were, however, observed in Cas~A
\citep{vandenBergh71,TFVDB01}, which is probably the remnant of a
Type~Ib explosion \citep{FB91}, and clump velocities of up to $\sim
6000$\,km/s and $\sim 4000$\,km/s were measured shortly after the
explosion in the Type~Ib SN 1987\,F \citep{Filippenko_Sargent89} and
the Type~IIb SN~1993\,J \citep{Spyromilio94}, respectively.  All of
these objects are assumed to be connected to progenitors with small or
lacking hydrogen envelopes. On the other hand \cite{Aschenbach02} and
\cite{AET95} deduce low mean expansion velocities of about 2000\,km/s
for the clumps in the Vela supernova remnant (using their angular
distance to the pulsar, the age of the latter and the distance of the
remnant), that are in reasonably good agreement with our Type~II Model
T310a. In fact, the Vela clumps show Mach cones and appear to move at
present as slow as $\sim 500$\,km/s \citep{Aschenbach02,AET95},
indicating that supersonic drag plays an important role even in the
supernova remnant phase.

\begin{acknowledgements}
We thank S.~E.~Woosley and S.~Bruenn for providing us with the
progenitor and post-bounce models, respectively, that we have used to
construct our initial data, and our referee, C. Fryer, for his
comments on the manuscript. KK and HTJ are grateful for support by the
Sonderforschungsbereich 375 on ``Astroparticle Physics'' of the
Deutsche Forschungsgemeinschaft. The work of TP was supported in part
by the US Department of Energy under Grant No. B341495 to the Center
of Astrophysical Thermonuclear Flashes at the University of Chicago,
and in part by grant 2.P03D.014.19 from the Polish Committee for
Scientific Research.  The simulations were performed on the CRAY J916,
NEC SX-4B and NEC SX-5/3C computers of the Rechenzentrum Garching, and
on the IBM Night Hawk\,II of the Max-Planck-Institut f\"ur
Astrophysik.
\end{acknowledgements}

\bibliography{paper}

\begin{thebibliography}{92}
\expandafter\ifx\csname natexlab\endcsname\relax\def\natexlab#1{#1}\fi

\bibitem[{{Arnett} {et~al.}(1989{\natexlab{a}}){Arnett}, {Bahcall}, {Kirshner},
  \& {Woosley}}]{ABKW89}
{Arnett}, W.~D., {Bahcall}, J.~N., {Kirshner}, R.~P., \& {Woosley}, S.~E.
  1989{\natexlab{a}}, \araa, 27, 629

\bibitem[{{Arnett} {et~al.}(1989{\natexlab{b}}){Arnett}, {Fryxell}, \&
  {M\"uller}}]{AFM89}
{Arnett}, W.~D., {Fryxell}, B., \& {M\"uller}, E. 1989{\natexlab{b}}, \apjl,
  341, L63

\bibitem[{{Aschenbach}(2002)}]{Aschenbach02}
{Aschenbach}, B. 2002, in Neutron Stars, Pulsars, and Supernova Remnants.
  Proceedings of the 270th WE-Heraeus Seminar, ed. W.~Becker, H.~Lesch, \&
  J.~Tr{\"u}mper (Garching: Max-Planck-Institut f\"ur extraterrestrische
  Physik), 13

\bibitem[{{Aschenbach} {et~al.}(1995){Aschenbach}, {Egger}, \&
  {Tr\"umper}}]{AET95}
{Aschenbach}, B., {Egger}, R., \& {Tr\"umper}, J. 1995, \nat, 373, 587

\bibitem[{Berger \& Colella(1989)}]{BC89}
Berger, M.~J. \& Colella, P. 1989, J. Comput. Phys., 82, 64

\bibitem[{Bruenn(1993)}]{Bruenn93}
Bruenn, S.~W. 1993, in Nuclear Physics in the Universe, ed. M.~W. Guidry \&
  M.~R. Strayer (Bristol: IOP), 31

\bibitem[{{Burrows} \& {Fryxell}(1993)}]{Burrows_Fryxell93}
{Burrows}, A. \& {Fryxell}, B.~A. 1993, \apjl, 418, L33

\bibitem[{{Burrows} {et~al.}(1995){Burrows}, {Hayes}, \& {Fryxell}}]{BHF95}
{Burrows}, A., {Hayes}, J., \& {Fryxell}, B.~A. 1995, \apj, 450, 830

\bibitem[{{Chevalier}(1976)}]{Chevalier76}
{Chevalier}, R.~A. 1976, \apj, 207, 872

\bibitem[{Colella \& Glaz(1985)}]{CG85}
Colella, P. \& Glaz, H.~M. 1985, J. Comput. Phys., 59, 264

\bibitem[{Colella \& Woodward(1984)}]{CW84}
Colella, P. \& Woodward, P.~R. 1984, J. Comput. Phys., 54, 174

\bibitem[{{Drake} {et~al.}(2002){Drake}, {Robey}, {Hurricane}, {Zhang},
  {Remington}, {Knauer}, {Glimm}, {Arnett}, {Kane}, {Budil}, \&
  {Grove}}]{Drake+02}
{Drake}, R.~P., {Robey}, H.~F., {Hurricane}, O.~A., {et~al.} 2002, \apj, 564,
  896

\bibitem[{Einfeldt(1988)}]{Einfeldt88}
Einfeldt, B. 1988, SIAM J. Num. Anal., 25, 294

\bibitem[{{Fassia} {et~al.}(1998){Fassia}, {Meikle}, {Geballe}, {Walton},
  {Pollacco}, {Rutten}, \& {Tinney}}]{F+98}
{Fassia}, A., {Meikle}, W. P.~S., {Geballe}, T.~R., {et~al.} 1998, \mnras, 299,
  150

\bibitem[{{Fesen} \& {Becker}(1991)}]{FB91}
{Fesen}, R.~A. \& {Becker}, R.~H. 1991, \apj, 371, 621

\bibitem[{{Filippenko} \& {Sargent}(1989)}]{Filippenko_Sargent89}
{Filippenko}, A.~V. \& {Sargent}, W. L.~W. 1989, \apjl, 345, L43

\bibitem[{{Fishbine}(2002)}]{Fishbine02}
{Fishbine}, B. 2002, Los Alamos Research Quarterly, 3, 6

\bibitem[{{Fryxell} {et~al.}(1991){Fryxell}, {M\"uller}, \& {Arnett}}]{FMA91}
{Fryxell}, B., {M\"uller}, E., \& {Arnett}, W.~D. 1991, \apj, 367, 619

\bibitem[{{Hachisu} {et~al.}(1990){Hachisu}, {Matsuda}, {Nomoto}, \&
  {Shigeyama}}]{HMNS90}
{Hachisu}, I., {Matsuda}, T., {Nomoto}, K.~I., \& {Shigeyama}, T. 1990, \apjl,
  358, L57

\bibitem[{{Hachisu} {et~al.}(1991){Hachisu}, {Matsuda}, {Nomoto}, \&
  {Shigeyama}}]{HMNS91}
---. 1991, \apjl, 368, L27

\bibitem[{{Hachisu} {et~al.}(1992){Hachisu}, {Matsuda}, {Nomoto}, \&
  {Shigeyama}}]{HMNS92}
---. 1992, \apj, 390, 230

\bibitem[{{Hachisu} {et~al.}(1994){Hachisu}, {Matsuda}, {Nomoto}, \&
  {Shigeyama}}]{HMNS94}
---. 1994, \aaps, 104, 341

\bibitem[{{Hamuy}(2003)}]{Hamuy03}
{Hamuy}, M. 2003, \apj, 582, 905

\bibitem[{{Harkness} {et~al.}(1987){Harkness}, {Wheeler}, {Margon}, {Downes},
  {Kirshner}, {Uomoto}, {Barker}, {Cochran}, {Dinerstein}, {Garnett}, \&
  {Levreault}}]{Harkness+87}
{Harkness}, R.~P., {Wheeler}, J.~C., {Margon}, B., {et~al.} 1987, \apj, 317,
  355

\bibitem[{{Herant} \& {Benz}(1991)}]{HB91}
{Herant}, M. \& {Benz}, W. 1991, \apjl, 370, L81

\bibitem[{Herant \& Benz(1992)}]{HB92}
Herant, M. \& Benz, W. 1992, \apj, 387, 294

\bibitem[{{Herant} {et~al.}(1992){Herant}, {Benz}, \& {Colgate}}]{HBC92}
{Herant}, M., {Benz}, W., \& {Colgate}, S. 1992, \apj, 395, 642

\bibitem[{Herant {et~al.}(1994)Herant, Benz, Hix, Fryer, \& Colgate}]{HBFC94}
Herant, M., Benz, W., Hix, W.~R., Fryer, C.~L., \& Colgate, S.~A. 1994, \apj,
  435, 339

\bibitem[{{Herant} \& {Woosley}(1994)}]{HW94}
{Herant}, M. \& {Woosley}, S.~E. 1994, \apj, 425, 814

\bibitem[{{Hillebrandt} \& {H\"oflich}(1989)}]{Hillebrandt_Hoeflich89}
{Hillebrandt}, W. \& {H\"oflich}, P. 1989, Rep. Prog. Phys., 52, 1421

\bibitem[{Hungerford {et~al.}(2003)Hungerford, Fryer, \&
  Warren}]{Hungerford+03}
Hungerford, A.~L., Fryer, C.~L., \& Warren, M.~S. 2003, \apj, submitted

\bibitem[{{Ivanova} {et~al.}(2002){Ivanova}, {Podsiadlowski}, \&
  {Spruit}}]{IPS02}
{Ivanova}, N., {Podsiadlowski}, P., \& {Spruit}, H. 2002, \mnras, 334, 819

\bibitem[{{Iwamoto} {et~al.}(1997){Iwamoto}, {Young}, {Nakasato}, {Shigeyama},
  {Nomoto}, {Hachisu}, \& {Saio}}]{Iwamoto+97}
{Iwamoto}, K., {Young}, T.~R., {Nakasato}, N., {et~al.} 1997, \apj, 477, 865

\bibitem[{Janka {et~al.}(2003)Janka, Buras, Kifonidis, Plewa, \&
  Rampp}]{Janka+02}
Janka, H.-T., Buras, R., Kifonidis, K., Plewa, T., \& Rampp, M. 2003, in From
  Twilight to Highlight: The Physics of Supernovae, ed. W.~Hillebrandt \&
  B.~Leibundgut (Berlin: Springer)

\bibitem[{{Janka} {et~al.}(2001){Janka}, {Kifonidis}, \& {Rampp}}]{JKR01}
{Janka}, H.-T., {Kifonidis}, K., \& {Rampp}, M. 2001, in LNP Vol. 578: Physics
  of Neutron Star Interiors, ed. D.~Blaschke, N.~Glendenning, \& A.~Sedrakian
  (Berlin: Springer), 333

\bibitem[{Janka \& M\"uller(1996)}]{JM96}
Janka, H.-T. \& M\"uller, E. 1996, \aap, 306, 167

\bibitem[{{Kane} {et~al.}(2000){Kane}, {Arnett}, {Remington}, {Glendinning},
  {Baz\a'an}, {M\"uller}, {Fryxell}, \& {Teyssier}}]{Kane+00}
{Kane}, J., {Arnett}, W.~D., {Remington}, B.~A., {et~al.} 2000, \apj, 528, 989

\bibitem[{{Kane} {et~al.}(2001){Kane}, {Robey}, {Remington}, {Drake}, {Knauer},
  {Ryutov}, {Louis}, {Teyssier}, {Hurricane}, {Arnett}, {Rosner}, \&
  {Calder}}]{Kane+01}
{Kane}, J.~O., {Robey}, H.~F., {Remington}, B.~A., {et~al.} 2001, \pre, 63,
  55401

\bibitem[{{Keil} {et~al.}(1996){Keil}, {Janka}, \& M{\"u}ller}]{Keil+96}
{Keil}, W., {Janka}, H.-T., \& M{\"u}ller, E. 1996, \apjl, 473, L111

\bibitem[{{Khokhlov} {et~al.}(1999){Khokhlov}, {H{\" o}flich}, {Oran},
  {Wheeler}, {Wang}, \& {Chtchelkanova}}]{Khokhlov+99}
{Khokhlov}, A.~M., {H{\" o}flich}, P.~A., {Oran}, E.~S., {et~al.} 1999, \apjl,
  524, L107

\bibitem[{{Kifonidis} {et~al.}(2000){Kifonidis}, {Plewa}, {Janka}, \&
  {M\"uller}}]{Kifonidis+00}
{Kifonidis}, K., {Plewa}, T., {Janka}, H.-T., \& {M\"uller}, E. 2000, \apjl,
  531, L123

\bibitem[{{Kifonidis} {et~al.}(2001){Kifonidis}, {Plewa}, \& {M{\"
  u}ller}}]{KPM01}
{Kifonidis}, K., {Plewa}, T., \& {M{\" u}ller}, E. 2001, in AIP Conf. Proc.
  561: Tours Symposium on Nuclear Physics IV, ed. M.~Arnould, M.~Lewitowicz,
  Y.~T. Oganessian, H.~Akimune, M.~Ohta, H.~Utsunomiya, T.~Wada, \& T.~Yamagata
  (Melville, New York: American Institute of Physics), 21

\bibitem[{{Klein} {et~al.}(2003){Klein}, {Budil}, {Perry}, \&
  {Bach}}]{Klein+03}
{Klein}, R.~I., {Budil}, K.~S., {Perry}, T.~S., \& {Bach}, D.~R. 2003, \apj,
  583, 245

\bibitem[{{Klein} {et~al.}(1994){Klein}, {McKee}, \& {Colella}}]{KMC94}
{Klein}, R.~I., {McKee}, C.~F., \& {Colella}, P. 1994, \apj, 420, 213

\bibitem[{LeVeque(1998)}]{LeVeque}
LeVeque, R.~J. 1998, in Computational Methods for Astrophysical Fluid Flow, ed.
  O.~Steiner \& A.~Gautschy (Berlin: Springer), 1

\bibitem[{Liou(2000)}]{Liou2000}
Liou, M.-S. 2000, J. Comput. Phys., 160, 623

\bibitem[{Lucy(1991)}]{Lucy91}
Lucy, L. 1991, \apj, 383, 308

\bibitem[{{McCray}(1993)}]{MCCray93}
{McCray}, R. 1993, \araa, 31, 175

\bibitem[{{Mezzacappa} {et~al.}(1998){Mezzacappa}, {Calder}, {Bruenn},
  {Blondin}, {Guidry}, {Strayer}, \& {Umar}}]{Mezzacappa+98}
{Mezzacappa}, A., {Calder}, A.~C., {Bruenn}, S.~W., {et~al.} 1998, \apj, 495,
  911

\bibitem[{Miller {et~al.}(1993)Miller, Wilson, \& Mayle}]{Miller+93}
Miller, D.~S., Wilson, J.~R., \& Mayle, R.~W. 1993, \apj, 415, 278

\bibitem[{{Mitchell} {et~al.}(2001){Mitchell}, {Baron}, {Branch}, {Lundqvist},
  {Blinnikov}, {Hauschildt}, \& {Pun}}]{MBB01}
{Mitchell}, R.~C., {Baron}, E., {Branch}, D., {et~al.} 2001, \apj, 556, 979

\bibitem[{M{\"u}ller(1998)}]{Mueller98}
M{\"u}ller, E. 1998, in Computational Methods for Astrophysical Fluid Flow, ed.
  O.~Steiner \& A.~Gautschy (Berlin: Springer), 371

\bibitem[{{M\"uller} {et~al.}(1991{\natexlab{a}}){M\"uller}, {Fryxell}, \&
  {Arnett}}]{MFA2Ds_91}
{M\"uller}, E., {Fryxell}, B., \& {Arnett}, W.~D. 1991{\natexlab{a}}, in
  Supernova 1987\,A and Other Supernovae, ed. J.~Danziger \& K.~Kj\"ar
  (Garching: European Southern Observatory), 99

\bibitem[{{M\"uller} {et~al.}(1991{\natexlab{b}}){M\"uller}, {Fryxell}, \&
  {Arnett}}]{MFA91}
{M\"uller}, E., {Fryxell}, B., \& {Arnett}, W.~D. 1991{\natexlab{b}}, \aap,
  251, 505

\bibitem[{{M\"uller} {et~al.}(1991{\natexlab{c}}){M\"uller}, {Fryxell}, \&
  {Arnett}}]{MFA3D_91}
{M\"uller}, E., {Fryxell}, B., \& {Arnett}, W.~D. 1991{\natexlab{c}}, in Elba
  Workshop on Chemical and Dynamical Evolution of Galaxies, ed. F.~Federini,
  J.~Franco, \& F.~Matteucci (Pisa: ETS Editrice), 394

\bibitem[{M{\"u}ller \& Steinmetz(1995)}]{MS95}
M{\"u}ller, E. \& Steinmetz, M. 1995, Comput. Phys. Commun., 89, 45

\bibitem[{{Nagataki} {et~al.}(1997){Nagataki}, {Hashimoto}, {Sato}, \&
  {Yamada}}]{NHSY97}
{Nagataki}, S., {Hashimoto}, M.-A., {Sato}, K., \& {Yamada}, S. 1997, \apj,
  486, 1026

\bibitem[{{Nagataki} {et~al.}(1998){Nagataki}, {Shimizu}, \& {Sato}}]{NSS98}
{Nagataki}, S., {Shimizu}, T.~M., \& {Sato}, K. 1998, \apj, 495, 413

\bibitem[{{Nomoto} {et~al.}(1988){Nomoto}, {Shigeyama}, {Kumagai}, \&
  {Hashimoto}}]{Nomoto+88}
{Nomoto}, K.~I., {Shigeyama}, T., {Kumagai}, S., \& {Hashimoto}, M.-A. 1988,
  Proceedings of the Astronomical Society of Australia, 7, 490

\bibitem[{{Nomoto} {et~al.}(1994){Nomoto}, {Shigeyama}, {Kumagai}, {Yamaoka},
  \& {Suzuki}}]{Nomoto+94}
{Nomoto}, K.~I., {Shigeyama}, T., {Kumagai}, S., {Yamaoka}, H., \& {Suzuki}, T.
  1994, in {Supernovae, Les Houches Session LIV}, ed. S.~A. {Bludman},
  R.~{Mochkovitch}, \& J.~{Zinn-Justin} (Amsterdam: Elsevier/North-Holland)

\bibitem[{Plewa \& M\"uller(1999)}]{PM99}
Plewa, T. \& M\"uller, E. 1999, \aap, 342, 179

\bibitem[{Plewa \& M\"uller(2001)}]{PM01}
---. 2001, Comput. Phys. Commun., 138, 101

\bibitem[{{Podsiadlowski}(1992)}]{Podsiadlowski92}
{Podsiadlowski}, P. 1992, \pasp, 104, 717

\bibitem[{Quirk(1994)}]{Quirk94}
Quirk, J.~J. 1994, Int. J. Num. Meth. Fluids, 18, 555

\bibitem[{Quirk(1997)}]{Quirk94_reprinted}
Quirk, J.~J. 1997, in Upwind and High-Resolution Schemes, ed.
  M.~Yousuff~Hussaini, B.~van Leer, \& J.~Van~Rosendale (Berlin: Springer), 550

\bibitem[{{Rampp} \& {Janka}(2002)}]{Rampp_Janka02}
{Rampp}, M. \& {Janka}, H.-T. 2002, \aap, 396, 361

\bibitem[{{Robey} {et~al.}(2001){Robey}, {Kane}, {Remington}, {Drake},
  {Hurricane}, {Louis}, {Wallace}, {Knauer}, {Keiter}, {Arnett}, \&
  {Ryutov}}]{Robey+01}
{Robey}, H.~F., {Kane}, J.~O., {Remington}, B.~A., {et~al.} 2001, Physics of
  Plasmas, 8, 2446

\bibitem[{Sedov(1959)}]{Sedov}
Sedov, L.~I. 1959, Similarity and dimensional methods in mechanics (London:
  Infosearch)

\bibitem[{{Shigeyama} {et~al.}(1996){Shigeyama}, {Iwamoto}, {Hachisu},
  {Nomoto}, \& {Saio}}]{SIHNS96}
{Shigeyama}, T., {Iwamoto}, K., {Hachisu}, I., {Nomoto}, K.~I., \& {Saio}, H.
  1996, in Supernovae and supernova remnants. Proceedings of the International
  Astronomical Union Colloquium 145, ed. R.~McCray \& Z.~Wang (Cambridge; UK:
  Cambridge University Press), 129

\bibitem[{{Shigeyama} {et~al.}(1990){Shigeyama}, {Nomoto}, {Tsujimoto}, \&
  {Hashimoto}}]{Shigeyama+90}
{Shigeyama}, T., {Nomoto}, K.~I., {Tsujimoto}, T., \& {Hashimoto}, M.-A. 1990,
  \apjl, 361, L23

\bibitem[{{Spyromilio}(1991)}]{Spyromilio91}
{Spyromilio}, J. 1991, \mnras, 253, 25P

\bibitem[{{Spyromilio}(1994)}]{Spyromilio94}
---. 1994, \mnras, 266, L61

\bibitem[{{Stone} \& {Norman}(1992)}]{SN92}
{Stone}, J.~M. \& {Norman}, M.~L. 1992, \apjl, 390, L17

\bibitem[{{Swartz} {et~al.}(1993){Swartz}, {Filippenko}, {Nomoto}, \&
  {Wheeler}}]{Swartz+93}
{Swartz}, D.~A., {Filippenko}, A.~V., {Nomoto}, K.~I., \& {Wheeler}, J.~C.
  1993, \apj, 411, 313

\bibitem[{Thielemann {et~al.}(1996)Thielemann, Nomoto, \& Hashimoto}]{TNH96}
Thielemann, F.-K., Nomoto, K.~I., \& Hashimoto, M. 1996, \apj, 460, 408

\bibitem[{{Thorstensen} {et~al.}(2001){Thorstensen}, {Fesen}, \& {van den
  Bergh}}]{TFVDB01}
{Thorstensen}, J.~R., {Fesen}, R.~A., \& {van den Bergh}, S. 2001, \aj, 122,
  297

\bibitem[{{van den Bergh}(1971)}]{vandenBergh71}
{van den Bergh}, S. 1971, \apj, 165, 457

\bibitem[{Van~Riper(1979)}]{VanRiper79}
Van~Riper, K.~A. 1979, \apj, 232, 558

\bibitem[{{Wang} \& {Hu}(1994)}]{WH94}
{Wang}, L. \& {Hu}, J. 1994, \nat, 369, 380

\bibitem[{{Wang} {et~al.}(2002){Wang}, {Wheeler}, {H{\" o}flich}, {Khokhlov},
  {Baade}, {Branch}, {Challis}, {Filippenko}, {Fransson}, {Garnavich},
  {Kirshner}, {Lundqvist}, {McCray}, {Panagia}, {Pun}, {Phillips}, {Sonneborn},
  \& {Suntzeff}}]{Wang+02}
{Wang}, L., {Wheeler}, J.~C., {H{\" o}flich}, P., {et~al.} 2002, \apj, 579, 671

\bibitem[{{Wheeler} {et~al.}(2002){Wheeler}, {Meier}, \& {Wilson}}]{Wheeler+02}
{Wheeler}, J.~C., {Meier}, D.~L., \& {Wilson}, J.~R. 2002, \apj, 568, 807

\bibitem[{{Witti} {et~al.}(1994){Witti}, {Janka}, \& {Takahashi}}]{Witti+94}
{Witti}, J., {Janka}, H.-T., \& {Takahashi}, K. 1994, \aap, 286, 841

\bibitem[{{Wooden}(1997)}]{Wooden97}
{Wooden}, D.~H. 1997, in Astrophysical Implications of the Laboratory Study of
  Presolar Materials, ed. T.~J. Bernatowicz \& E.~Zinner (Woodbury: American
  Institute of Physics), 317

\bibitem[{Woosley \& Eastman(1997)}]{WE97}
Woosley, S.~E. \& Eastman, R. 1997, in Thermonuclear Supernovae, ed.
  P.~Ruiz-LaPuente, R.~Canal, \& J.~Isern (Dordrecht: Kluwer), 821

\bibitem[{{Woosley} {et~al.}(1994){Woosley}, {Eastman}, {Weaver}, \&
  {Pinto}}]{WEWP94}
{Woosley}, S.~E., {Eastman}, R.~G., {Weaver}, T.~A., \& {Pinto}, P.~A. 1994,
  \apj, 429, 300

\bibitem[{Woosley {et~al.}(1997)Woosley, Heger, Weaver, \& Langer}]{WHLW97}
Woosley, S.~E., Heger, A., Weaver, T.~A., \& Langer, N. 1997, MPA Preprint,
  1024

\bibitem[{{Woosley} {et~al.}(1995){Woosley}, {Langer}, \& {Weaver}}]{WLW95}
{Woosley}, S.~E., {Langer}, N., \& {Weaver}, T.~A. 1995, \apj, 448, 315

\bibitem[{{Woosley} {et~al.}(1988){Woosley}, {Pinto}, \& {Ensman}}]{WPE88}
{Woosley}, S.~E., {Pinto}, P.~A., \& {Ensman}, L. 1988, \apj, 324, 466

\bibitem[{{Woosley} \& {Weaver}(1995)}]{Woosley_Weaver95}
{Woosley}, S.~E. \& {Weaver}, T.~A. 1995, \apjs, 101, 181

\bibitem[{{Yamada} \& {Sato}(1990)}]{YS90}
{Yamada}, S. \& {Sato}, K. 1990, \apjl, 358, L9

\bibitem[{{Yamada} \& {Sato}(1991)}]{YS91}
---. 1991, \apj, 382, 594

\bibitem[{Zoldi(2002)}]{Zoldi_PhD}
Zoldi, C. 2002, PhD thesis, State University of New York at Stony Brook

\end{thebibliography}

\end{document}